%% file: main.tex
\documentclass[twocolumn]{aastex63}
\usepackage{verbatim}
\usepackage{comment} 
\usepackage{amsmath}
\include{definitions}

\usepackage{enumitem}
\makeatletter
\def\@hex@@Hex#1%
 {\if a#1A\else \if b#1B\else \if c#1C\else \if d#1D\else
  \if e#1E\else \if f#1F\else #1\fi\fi\fi\fi\fi\fi \@hex@Hex}
\makeatother

\newcommand{\sfe}{\epsilon_{\star}}
\newcommand{\yd}{y_d}
\newcommand{\MUV}{M_{\rm UV}}

\revised{\today}
\shorttitle{}
\shortauthors{}
\graphicspath{{./}{figures/}}

\begin{document}


\title{Blue Monsters and Dusty Descendants: Reconciling UV and IR Emission from Galaxies from $z\sim7$, up to $z=14$}

\shorttitle{Blue Monsters and Dusty Descendants}
\shortauthors{Sommovigo et al.}

\correspondingauthor{Laura Sommovigo}
\email{laura.sommovigo.work@gmail.com}

\author[0000-0002-2906-2200]{Laura Sommovigo}
\affiliation{Department of Astronomy, Columbia University, 550 W 120th St, New York, NY 10025, USA}
\affiliation{Center for Computational Astrophysics, Flatiron Institute, 162 5th Ave, New York, NY 10010, USA}

\author[0000-0002-0041-4356]{Lachlan Lancaster}
\altaffiliation{Simons Fellow}
\affiliation{Center for Computational Astrophysics, Flatiron Institute, 162 5th Ave, New York, NY 10010, USA}
\affiliation{Department of Astronomy, Columbia University, 550 W 120th St, New York, NY 10025, USA}
\author[0000-0001-5944-291X]{Shyam H. Menon}
\affiliation{Center for Computational Astrophysics, Flatiron Institute, 162 5th Ave, New York, NY 10010, USA}
\affiliation{Department of Physics and Astronomy, Rutgers University, 136 Frelinghuysen Road, Piscataway, NJ 08854, USA}

\author[0000-0002-5157-9222]{Joseph A. O'Leary}
\affiliation{Carl Zeiss AG, Kistlerhofstrasse 75, 81379 Munich, Germany}

\author[0000-0002-6748-6821]{Rachel S. Somerville}
\affiliation{Center for Computational Astrophysics, Flatiron Institute, 162 5th Ave, New York, NY 10010, USA}
\affiliation{Space Telescope Science Institute, 3700 San Martin Drive, Baltimore, MD 21218, USA}

\author[0000-0003-2630-9228]{Greg L. Bryan}
\affiliation{Department of Astronomy, Columbia University, 550 W 120th St, New York, NY 10025, USA}

\begin{abstract}
Recent \textit{JWST} observations reveal massive, UV-bright galaxies at $z>10$ with little apparent dust attenuation, whereas \textit{ALMA} detections at $z\simeq7$ show similarly massive systems that are already dust-rich and IR-luminous. This raises a fundamental question: can a single physical model for star formation and dust production explain both populations across cosmic time?
We address this using a minimal, physically motivated framework with only two free parameters—the instantaneous star formation efficiency ($\sfe$) and the dust yield per Type~II supernova ($y_{\rm d}$)—and predict the rest-frame UV and IR luminosity functions (LFs) from $z\simeq14$ to 7. For a uniform ISM, we find a UV--IR tension at the bright-end of the LFs at $z\!\ge\!7$. The UV LF requires low dust yields ($y_{\rm d}\!\lesssim\!0.01\,{\rm M_\odot}$), while the $z=7$ IRLF require high yields ($y_{\rm d}\!\sim\!0.1\,{\rm M_\odot}$) unless the star formation efficiency is boosted above $\sfe\!\approx\!5$--10\%.
We show that incorporating a porous, turbulent ISM largely resolves this tension: turbulence opens low--column-density sightlines that enhance the UV escape fraction while leaving the total absorbed energy---and thus the IR luminosity---nearly unchanged once radiative--transfer--induced flattening of the attenuation curve is included. Large-grain dust distributions, while reducing UV opacity, become secondary once ISM porosity and radiative transfer are taken into account.
At $z>10$, however, even strong turbulence cannot reproduce the bright-end of the UVLF at high dust yield. This could be resolved by efficient dust removal in early massive systems or substantial ISM dust growth by $z\simeq7$. Our results highlight dust physics as a key lever for interpreting the rapidly growing UV and IR observational constraints within the broader context of early galaxy formation.
\end{abstract}

\section{Introduction}

In just two years of operations, the \textit{James Webb Space Telescope} (\textit{JWST}) has transformed our view of the early Universe. Deep NIRCam surveys have uncovered large populations of massive, UV-bright galaxies at $z>10$ \citep[e.g.][]{Naidu22, Arrabal23, Hsiao23, Wang23, Fujimoto23b, Atek22, Curtis23, Robertson23, Bunker23, Tacchella23, Finkelstein23, Castellano24, Zavala24, Helton24, Carniani24a, Robertson24}, including a record-holder at $z=14.44$ \citep{Naidu25}. These \quotes{Blue Monsters} \citep{Ziparo23}, characterized by compact morphologies ($r_{\rm e}\!\lesssim\!200$ pc; \citealt{Baggen_2023, Morishita_2024, Carniani24a}), large stellar masses ($M_\star\!\sim\!10^9\,M_\odot$), and steep UV slopes ($\beta_{\rm UV}\!\lesssim\!-2.2$; \citealt{Topping_2022, Cullen_2024, Morales_2024}), appear far more numerous and luminous than predicted by pre-\textit{JWST} galaxy formation models.

This apparent \quotes{overabundance problem} has spurred an intense debate about whether our understanding of early galaxy formation—or the underlying cosmological framework itself—requires revision. While some works have speculated about exotic solutions such as alternative cosmologies \citep[see e.g.][]{Boylan23} or highly accreting and thus UV luminous primordial black holes \citep{Liu23,Matteri25}, most recent studies have reduced the extent of the too many, too blue $z>10$ galaxies tension, finding that it might be the outcome of baryonic physics in the distinct physical conditions realized in high-z galaxies. Several mechanisms have been proposed to boost the LFs of early galaxies: (i) high star formation efficiencies (SFE) driven by dense gas reservoirs \citep{Dekel23, Somerville25}; (ii) bursty or stochastic star formation histories (SFHs) that temporarily elevate UV luminosities \citep{Mason23, Sun23, Pallottini23,Basu25}; (iii) top-heavy stellar initial mass functions (IMFs) increasing the UV output per unit SFR \citep{Inayoshi22, Wang_2023, Trinca24, Yung24}; or (iv) dust-poor conditions (\citealt{Ferrara23a, Ziparo23, Ferrara24a, Ferrara24b}), where radiation pressure efficiently expels dust and lowers effective attenuation even in metal-rich systems. In all cases, the emerging consensus is that bright UV emission at $z>10$ reflects either exceptionally (but not unphysically, \citealt{Yung25}) efficient and/or unobscured star formation.

A handful of the brightest JWST spectroscopically confirmed sources at $z>10$ have been followed up with \textit{ALMA}. Several have yielded detections of bright far-infrared fine-structure lines—[O\,\textsc{iii}]\,88\,$\mu$m in particular \citep{Fujimoto23b,Popping22,Bakx22}, and one case of [C\,\textsc{ii}]\,158\,$\mu$m \citep{Schouws24,Carniani24_z14}—but no dust-continuum detections to date. The current redshift record holder for dust continuum detection is Y1 at $z=8.3$ \citep{Bakx25,Bakx20,Tamura19}, which crucially shows little dust content and was selected as UV bright, and yet has been recently confirmed to be classifiable as an ultra luminous infrared galaxy (the earliest discovered so far, see \citealt{Bakx25}) due to its hot dust temperature $T_{\rm d}=90\ \mathrm{K}$ (reminder: $L_{\rm IR}\propto M_{\rm d}\ T_{d}^6$ in a simple greybody model). 
Further, \citet{Bakx25_PIXIE} stacks nearly 200 hours of \textit{ALMA} and NOEMA observations across ten $z>8$ galaxies, reaching 1$\sigma$ depths of 1.5–2\,$\mu$Jy at rest-frame 88–158\,$\mu$m. No significant dust emission is detected, implying stringent $3\sigma$ limits of $M_{\rm d}<10^5\,M_\odot$ and $M_{\rm d}/M_\star<4\times10^{-4}$ (for $T_{\rm d}=50$\,K and $\beta_{\rm d}=2.0$). This stacking experiment suggests inefficient dust buildup in the $z>8$ Universe, consistent with the blue rest-UV slopes inferred from JWST data.

At first glance, this is consistent with the young ages of these systems and their early cosmic epochs. However, \textit{ALMA} observations have revealed a more complex picture at slightly later times. At $z\!\simeq\!7$, just $\sim 300\ \mathrm{Myr}$ later, deep millimeter surveys such as \textit{REBELS} \citep{Inami22}, have shown that massive galaxies ($M_{\star}>10^{8.5}\ \mathrm{M_{\odot}}$, \citealt{Topping_2022}) can already be as dust-rich and IR-luminous as local galaxies \citep{Rowland25,Algera25}, with dust-to-stellar mass ratios of $M_{\rm d}/M_\star\!\sim\!0.01$ and $L_{\rm IR}\!\gtrsim\!10^{11.5}L_\odot$ \citep{Algera25,Sommovigo22_z7}. The combination of high metal content, intense IR emission, and relatively blue UV slopes suggests complex dust geometries or turbulent, multiphase ISM structures in which UV light can escape through low-density channels \citep{Sommovigo20, Sommovigo22_z57, Ferrara22a,Dayal22,Menon_25}. These findings highlight that, while early systems may be compact, they are not necessarily free of metals and dust.
This growing sample underscores the need for a physically consistent picture that simultaneously explains the lack of dust attenuation/emission at cosmic dawn, but also the emergence of dusty and metal-rich, yet still UV-bright sources at the epoch of Reionization. 

In parallel, a new generation of joint \textit{ALMA}–\textit{JWST} programs aims to bridge this gap by characterizing both the obscured and unobscured star-formation channels across redshift. At $z\!\simeq\!4$–7, programs such as \textit{REBELS–IFU} \citep{Rowland25, Fisher25, Algera25}, and \textit{ALPINE-CRISTAL–JWST} \citep{Faisst25} have demonstrated the power of combining rest-frame optical and FIR data to constrain ISM enrichment, geometry, and feedback. Extending this approach to earlier epochs, the recently approved \textit{ALMA} Large Program \textit{PHOENIX} (2025.1.01606.L; PI S.~Schouws) will target the diagnostic [O\,\textsc{iii}]\,88\,$\mu$m line and dust continuum emission in 15 intrinsically luminous galaxies at $8<z<15$, promising a leap forward in our understanding of dust and metal build-up at cosmic dawn. 
The synergy between all the previously listed programs emphasizes the urgency of developing theoretical models that remain predictive across cosmic time and wavelength, avoiding fine-tuned prescriptions that explain one redshift regime at the expense of another.

In this context, we develop a minimal, physically motivated framework to connect the dust-poor, UV-bright galaxies at $z>10$ to the dust-rich, IR-luminous systems observed at $z\simeq7$. 
Our approach allows us to assess whether a single set of parameters $(\sfe, y_{\rm d})$ can simultaneously explain the apparent lack of dust attenuation at $z>10$ and the strong IR emission seen at $z\simeq7$. By confronting the model with multiwavelength observables, we probe the roles of ISM turbulence, grain size, and dust yield in governing the visibility of the earliest galaxies. Our analysis provides a quantitative framework to interpret the forthcoming PHOENIX dataset, as well as future large ALMA and \textit{JWST} observations targeting massive systems at the epoch of Reionization, and thus the transition from the first UV-luminous to dust-enshrouded galaxies.

The paper is organized as follows. In Section~\ref{sect:model_mass}, we present our baseline model for galaxy mass build–up and intrinsic properties.  Section~\ref{sect:emission_model} describes the emission modeling, with a focus on the treatment of dust in Section~\ref{subsec:dust}.  In Section~\ref{sect:turbulent_ism} we introduce a minimal prescription for turbulence to account for ISM porosity, and its effect on attenuation. We then combine all model variations in Section~\ref{sect:uv_ir_lfs} to compute the UV and IR luminosity functions under the different dust and ISM assumptions.  
Finally, we discuss caveats, limitations, and future extensions in 
Section~\ref{sect:Cav_Disc}, and summarize our conclusions in 
Section~\ref{sect:summary}.

\begin{figure}
    \centering
    \includegraphics[width=0.495\textwidth]{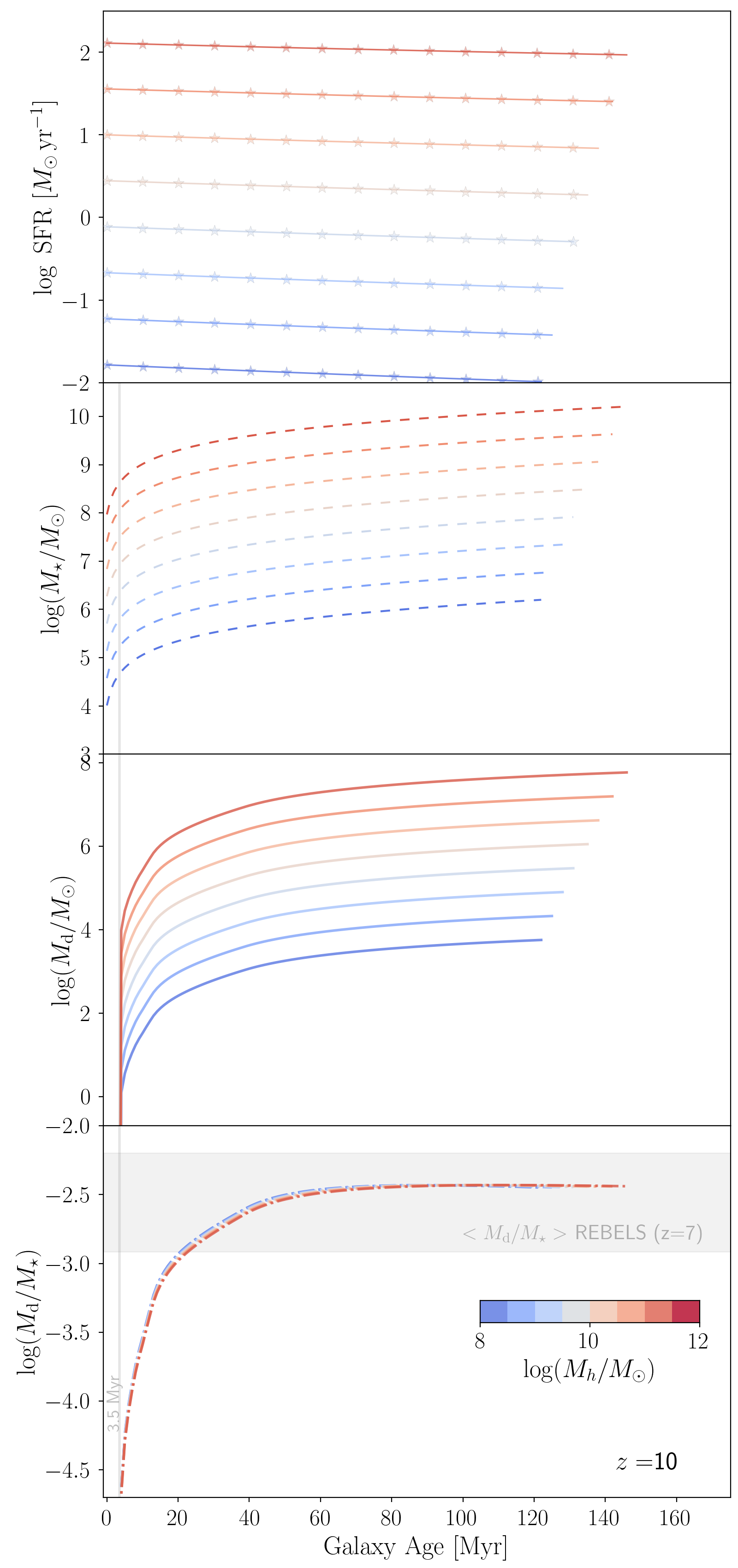}
    \caption{From top to bottom: Logarithm of the SFR, stellar (dashed), dust mass (solid), and dust to stellar mass ratio build up as a function of galaxy age for halos in the mass range $10^7-10^{12}\ \mathrm{M_{\odot}}$ at $z=10$ (see colorbar) according to the minimal model described in eq.~\ref{som22} - \ref{md}. Here we assume $\epsilon_{\star}=0.1$ and $\yd=0.1\ \mathrm{M_{\odot}}$ per SN event. The grey shaded band in the bottom panel shows the dust-to-stellar mass ratio compared to that inferred for REBELS galaxies at $z=7$ in \citet{Sommovigo22_z7}. The vertical line across the second to last panel marks 3.5 Myr.}
    \label{Fig1}
\end{figure}

\section{A Mass Model for $z\sim 10$ Galaxies} \label{sect:model_mass}
Following the approach in \citet{Sommovigo22_z7}, at a given redshift $z$ we express the stellar mass $M_{\star}$ and instantaneous SFR of a galaxy in terms of its halo mass $M_h$ and mean dark matter accretion rate $\langle dM_h/dt \rangle$:
\begin{align}
        M_{\star} & = f_{\rm b}\, \epsilon_{\star,{\rm int}}\, M_h, \label{som22}\\
        {\rm SFR} & = f_{\rm b}\, \epsilon_{\star}\, \left\langle \frac{dM_h}{dt} \right\rangle, \label{sfr}
\end{align}
where $\epsilon_{\star}$ and $\epsilon_{\star,{\rm int}}$ are the \emph{instantaneous} and \emph{integrated} star formation efficiencies, respectively, and $f_{\rm b}=\Omega_{\rm b}/\Omega_{\rm m}$ is the cosmic baryon fraction.

In this first implementation, we assume a \emph{$\delta$-function switch-on} of star formation: the efficiency $\epsilon_{\star}$ rises instantaneously from zero to a constant value at a specified time (corresponding to the oldest time bin in the SFH; see details below). This ensures a direct mapping between the halo’s instantaneous accretion rate and its current SFR, while the integrated efficiency $\epsilon_{\star,{\rm int}}$ reflects the cumulative conversion of accreted baryons into stars since that formation time. 

We use the mean halo accretion rate $\langle dM_h/dt\rangle$ derived from the dark-matter–only \textit{Gadget at Ultrahigh Redshift with Extra-Fine Timesteps} (\textsc{GUREFT}) cosmological simulations, as fit by \citet{Yung23}:
\begin{equation}\label{dmdt}
    \langle\frac{dM_h}{dt}\rangle  = \beta(z)\ \left( M_{\rm 12}\ E(z) \right)^{\alpha(z)}\  \frac{M_{\odot}}{\rm yr},
\end{equation}
where $M_{\rm 12}=M_h/10^{12} M_\odot$, 
\begin{align*}
    \log \beta (z) & = 2.578 - 0.989 (1+z)^{-1} - 1.545 (1+z)^{-2},\\
    E(z) & = \sqrt{\Omega_{m,0}(1+z)^3 + \Omega_{\Lambda,0}},\\
    \alpha (z) & = 0.858 + 1.554 (1+z)^{-1} - 1.176 (1+z)^{-2},
\end{align*}
and $\Omega_{m,0}$ and $\Omega_{\Lambda,0}$ denote the present-day matter and dark energy density parameters\footnote{We caution that in \citet{Sommovigo22_z7} we relied on a different dark matter only simulation \citep{Fakhouri10,Correa15}. 
 The resulting difference in the mean halo accretion rate is modest at the instantaneous level and largely uniform across
redshift (with the GUREFT median $dM_h/dt$ being $\sim$40\% higher at $z=10$), and does not qualitatively affect the results discussed here.}.

For each halo, we then construct the SFH starting at $z_i=10$, proceeding backward in time and adding instantaneous bursts of star formation (following Eq.~\ref{sfr}) at fixed timesteps of width $\Delta t$. The cumulative stellar mass at each epoch is computed as:
\begin{equation}\label{SFH_const_mst}
    M_{\star}(t_i)= \int_0^{t_i} {\rm SFR}(t')\ dt'\ ,
\end{equation}
ensuring consistency with Eq.~\ref{som22}. 
 This procedure effectively produces an almost constant SFH, since the timestep $\Delta t=1$ Myr is much shorter than the timescale over which the mean halo accretion rate varies for a fixed halo mass (changing by order unity over $\sim$0.2–0.4 Gyr across $10<z<15$, with faster evolution at higher redshift).

Nevertheless, this parametrization as a series of bursts allows easy adaptation to more stochastic SFHs, which will be explored in future work. We note that for the purpose of dust production, a burst more recent than the typical SN lifetime ($\sim3$ Myr) does not affect the dust mass budget.

With these SFHs we then compute the cumulative dust mass, considering core collapse SNe as the sole dust production mechanism.  
Observational evidence suggests that dust, at the epochs we consider, might have a carbonaceous component: for example, tentative polycyclic aromatic hydrocarbon (PAH) absorption features have been identified in NIRSpec observations of a $z=6.7$ \citep{Witstok23} and a $z=7.1$ galaxy \citep{Ormerod25}. However, AGB stars require $\sim 300$ Myr before they can contribute significantly to the production of carbonaceous dust \citep[for a recent review see][]{Schneider23}, which is too long compared to the ages of early galaxies. This justifies our assumption that supernovae are the dominant stellar dust source in the early Universe, as they are efficient producers of carbon dust \citep{Slavin20,Ferrara00,Schneider23}.
Dust growth can play a role at later epochs, as galaxies get progressively metal enriched. However, the extreme ISM conditions of $z>10$ bursty and likely metal-poor (down to $Z/Z_{\odot} \sim 0.07 $, see e.g. \citealt{Curti24}) galaxies are expected to slow down the timescales of dust growth to $\sim 400$ Myr characteristic timescales, implying a minor contribution to the dust budget at least at $z\sim 10$ \citep{Ferrara16,Dayal22}. Further discussion on including the effect of growth can be found in Sec.~\ref{sect:Cav_Disc}

We obtain the IMF--averaged delay--time distribution of the supernova rate per unit stellar mass, $\dot{n}_{\rm SN}(t)$, using the stellar population synthesis code \textsc{STARBURST99} \citep{Leitherer99}\footnote{We assume a Salpeter IMF (1--100~$M_\odot$) and a fixed stellar metallicity $Z_\star = 0.1\,Z_\odot$.}. The code provides the supernova rate as a function of time for a single instantaneous burst forming $10^{6}\,{\rm M_\odot}$ of stars. We therefore divide this rate by the total stellar mass of the burst to obtain $\dot{n}_{\rm SN}(t)$, which has units of supernovae per unit time per unit stellar mass.

For a general star--formation history $\mathrm{SFR}(t)$, the instantaneous supernova rate at time $t_i$ is obtained by convolving this delay--time distribution with the past star--formation history,
\begin{equation}\label{eqNSN}
\dot{N}_{\rm SN}(t_i) =
\int_0^{t_i} \dot{n}_{\rm SN}(t_i - t')\,
\mathrm{SFR}(t')\,dt'.
\end{equation}
The cumulative number of supernovae up to time $t_i$, $N_{\rm SN}(t_i)$, is then obtained by integrating (or, in practice, summing over discrete time bins of $\delta t=1$ Myr) this rate over time.
We can now write the dust content accumulated in a galaxy at a given time $t_i$ as:
\begin{equation}\label{md}
    \rm{M_d (t_i)} = \yd\ N_{\rm SN}(t_i)\ ,
\end{equation}
where $\yd$ is the effective dust yield resulting from each SN explosion. 
The dust mass will effectively start to build up after $\sim 3$ Myr, i.e. after the SN formed within the first burst of SF have exploded.

We treat the dust yield per core-collapse supernova, $\yd$, as an effective parameter that encapsulates both dust formation in the ejecta and subsequent destruction by the reverse shock. Theoretical models predict intrinsic dust yields spanning $y_{\rm d}\sim0.02$--$1\ \rm{M_\odot}$, but differ widely in their predictions for dust survival, with destruction efficiencies ranging from $\sim40\%$ to $\gtrsim90\%$ \citep{Todini01,Schneider04,Nozawa10,Sarangi_18,Sarangi22,Sarangi_2025,Schneider24,Valiante09,Slavin20,Ferrara21}. Observational estimates of SN dust masses also span several orders of magnitude ($\sim10^{-5}$--$0.7\ \rm{M_\odot}$), reflecting uncertainties related to remnant age, wavelength coverage, and reverse-shock processing \citep{Kotak09,Inserra11,Gallagher12,Gomez12,Matsuura15,Bocchio16,Shahbandeh23,Zsíros24,Szalai25}. We therefore in the following we adopt a conservative range $\yd=0.02$--$0.3\ \rm{M_\odot}$, corresponding to physically plausible effective yields after partial dust destruction. In this framework, $\yd$ should be interpreted as an effective retained yield: low values can equally represent intrinsically high condensation efficiencies coupled to efficient reverse-shock destruction or to large-scale dust redistribution/removal (e.g. radiation pressure or approximately uniform outflows) that lowers the dust column density within the star-forming regions.

In Fig.~\ref{Fig1} we plot the build-up of the SFH (dotted lines), stellar (dashed lines), and dust mass (solid lines) for halo masses in the range $10^8-10^{12}\ \mathrm{M_{\odot}}$ at $z=10$. As a reference, assuming\footnote{If not specified otherwise, these values of $(\epsilon_{\star},\yd)=(0.1,0.1\ \mathrm{M_{\odot}})$
are assumed as fiducial.} $\epsilon_{\star}=0.1$ and $\yd=0.1\ \mathrm{M_{\odot}}$, for a halo mass of $M_h=10^{10.9}\ \mathrm{M_{\odot}}$ we infer a stellar mass of $M_{\star}=10^{9.0}\ \mathrm{M_{\odot}}$ and dust mass $M_{\rm d} = 10^{6.6}\ \mathrm{M_{\odot}}$ accumulated over $138$ Myr ($134$ Myr for the dust mass due to the initial delay of $4$ Myr). This implies a dust-to-stellar mass ratio of $M_{\rm d}/M_{\star} = 0.003$.

For the galaxy radius $r_d$ (which we will use for the dust distribution) we use the prescription for the cold disk gas radius provided in \citep{Ferrara00b}:
\begin{align}\label{rd}
    & r_d  = 4.5 \Lambda r_{\rm vir}\\
    & r_{\rm vir} = \frac{0.784}{h}\ \left( \frac{M_h}{10^8 h^{-1} \msun} \right)^{1/3} \left[ \frac{\Omega_{m,0}\ \Delta_c}{\Omega_{m,z}\ 18 \pi^2} \right]^{-1/3} \left( \frac{1+z}{10} \right)^{-1} 
\end{align}
where  $r_{\rm vir}$ is expressed in kpc, and $\Delta_c$ is the overdensity at virialization:
\begin{equation}
    \Delta_c = 18 \pi^2 + 82 (\Omega_{m,z}-1) -39 (\Omega_{m,z}-1)^2\ ,
\end{equation}
and $\Lambda$ is the spin parameter. For the latter, we consider the best-fit distribution from the GUREFT simulation, a log-normal distribution with mean value $\bar{\Lambda} = 0.027$ and standard deviation $\sigma_{\Lambda} = 0.5390$ (implying a 16th and 84th percentile of $0.015,0.049$). This distribution does not vary significantly with redshift or halo mass.

For our reference halo mass of $M_h = 10^{10.86}\ \mathrm{M_{\odot}}$, the corresponding dust radius is $r_d = 1.74^{+1.00}_{-0.91}$ kpc. We note that our predicted disk/dust sizes ($100$–$2000$\,pc for the halo masses shown in Fig.~\ref{Fig1}) are larger than the typical UV–light radii measured for the aforementioned sources, which span $70$–$700$\,pc. However, once we account for the dispersion introduced by the spin–parameter distribution, our predicted sizes remain fully consistent with the observed range, lying within the $1\sigma$ scatter of the measurements. Given the substantial uncertainties in size estimates, and the unclear degree to which the UV radius traces the spatial distribution of gas and dust-- while the
effective optical depth is directly constrained by SED fitting --, we refrain from introducing any ad–hoc rescaling of the dust size. This choice reflects the overall philosophy of this work: rather than tuning parameters to fit observations, we aim to show where the model naturally lies once different physical ingredients are included.

\begin{figure}
    \vspace{6pt}
    \centering
    \includegraphics[width=0.499\textwidth]{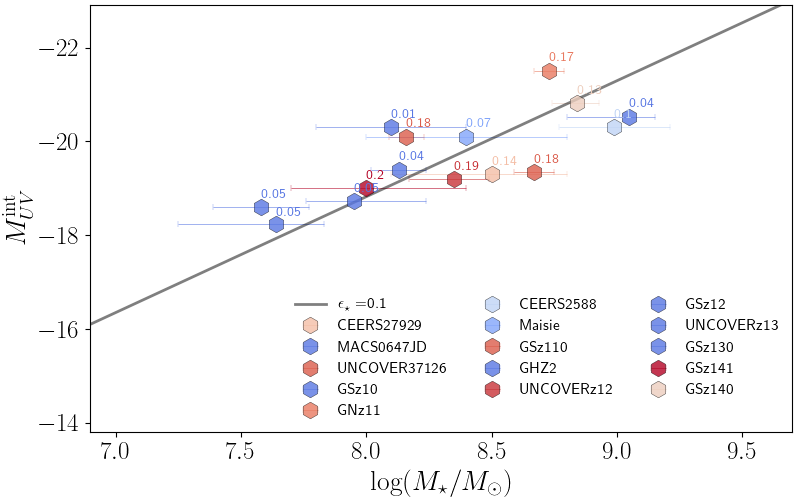}
    \caption{Intrinsic UV magnitude $\rm \MUV^{\rm int}$ vs. stellar mass ($\rm M_\star$) at $z=10$, assuming $\sfe = 0.1$. Colored points represent spectroscopically confirmed $z>10$ galaxies (see text for detailed references), color-coded by inferred $\tau_V$ (increasing from blue to red; values labeled).}
    \label{Fig_Muv-Mstar}
\end{figure}
\section{Emission Model}\label{sect:emission_model}
Having established the intrinsic properties of each galaxy, we now proceed to model their emergent emission across different wavelength regimes. This requires modeling both the intrinsic stellar emission and the contribution of dust, which attenuates the emission at short wavelengths and re-emits thermally in the rest-frame FIR.  

\begin{figure*}
    \centering
    \includegraphics[width=0.435\textwidth]{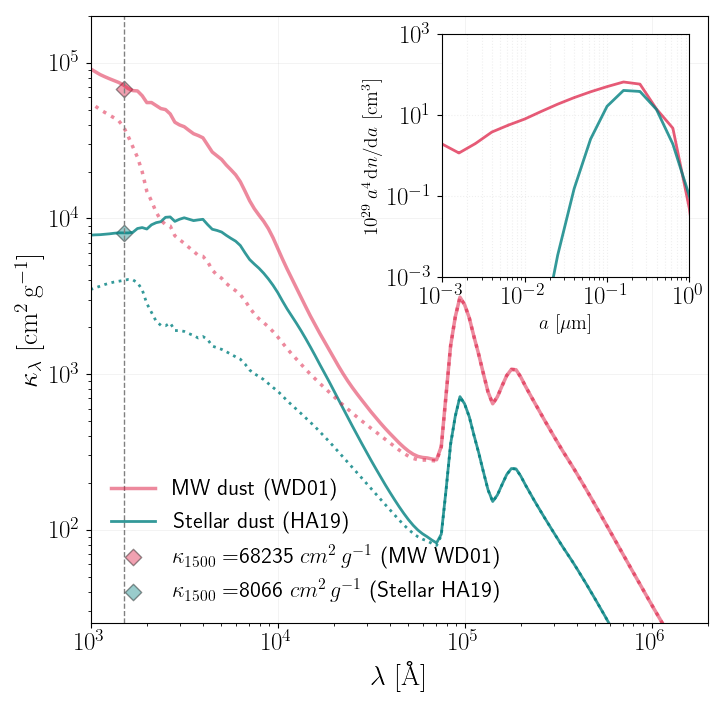}
    \includegraphics[width=0.49\textwidth]{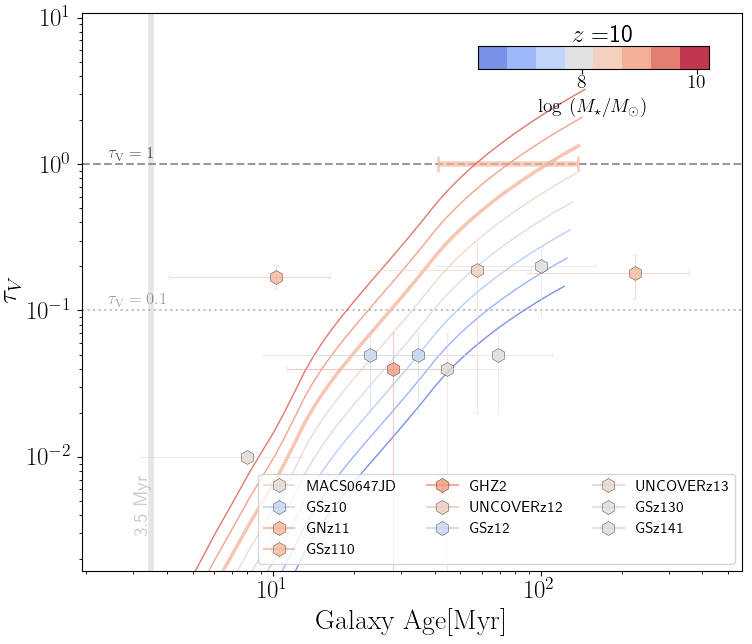}
    \caption{\textbf{Left:} Dust opacity $\kappa_\lambda$ for Milky Way (\citealt{Draine03}; pink) and stellar dust (\citealt{Hirashita19}; teal). Solid lines show total extinction and dotted lines absorption only; differences arise solely from the adopted grain–size distributions (inset).  Diamonds mark the UV opacity at $1500$\,\AA\ used in our RT calculations.
    \textbf{Right:} Time evolution of the V–band optical depth $\tau_V$ for the same halos shown in Fig.~\ref{Fig1}, assuming Milky Way dust, $\sfe = 0.1$, and a dust yield of $\yd = 0.1\,{\rm M_\odot}$ per SN. The horizontal bar illustrates the uncertainty in the time at which the galaxy becomes optically thick for a halo of mass $M_h = 10^{10.86}\,{\rm M_\odot}$, corresponding to the $16$–$84$th percentile range in $\tau_V$ arising from galaxy-to-galaxy variations in size. Horizontal dashed and dotted grey lines mark $\tau_V = 1$ and $0.1$.}
    \label{Fig_opacity_tau}
\end{figure*}

Consistent with our treatment of the supernova delay–time distribution, we compute the intrinsic rest–frame UV luminosity at $1500\,\text{\AA}$ using the \textsc{Starburst99} monochromatic luminosity $L_{\rm UV,inst}^{\rm SB99}(t)$, tabulated for an instantaneous burst forming $10^6\,{\rm M_\odot}$ of stars. For a general star–formation history, the intrinsic luminosity at time $t_i$ is therefore obtained by convolving this luminosity–age response function with the past SFH,
\begin{equation}
L_{\rm UV}^{\rm int}(t_i) = \int_0^{t_i} L_{\rm UV,inst}^{\rm SB99}(t_i - t')\ {\rm SFR} (t')\,dt'\ .
\end{equation}
In practice, as for the SN delay-time distribution, we evaluate this convolution over our discretized SFH with $\Delta t = 1$ Myr. The differences between this approach and adopting a simple proportionality between SFR and UV luminosity are discussed in Appendix~\ref{SF_to_LUV_appx}.

We show the intrinsic $\MUV$ as a function of stellar mass ($\rm M_{\star}$) for our model galaxies at $z=10$ in Fig.~\ref{Fig_Muv-Mstar} (solid line). For comparison, we also include observational estimates for spectroscopically confirmed galaxies at $z \geq 10$ (MACS0647--JD from \citealt{Hsiao23}, 
GS--z10 and GS--z11 from \citealt{Curtis23,Robertson23}, 
GN--z11 from \citealt{Bunker23,Tacchella23}, 
GHZ2 from \citealt{Castellano24}, UNCOVER--z12 and UNCOVER--z13 from \citealt{Wang23,Fujimoto23b,Atek22},  and finally GS--z12, GS--z13,  and GS--z14 from \citealt{Carniani24a,Carniani24_z14,Schouws24,Robertson24}).

We can see that a standard $\epsilon_{\star}=0.1$ is sufficient to broadly reproduce the observed $\MUV$–$M_\star$, although some galaxies appear more UV-luminous than our predictions, even in the dust-free case. We note that, in our prescription, varying $\epsilon_\star$ does not shift galaxies vertically in the $M_{\rm UV}$–$M_\star$ plane: both the UV luminosity and the stellar mass grow proportionally with $\epsilon_\star$, so the locus moves diagonally but $M_{\rm UV}$ at fixed $M_\star$ remains unchanged.

\subsection{Dust Model}\label{subsec:dust}
In order to obtain the \textit{observed} UV magnitudes, we must account for dust attenuation. This requires two ingredients. First, starting from an assumed dust composition and grain–size distribution, we compute the wavelength-dependent dust opacity and hence the dust optical depth, which defines the underlying extinction curve (see Sec.~\ref{subsect-extinct}). Second, to translate this extinction curve into an attenuation curve, we must -- attempt to -- incorporate the effects of dust–star geometry and scattering (see Sec.~\ref{subsec-atten}).

\subsubsection{Extinction curve}\label{subsect-extinct}
In the following, we consider two physically motivated dust models and derive their associated opacities. The first is the classical Milky Way dust model of \citet{Weingartner01} and \citet{Draine03}, widely used in galactic and extragalactic studies \citep{Sommovigo20,Ferrara22a,Bakx20,Algera25} because it captures the ISM-processed grain population typical of the local Universe. Such a mixture may not be appropriate for very young, metal-poor systems---such as the Blue Monsters---where dust is expected to be dominated by
freshly condensed supernova grains rather than by the small-grain–rich population produced by ISM processing \citep[e.g.][]{Schneider24,Narayanan25}. For this reason, we also consider the \quotes{stellar dust} model of \citet{Hirashita19}, in which dust is composed of carbonaceous and silicate grains whose size distribution follows a log-normal peaked at $a \sim 0.1$--$0.2\,\mu{\rm m}$, as expected for unprocessed SN dust. We adopt the same carbonaceous-to-silicate mass ratio as in the MW case, as in \citet{Hirashita19}.

For both dust models, the opacity $\kappa_\lambda$ is computed by integrating the wavelength-dependent absorption and scattering efficiencies, $Q_{\rm abs}(a,\lambda)$ and $Q_{\rm sca}(a,\lambda)$, over the respective grain–size distribution. We use the Mie efficiencies tabulated by \citet{Weingartner01} and \citet{Draine03}\footnote{Specifically, we rely on the optical–constant
tables publicly available at \url{https://www.astro.princeton.edu/~draine/dust/dust.diel.html}, using the carbonaceous grains and the smoothed UV astronomical silicate datasets.}. This yields both the total opacity $\kappa_{1500}$ and the absorption-only component $\kappa_{1500,abs}$ for each dust model.

The resulting MW and stellar dust opacities differ substantially from the UV to the FIR. The MW model yields $\kappa_V = 20531~{\rm cm^2\,g^{-1}}$ at $0.547\,\mu{\rm m}$ and $\kappa_{1500} = 68235~{\rm cm^2\,g^{-1}}$ at 1500\,\AA ($\kappa_{1500,\mathrm{abs}} = 38880~{\rm cm^2\,g^{-1}}$), 
reflecting the strong contribution of small grains that steepen the UV extinction curve. 
In contrast, the stellar dust model of \citet{Hirashita19} produces a significantly flatter and overall lower UV opacity curve, with $\kappa_V = 7444~{\rm cm^2\,g^{-1}}$ and 
$\kappa_{1500} = 8067~{\rm cm^2\,g^{-1}}$ ($\kappa_{1500,\mathrm{abs}} = 3995~{\rm cm^2\,g^{-1}}$), 
a consequence of the dominance of large SN-condensed grains and the relative lack of very small grains. 
The corresponding grain--size distributions and opacities for both dust models are shown in Fig.~\ref{Fig_opacity_tau}.

\begin{figure*}
    \vspace{6pt}
    \centering
    \includegraphics[width=0.95\textwidth]{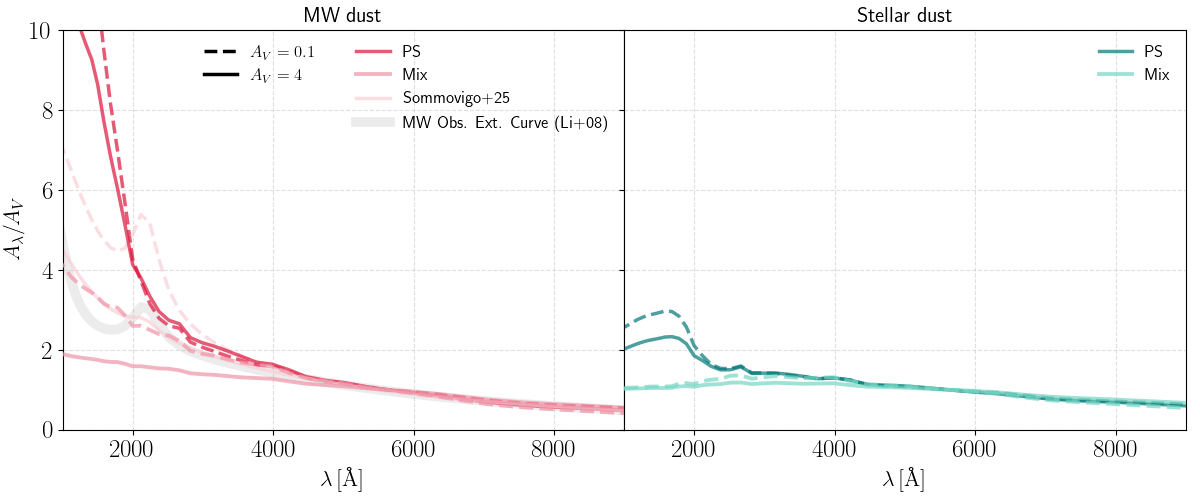}  
    \caption{Attenuation curves obtained for Milky Way dust (left) and stellar–dust grain–size distributions (right).  Different shades correspond to different prescriptions for the effective transmission, from darker to lighter: point–source geometry (Eq.~\ref{T1500_sphere}); stars and dust mixed in a spherical configuration (Eq.~\ref{T1500_mix}); and the best–fit median attenuation curves (for given $A_V$) from \citet{Sommovigo25_TNG}, summarised in their Eqs.~7–9 by post–processing $\sim 6500$ TNG50/TNG100 \textit{local} galaxies with {\sc SKIRT} over 51 lines of sight per source.  Solid (dashed) lines correspond to the optically thick (dashed) cases with $A_V = 4$ ($A_V = 0.1$), respectively. These values bracket the $\tau_V$ values that we predict for our massive halos in Fig.~\ref{Fig_opacity_tau}.}
    \label{Fig_att_curve}
\end{figure*}

Having obtained the wavelength-dependent dust opacities for both dust models described above, we
now compute the corresponding optical depths. Assuming a spherical geometry with radius $r_d$
(eq.~\ref{rd}), the dust optical depth at wavelength $\lambda$ and time $t_i$ is given by
\begin{equation}\label{tau_v}
    \tau_{\lambda}(t_i)
    = \kappa_{\lambda}\,\Sigma_{\rm d}
    = \kappa_{\lambda}\,\frac{M_{\rm d}(t_i)}{f_{\mu}\,\pi r_d^2}
    = \kappa_{\lambda}\,\frac{y_{\rm d}\,N_{\rm SN}(t_i)}{f_{\mu}\,\pi r_d^2},
\end{equation}
where the geometrical factor $f_{\mu}$ accounts for the spatial configuration of dust relative to the UV-emitting sources. In this work we adopt the the spherical configuration ($f_{\mu}=4/3$) as our fiducial case, as this setup more closely reflects the internal structure of $z\sim10$ galaxies, where dust and stars are likely to be co-spatial within GMC-like regions \citep{Somerville25}. We revisit this approximation in Sec.~\ref{sect:turbulent_ism}.

In Fig.~\ref{Fig_opacity_tau} we show the resulting optical depth for MW dust as a function of galaxy age for the $z=10$ systems whose mass assembly histories were presented in Fig.~\ref{Fig1}. The epoch at which a galaxy becomes optically thick ($\tau_V \rightarrow 1$) depends sensitively on $r_d$, and therefore on the halo spin parameter. For the reference halo mass, the transition to $\tau_V>1$ occurs after $106^{+32}_{-65}$ Myr from the onset of star 
formation. Since the galaxy age at $z_i=10$ is $138$ Myr, this implies that it becomes optically 
thick at $z=10.53$. In other words, a galaxy with $M_\star = 10^{9.1}\,{\rm M_\odot}$ at $z=10$ would have been optically thin in the V band if observed at earlier times. Lower-mass systems ($M_\star \leq 10^8\,{\rm M_\odot}$) remain in the range $0.1 < \tau_V < 1$ at $z=10$: although not fully optically thick, these values are still larger than those typically inferred—albeit with large uncertainties—from current SED-fitting analyses.

\subsubsection{Attenuation Curve}\label{subsec-atten}
Now, to move from opacity or \quotes{physical} optical depth to attenuation (and thus \quotes{effective} optical depth), we need to make an assumption on the dust-to-star geometry. In an analytical framework, this requires relying on transmission functions $T_{\lambda}(\tau_{\lambda})$, so that the attenuation at a given wavelength is $A_{\lambda} = -2.5\,\log T_{\lambda}(\tau_{\lambda})$. 

In the classic spherical–shell foreground screen approximation, where $T_\lambda = e^{-\tau_\lambda}$, the attenuation curve is identical in shape to the extinction curve, as no geometric or scattering effects are included. However, in a realistic radiative transfer treatment, scattering along and across the line of sight \citep[see e.g.][]{Matsumoto25} leads to flatter attenuation curves for high-density (high $\tau_{\lambda}$) sightlines and steeper ones for low-density (low $\tau_{\lambda}$) sightlines. This effect is well established in both hydrodynamical and idealized simulations with radiative transfer \citep{Chevallard13,SeonDraine16,Narayanan18,Trayford20,Lin21,Sommovigo25_TNG,Matsumoto25,Dubois24,DiMascia24}, as well as in observations at low \citep{Battisti2016,Barisic2020,Salim18,Salim20} and high redshift \citep{Markov24,Fisher25,Shivaei25}.

To account for these effects while preserving the simplicity of our analytical framework, we adopt two effective transmission functions motivated by classical radiative transfer solutions. For the case in which UV photons originate from a central point source and propagate through a surrounding spherical dusty medium, we use the analytic approximation from \citet{Code73}. This solution is derived from Monte Carlo radiative transfer calculations and includes both absorption and anisotropic scattering at different wavelengths:
\begin{equation}\label{T1500_sphere}
    T_{\lambda,\mathrm{ps}} = \frac{2}{(1 + \eta)\, e^{\xi \tau_{\lambda}} + (1 - \eta)\, e^{-\xi \tau_{\lambda}}},
\end{equation}
where
\begin{equation}
    \eta = \sqrt{\frac{1 - \omega}{1 - \omega g}}, \qquad 
    \xi = \sqrt{(1 - \omega)(1 - \omega g)}.
\end{equation}
Here, $\omega$ is the single-scattering grain-size-distribution averaged albedo and $g$ is the grain-size-distribution averaged scattering asymmetry parameter. 

To account for the possibility that UV sources are spatially mixed with the dust we can instead use the classical mixed-geometry attenuation solution of \citet{Városi_1999}. This expression neglects scattering but accounts for the reduction in effective attenuation due to the dust star geometry:
\begin{equation}\label{T1500_mix}
 T_{\lambda,\mathrm{mix}}  = \frac{3}{4\tau_{\lambda}} 
 \left( 
 1 - \frac{1}{2\tau_{\lambda}^2} 
 + 
 \left( \frac{1}{\tau_{\lambda}} + \frac{1}{2\tau_{\lambda}^2} \right) 
 e^{-2\tau_{\lambda}} 
 \right).
\end{equation}

Both expressions for $T_{\lambda}$ converge to the same limit for $\tau_{\lambda} \lesssim 1$ (but still provide a higher transmission than the simple screen approximation, $T_{\lambda} = e^{-\tau_{\lambda}}$, e.g. increasing the transmission from $55\%$ to $67\%$ at $\tau_{\lambda} =0.6$), but they differ in the optically thick regime. For example, at $\tau_{\lambda}=3$, the point source model yields $T_{\lambda,\mathrm{ps}} \approx 13\%$, while the mixed model gives $T_{\lambda,\mathrm{mix}} \approx 25\%$ compared to only $5\%$ for the classical screen.. 

We show the resulting attenuation curves $A_{\lambda}/A_V$ in Fig.~\ref{Fig_att_curve}. Their behavior differs markedly between the two dust mixtures and between the point--source and mixed geometries. For the point--source (PS) configuration, the wavelength dependence of $A_{\lambda}/A_V$ is driven primarily by the ratio $\tau_{\lambda}/\tau_V=\kappa_{\lambda}/\kappa_V$. Because we normalize by the RT--computed $A_V$, much of the explicit dependence on the absolute 
optical depth cancels out. As a result, for a fixed dust mixture the PS curves corresponding to different values of $A_V$ nearly overlap. A small residual dependence remains, arising from the wavelength dependence of the scattering parameters $\omega(\lambda)$ and $g(\lambda)$, which introduce mild optical--depth--dependent corrections.
The mixed geometry behaves differently. Here the attenuation depends non--linearly on $\tau_{\lambda}$. At low optical depth ($\tau_{\lambda}\lesssim 1$), the behavior resembles that of the PS case. However, as $\tau_{\lambda}$ increases, the transmission approaches the 
asymptotic scaling $T_{\lambda,\mathrm{mix}}\propto 1/\tau_{\lambda}$, which progressively flattens the normalized attenuation curve. The strength of this effect depends on how rapidly $\kappa_{\lambda}$ rises toward the UV.
For Milky Way dust, the steep UV opacity ($\kappa_{1500}/\kappa_V \gg 1$) drives $\tau_{1500}$ into the optically thick regime over the range of $A_V$ considered, producing a clear flattening of the mixed attenuation curves as $A_V$ increases. In contrast, the stellar--dust mixture exhibits a much greyer opacity curve, with $\kappa_{1500}/\kappa_V$ closer to unity. Consequently, $\tau_{1500}$ 
remains of order unity or below even at large $A_V$, and the mixed attenuation curves show only weak variation with optical depth or geometry.

A flatter attenuation curve at high gas surface densities is also seen in more realistic/reliable radiative–transfer simulations—where dust and UV sources coexist within compact, dense regions \citep{Matsumoto25,Dubois24,Narayanan18}. Thus, we adopt the mixed geometry as our fiducial prescription ($T_{\lambda, {\rm mix}}$, eq.\ref{T1500_mix}).
For comparison, we also include the \citet{Sommovigo25_TNG} best–fit relations for the attenuation curves obtained from {\sc SKIRT} post–processing of 6400 \textit{local} TNG50/TNG100 galaxies over 51 sightlines, assuming a MW dust mixture. While applying such scaling relations directly to our idealised high-$z$ galaxy model is not physically accurate, the fact that our point–source and mixed curves bracket their best–fitting relations in each $A_V$ bin is reassuring (see also Fig. 5 in \citealt{Sommovigo25_TNG} for a comparison between these simulations and local observations by \citealt{Salim18}). A more accurate  assessment will require dedicated high–redshift, high–resolution 
simulations with radiative transfer (see Sec.~\ref{sect:Cav_Disc} for a more detailed discussion).

\begin{figure}
    \vspace{6pt}
    \includegraphics[width=0.49\textwidth]{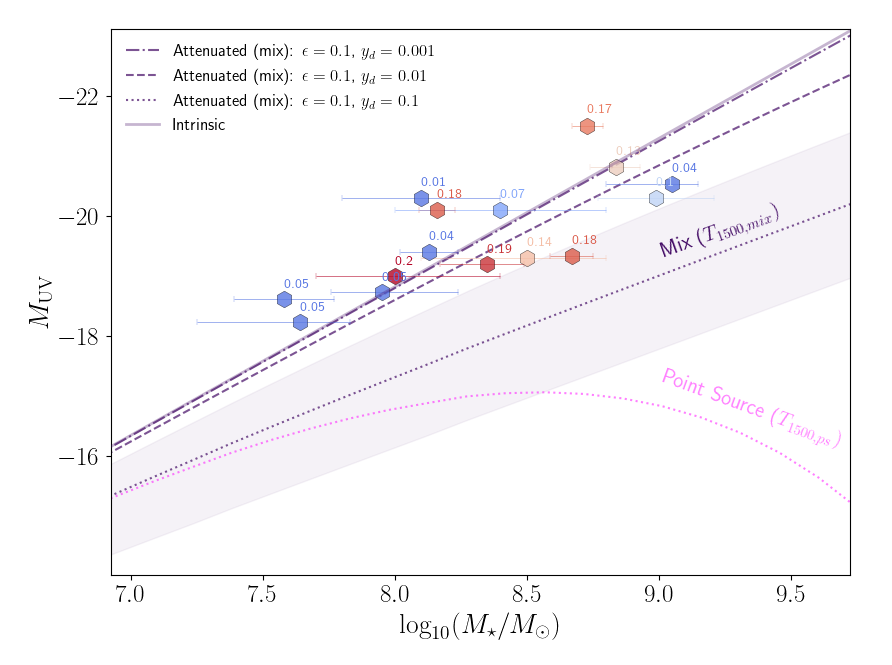}
    \caption{Attenuated UV magnitude ($\MUV$) versus stellar mass ($M_\star$) at $z=10$, assuming $\sfe=0.1$, MW dust, and varying SN dust yields $\yd = 10^{-3},\,10^{-2},\,0.1\,{\rm M_\odot}$ (dash–dotted, dashed, and dotted curves). The magenta dotted line shows the point–source case $T_{1500,{\rm ps}}$ (Eq.~\ref{T1500_sphere}), which corresponds to a steeper attenuation curve (see Fig.~\ref{Fig_att_curve}). The solid line indicates the intrinsic (unattenuated) $\MUV$–$M_\star$ relation. Colored symbols denote spectroscopically confirmed $z>10$ galaxies (see text for references), color–coded by the inferred $\tau_V$ (increasing from blue to red; values labeled). The shaded region around the dotted curve reflects the 16th–84th percentile scatter in attenuated $\MUV$ due to galaxy–to–galaxy variations in size.}
    \label{Fig_Muv-Mstar_att}
\end{figure}

For any transmission function $T_\lambda$, the attenuated UV luminosity is simply
\[
    L_{\rm UV}^{\rm att} = T_{1500}\,L_{\rm UV}^{\rm int}.
\]
Assuming energy balance, the total IR luminosity is then
\begin{equation}
\label{eq_LIR_unif}
    L_{\rm IR}
    = \left[1 - T_{1500}(\tau_{1500,\rm abs})\right] 
      L_{\rm UV}^{\rm int},
\end{equation}
where $\tau_{1500,\rm abs}$ is the \emph{absorption-only} opacity (unlike eq.~\ref{tau_v}, which includes scattering). It is worth noting that the dust optical depth $\tau_{1500}$ depends on the dust mass (eq.~\ref{tau_v} and therefore on the product $\epsilon_\star\,y_d$, such that different $(\epsilon_\star,y_d)$ pairs can yield similar $T_{1500}$ and hence similar ratios $L_{\rm IR}/L_{\rm UV}=\left[1 - T_{1500}(\tau_{1500,\rm abs})\right]/T_{1500}$. However, the absolute luminosities $L_{\rm UV}^{\rm att}$ and $L_{\rm IR}$ scale with the intrinsic UV output $L_{\rm UV}^{\rm int}\propto\epsilon_\star$, and are therefore not degenerate even when the partition between UV and IR is.

The attenuated $\MUV$ expected for different halo masses at $z=10$ relying on the fiducial $T_{1500,{\rm mix}}$ case as well as $T_{1500,{\rm ps}}$ are shown in Fig.~\ref{Fig_Muv-Mstar_att}. The different linestyles correspond to varying supernova dust yields, $\yd = (10^{-3}, 10^{-2}, 0.1)\ \mathrm{M_{\odot}}$, assuming a fixed star formation efficiency of $\epsilon_{\star} = 0.1$. Notably, when adopting the highest dust yield value ($\yd = 0.1\ \mathrm{M_\odot}$), our models systematically underpredict (especially for $T_{1500,{\rm ps}}$) the observed UV magnitudes for the largest haloes, which are more dust rich in the modeled halos ($\rm{\log (M_{\star}/M_{\odot})>8}$). This discrepancy persists even after accounting for the full distribution of galaxy sizes—modeled via 1000 random realizations of the halo spin parameter -- and thus dust spatial extension $r_d$ (see eq.~\ref{rd}) -- as shown by the shaded area around the dotted curve in Fig.~\ref{Fig_Muv-Mstar_att}.

At the highest redshifts, $z>8.5$, no secure dust–continuum detections are currently available to directly test our IR luminosity predictions; only upper limits exist for a handful of UV-bright systems \citep{Fudamoto23,Bakx22,Popping22,Carniani24_z14,Schouws24}. By contrast, at $z=7$ the predicted IR luminosities can be directly compared to the REBELS sample \citep{Inami22, Barrufet23}, which also underpins the IR luminosity function used later in this work. Crucially, REBELS galaxies were selected to be UV-bright and populate the bright end of the observed UV LF ($\MUV < -21$), so the UV–bright and IR–bright systems probed at this epoch substantially overlap. This motivates the need for a self-consistent dust treatment such as the one developed here.
A comparison between our predicted IR fluxes and available sub-mm measurements at $z=7$ is presented in Appendix~\ref{build-up_z7_appx}, while the corresponding predictions and current upper-limit constraints at $z\geq10$ are shown in Appendix~\ref{App_IR_z12}.

\section{A minimal single-zone, porous ISM}\label{sect:turbulent_ism}
\begin{figure*}[]
    \centering
    \includegraphics[width=0.95\textwidth]{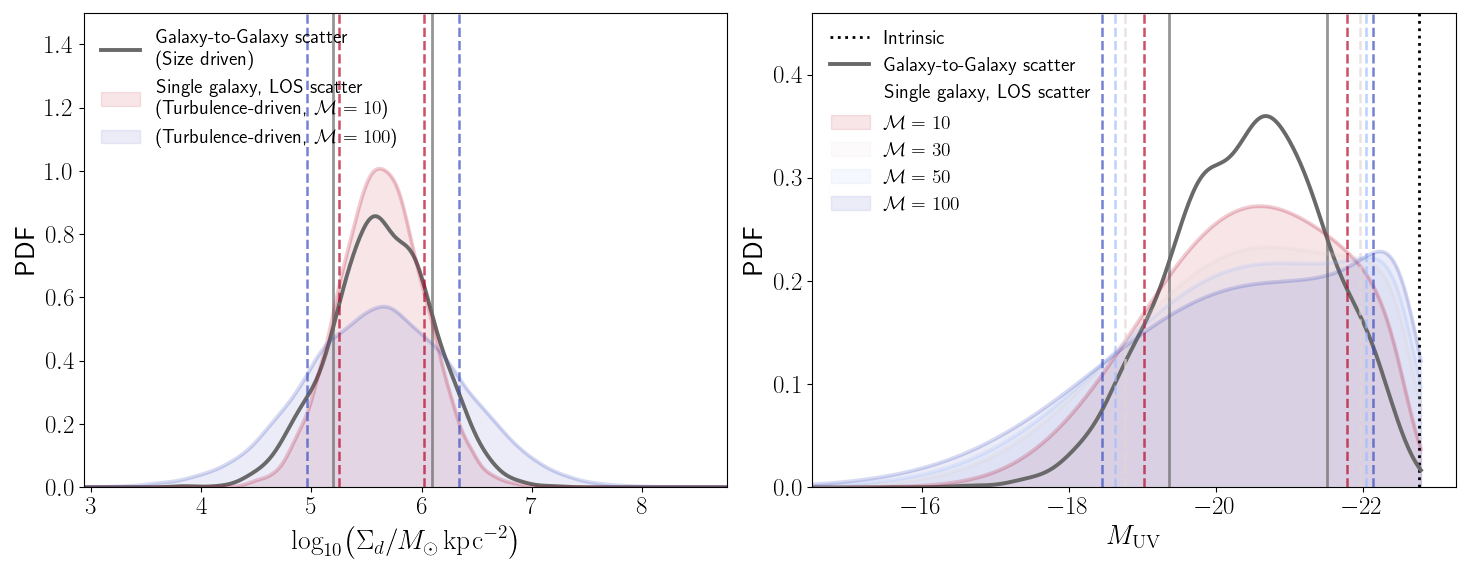}
    \caption{\textbf{Left:} Distribution of dust surface densities $\Sigma_d$ for the reference halo with $\log(M_h/M_\odot)=10.86$ at $z=10$, assuming $\yd=0.1$ and $\sfe=0.1$. The grey curve shows the \emph{galaxy-to-galaxy} scatter at fixed halo mass, driven solely by the spin–induced variation in galaxy size (eq.~\ref{tau_v}); vertical grey lines mark the 16th–84th percentiles. The red and blue curves show instead the \emph{within-galaxy} line-of-sight scatter expected from turbulence (eq.~\ref{eq:sigma_lnSigma}) for Mach numbers $\mathcal{M}=10$ and $\mathcal{M}=100$, respectively. Vertical lines again mark the 16th-84th percentiles. \textbf{Right:} Corresponding distributions of attenuated UV magnitudes. For the turbulent models (shaded curves), the spread reflects how the observed $\MUV$ varies depending on which line of sight crosses the clumpy ISM.  For comparison, the grey curve shows the uniform–ISM case, where the scatter instead arises from variations in galaxy size among halos of the same mass (i.e.\ galaxy-to-galaxy rather than LOS-to-LOS variation).}
    \label{Fig_TurbulentISM_Mh11}
\end{figure*}

So far, we have treated our toy high-redshift galaxies as uniform, spherical systems. This is, of course, a major simplification. In a more realistic, multi-phase ISM, the regions responsible for UV (low attenuation) and IR (high attenuation) emission can be non-co-spatial \citep{Sommovigo20, Sommovigo22_z57, Ferrara22a,Pallottini22,Dayal22}. Even within a single phase, turbulence naturally generates a broad distribution of column densities along different lines of sight.

In \citet{Ziparo23}, they argued that if UV and IR emission were not co-spatial in $z\sim10$ galaxies then these galaxies should exhibit \emph{higher IR fluxes} than implied by current ALMA and NOEMA upper limits \citep{Popping22,Bakx22,Fujimoto22,Kaasinen22,Yoon22}. This assessment was based on the \emph{molecular index threshold} argument proposed in \citet{Ferrara22a}. The molecular index—defined as the ratio between the IR (at rest-frame 158 $\mu$m)  and UV continuum fluxes, divided by the deviation of the UV slope from its intrinsic value,
\begin{equation}
I_{\rm m} = \frac{F_{158}/F_{1500}}{\beta_{\rm UV} - \beta_{\rm UV,int}}\ ,
\end{equation}
cannot exceed a critical value in a \emph{single–phase}, co-spatial ISM.
Within this single-phase, uniform-screen framework, adjusting the optical depth is the only available lever. Increasing it boosts $F_{158}$ and suppresses $F_{1500}$, but also reddens $\beta_{\rm UV}$; lowering it reduces the reddening and $F_{158}$, but only down to the CMB-imposed IR floor. No choice of a single optical depth can simultaneously satisfy all constraints once $I_{\rm m}^{\star}\gtrsim1120$. Such systems therefore require at least two phases with distinct optical depths, rather than a uniform medium.

In $z\sim4$--7 galaxies, a few systems that cross this threshold have already been observed \citep{Sommovigo22_z57,Ferrara22a}, and \citet{Inami22} tentatively reported a correlation between high molecular index and the spatial offset between the UV and IR emission peaks in the low–resolution REBELS ALMA and \emph{HST} data.
Applying this same threshold ($I_m\geq1120$) to the Blue Monsters, and reverse-engineering the $F_{158}$ from the measured $(F_{1500},\beta_{\rm UV})$, yields lower limits on IR fluxes \emph{above} the current ALMA/NOEMA upper limits for the several sources that remain undetected in the sub-mm \citep{Popping22, Bakx22, Fujimoto23, Carniani24_z14, Schouws24}. In other words, if these galaxies hosted a spatially segregated ISM similar to the systems at lower redshift, they should have been detected in the far–infrared --- yet they are not.  
This failure of the spatial–segregation scenario for explaining the extremely blue UV slopes of the Blue Monsters is not unexpected: their predicted sizes --- and the sizes tentatively inferred from UV imaging --- are $\ll 1\,\mathrm{kpc}$, consistent with objects resembling
\quotes{super} giant molecular clouds \citep{Somerville25} rather than extended, multi–phase ISM structures.

Motivated by this, we now explore a different approach, linking ISM \quotes{porosity} to the width of the dust surface density distribution generated in a highly turbulent medium, even within a single idealised region. We caution that this is a qualitative exercise and should not be interpreted as a quantitative method for inferring turbulence levels of individual galaxies from population statistics. Further, we address the limitations of extending this formalism to $z\sim 7$ in Sec.\ref{sect:Cav_Disc}.

For approximately isothermal turbulence, both simulations and analytic arguments show that the PDF of column (or surface) density is well described by a lognormal distribution \citep[e.g.][]{Semadeni2001, Ostriker2001, Federrath2010, Thompson16}.  
Following the formulation of \citet{Thompson16}, the logarithmic dispersion of the gas surface density, $\sigma_{\ln \Sigma_g}$, depends on the turbulent Mach number $\mathcal{M}=v_{\rm rms}/c_s$ (rms gas velocity to the sound speed) as
\begin{equation}
    \sigma_{\ln \Sigma_g}^{2} = 
    \ln\!\left[1 + \frac{R(\mathcal{M})\, \mathcal{M}^{2}}{4}\right]\ ,
    \label{eq:sigma_lnSigma}
\end{equation}
where $R(\mathcal{M})$ corresponds to
\begin{equation}
    R(\mathcal{M}) = 
    \frac{1}{2}\,\frac{3-\alpha}{2-\alpha}\,
    \frac{1 - \mathcal{M}^{2(2-\alpha)}}
         {1 - \mathcal{M}^{2(3-\alpha)}}\ ,
    \label{eq:R_M}
\end{equation}
and $\alpha$ is the slope of the turbulent velocity power spectrum, $P(k) \propto k^{-\alpha}$.  
For supersonic turbulence, we adopt their fiducial value $\alpha = 2.5$ \citep{Thompson16}, which implies $R(\mathcal{M}) = (2 \mathcal{M})^{-1}$. The \citet{Thompson16} functional form has been shown to reasonably describe the column density distributions around stellar clusters in radiation hydrodynamical simulations of turbulent self-gravitating clouds \citep{Raskutti_2017,Menon_25} and in the diffuse ISM of local spiral galaxies \citep{Pathak_2024}

We apply this formalism to the distribution of dust surface densities, $\Sigma_{\rm d}$ within galaxies, assuming that dust and gas are co-located. 
Because galaxies are treated as unresolved systems in our model, we do not track individual stars or star-forming regions. Instead, attenuation is computed by statistically marginalizing over the ensemble of emitting regions and lines of sight through the ISM via a probability distribution.
At fixed total dust mass $M_{\mathrm{d}}$ (set by $\yd$ and $\epsilon_{\star})$, we assign the mean $\langle \Sigma_{\mathrm{d}} \rangle$ to match in the uniform case of eq.~\ref{tau_v}, while the local fluctuations around this mean are drawn from the Mach number-dependent lognormal PDF defined by Eq.~(\ref{eq:sigma_lnSigma}). This provides a simple description of the inhomogeneities in $\Sigma_{\mathrm{d}}$ produced by turbulence within the ISM.  

It is important not to confuse this within-galaxy scatter with the galaxy-to-galaxy variation in $\Sigma_{\mathrm{d}}$ that arises from size differences among galaxies of the same halo mass and redshift, driven by the log-normal scatter in spin parameter $\Lambda$. Hence, variations in $\Lambda$ generate a distribution of mean surface densities across galaxies, whereas the Mach-dependent lognormal in Eq.~(\ref{eq:sigma_lnSigma}) introduces an additional internal dispersion within each galaxy due to turbulence. For the reference system shown in Fig.~\ref{Fig_TurbulentISM_Mh11}, the dispersion in $\Sigma_{d}$ produced by turbulence with $\mathcal{M}\!=\!10$ is already comparable to the galaxy-to-galaxy dispersion from $\Lambda$, while for $\mathcal{M}\!\gtrsim\!30$ it becomes significantly broader.

For each galaxy and Mach number, we draw a set of $\Sigma_{\mathrm{d}}$ values from the lognormal distribution to represent different lines of sight through the dusty ISM.  
Each sightline is mapped to an effective UV optical depth via the 
dust opacity and to an emergent luminosity through the transmission function\footnote{From now on, unless otherwise stated, the fiducial mixed star--dust geometry attenuation prescription is adopted. 
Accordingly, when no explicit subscript is given, the transmission functions $T_{1500}$ refer to the mixed–geometry solution (eq.~\ref{T1500_mix}).} $T_{1500}(\tau_{\rm 1500})$. Retaining the full distribution of $M_{\mathrm{UV}}^{\rm att}$ values captures 
the stochastic attenuation introduced by turbulence, representing a luminosity-weighted ensemble over many emitting regions within an unresolved galaxy, rather than the output of a single physical line of
sight.

Figure~\ref{Fig_TurbulentISM_Mh11} illustrates the resulting distributions of $\Sigma_{\mathrm{d}}$ and UV magnitudes for $\mathcal{M}=10$–100 in a $M_{\rm h}=10^{10.86}\,{\rm M_\odot}$ halo at $z=10$ (corresponding to $M_\star\simeq10^{9}\,{\rm M_\odot}$ for $\sfe=0.1$). 
As $\mathcal{M}$ increases, the bright tail of the $\MUV^{\rm att}$ distribution moves closer to the intrinsic (unattenuated) value.  Thus, higher Mach numbers systematically increase the probability of sightlines 
with low $\Sigma_{\rm d}$—and therefore low attenuation—helping to reconcile high dust yields with the relatively blue UV luminosity functions observed at $z\gtrsim 10$.

The adopted Mach number range ($\mathcal{M}=1$--100) is motivated by theoretical and numerical studies indicating that gas-rich high-redshift galaxies host highly supersonic turbulence \citep{Kazandjian16,Krumholz18,Ginzburg22}, with characteristic velocity dispersions of tens to a few hundred km\,s$^{-1}$, arising from gravitational instability and global gas transport within the galaxy \citep[see e.g.][]{Carniani24b,Algera25b,Xu25} This is also in line with observational studies finding an increase in turbulent velocities at higher redshifts \citep{Rizzo24}.

We conclude by noting that the corresponding emitted infrared luminosity for a porous ISM is obtained by integrating the absorbed fraction over the continuous distribution of surface densities:
\begin{equation}\label{eq_LIR_Mach}
    L_{\mathrm{IR}} = 
    L_{UV}^{\rm int}
    \int \left[1 - T_{1500}(\tau_{\rm 1500,abs}) \right]
    \ P(\Sigma_{\mathrm{d}}\,|\,M_h)\ {\rm d}\Sigma_{\mathrm{d}}\ .
\end{equation}  
As the Mach number increases, the broadening of the PDF $P(\Sigma_{\rm{d}})$ boosts the fraction of low–column sightlines that permit UV escape, while the densest, most optically thick regions continue to dominate the IR re–emission. Thus, a highly turbulent ISM simultaneously reduces the effective UV attenuation and maintains substantial IR luminosity.  

We stress that obtaining a distribution of $L_{\rm UV}^{\rm att}$ while computing a single $L_{\rm IR}$ is physically consistent: UV attenuation varies strongly with line of sight through a clumpy, turbulent ISM, whereas the IR luminosity is set by the total energy absorbed by dust and is therefore effectively insensitive to viewing angle.  This asymmetry is a well-established result of radiative–transfer calculations, as explicitly shown in e.g. the GMC plus radiative–transfer simulations of \citet[][their Fig.~5]{DiMascia24} or in the high-$z$ galaxy simulations post-processing of \citep{Cochrane_2024}, where the UV output depends markedly on orientation while the IR emission remains unchanged.

\section{Ultraviolet and Infrared Luminosity Functions at $z=7-14$} \label{sect:uv_ir_lfs}

\begin{figure} 
    \centering
    \includegraphics[width=0.39\textwidth]{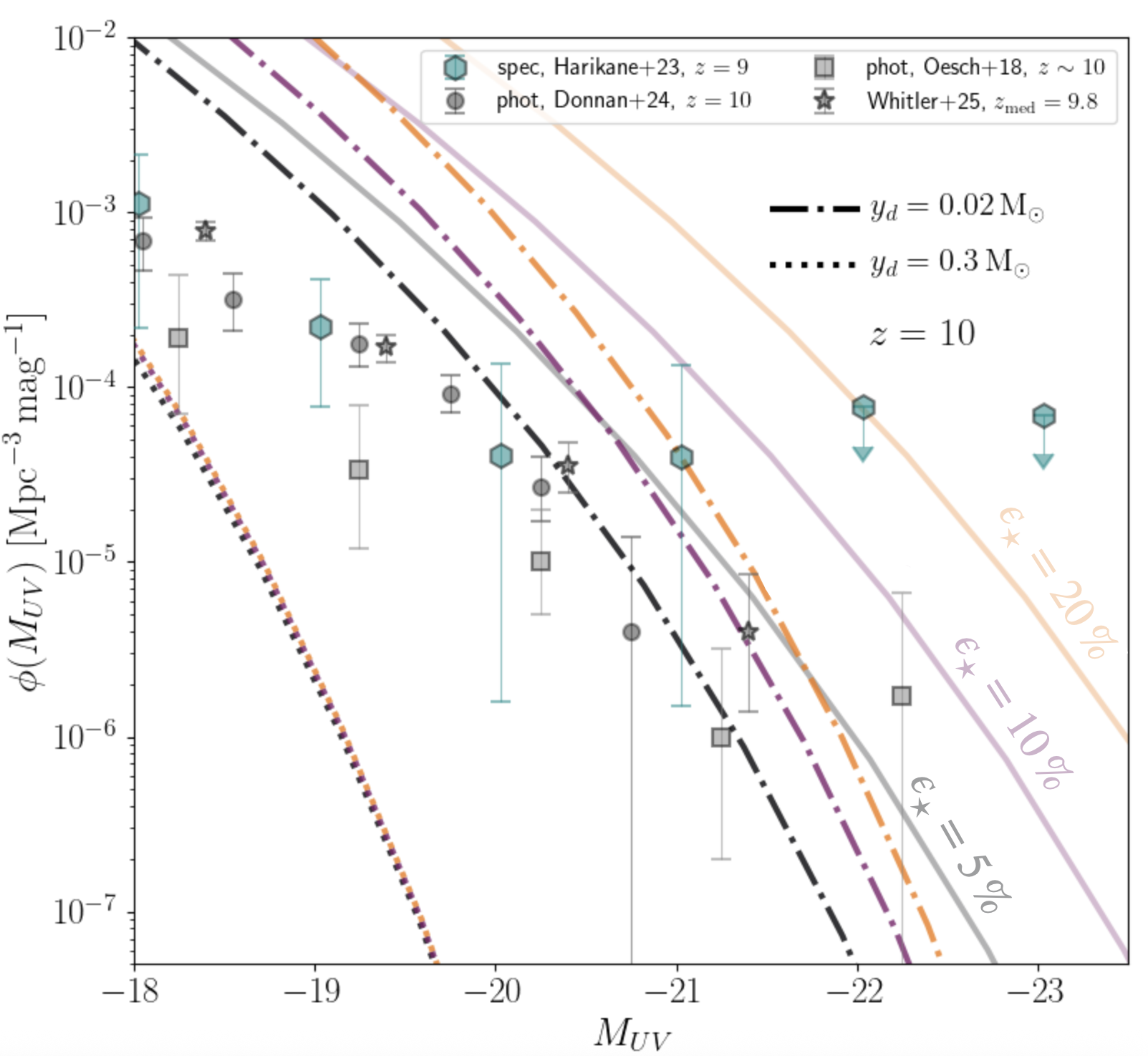}
    \includegraphics[width=0.39\textwidth]{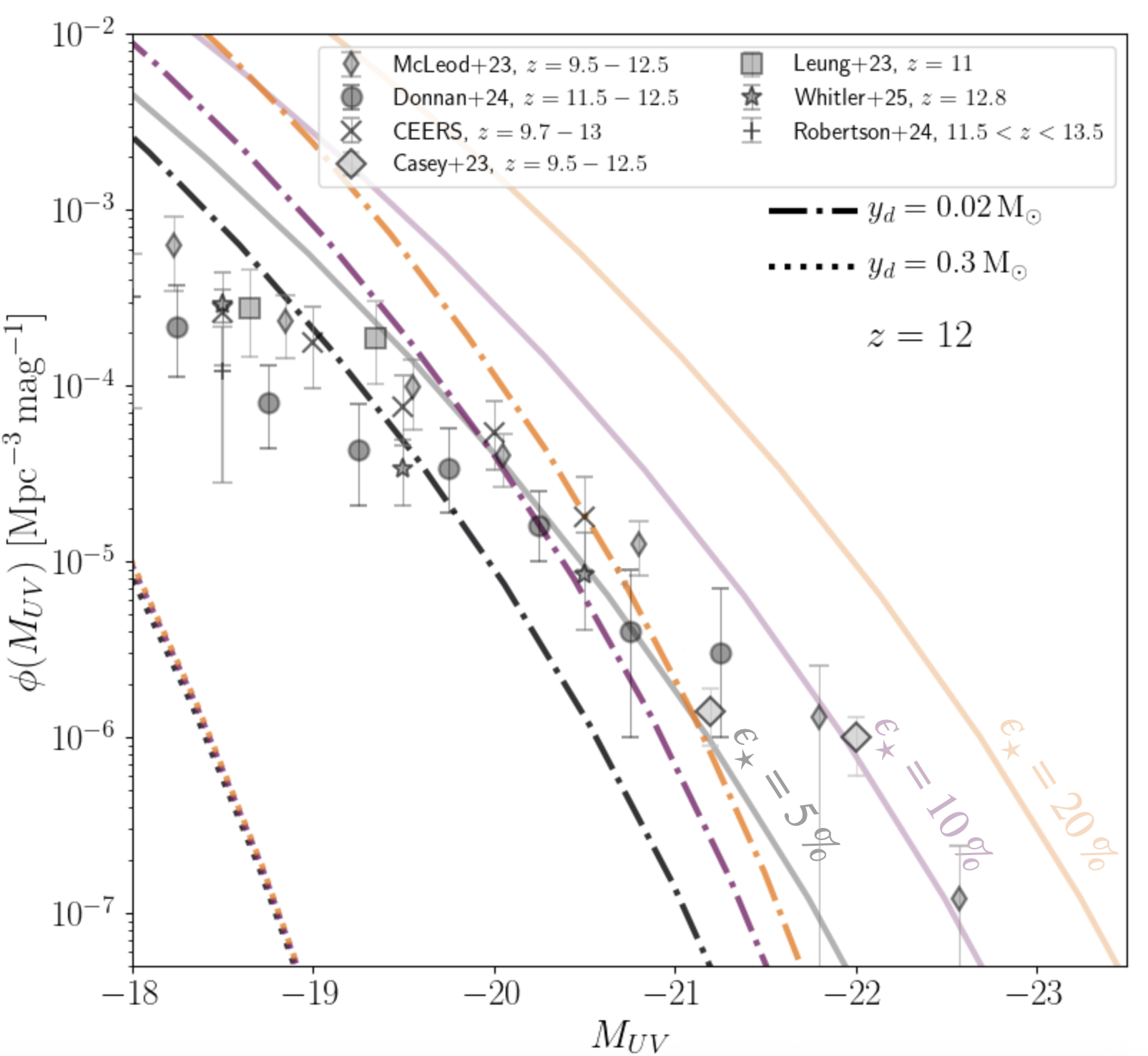}
    \includegraphics[width=0.393\textwidth]{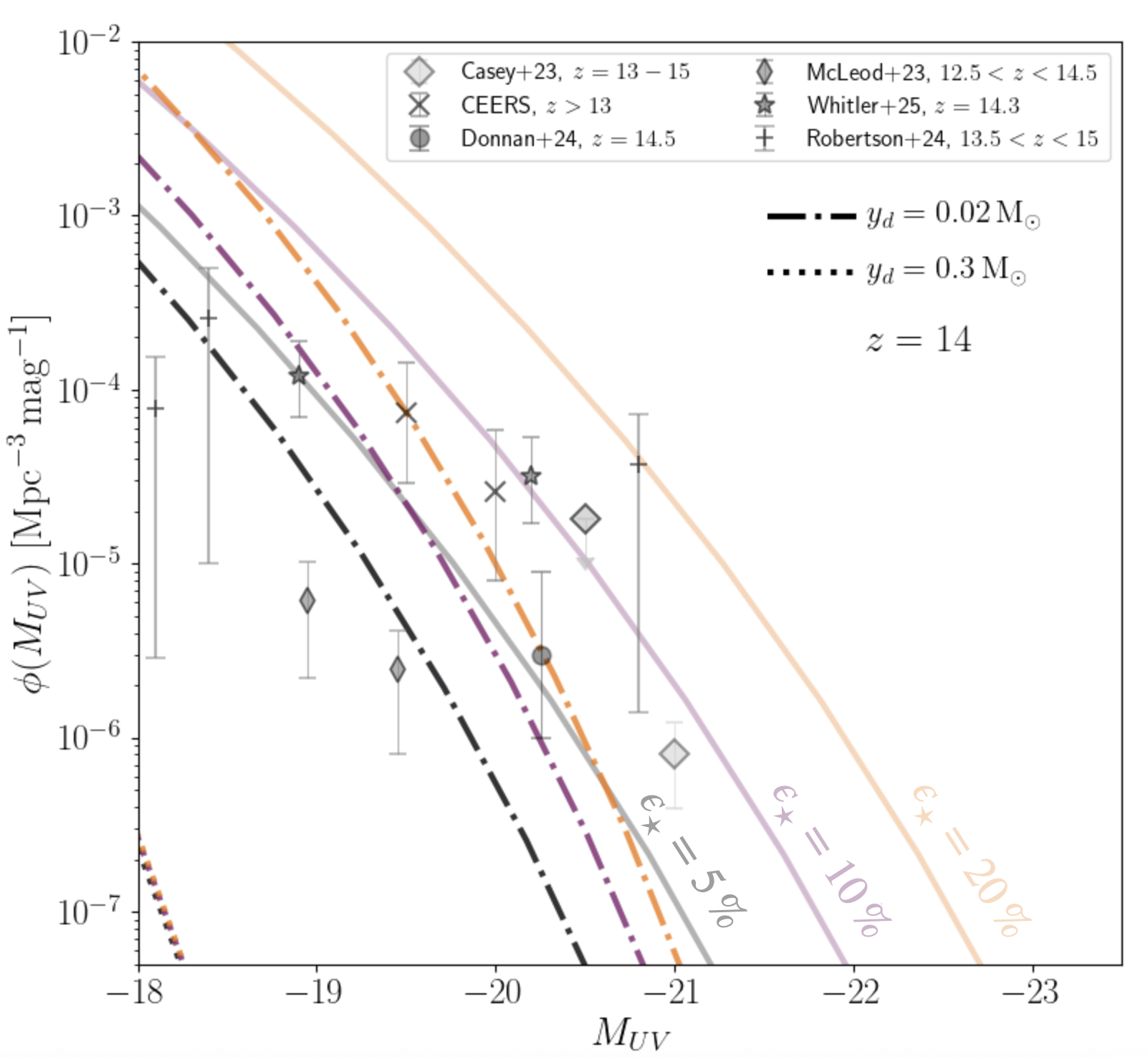}
    \caption{ 
    \textbf{From Top to bottom:} Predicted UV luminosity function at $z=10, 12, 14$ for the same set of model assumptions, obtained by forward-modeling attenuation on a galaxy-by-galaxy basis following the method described in the text.  The results are compared with current JWST constraints 
    (see references in the legend). The light, transparent curves show the intrinsic luminosity functions. 
    \textit{Overall: At $z \sim 10$, the bright end of the UV luminosity function can be reproduced with \quotes{normal} star formation efficiency ($\epsilon_\star \simeq 5\%$; black curves) provided the dust yield is low. High dust yields strongly suppress the UV luminosity at all redshifts for MW dust and uniform ISM. At $z \gtrsim 12$, galaxies with $M_{\rm UV} \lesssim -20$ require higher star formation efficiencies, and at $z = 14$ the brightest end remains underpredicted even for $\epsilon_\star = 20\%$ and the lowest dust yield.
}}
    \label{Fig_fobs-UVLF-z1014}
\end{figure}

\begin{figure*}[t]
\centering

\begin{minipage}[c]{0.05\textwidth}
  \centering
  \rotatebox{90}{\textbf{MW dust, Uniform ISM}}
\end{minipage}%
\hfill
\begin{minipage}[c]{0.45\textwidth}
  \centering
  \includegraphics[width=\linewidth]{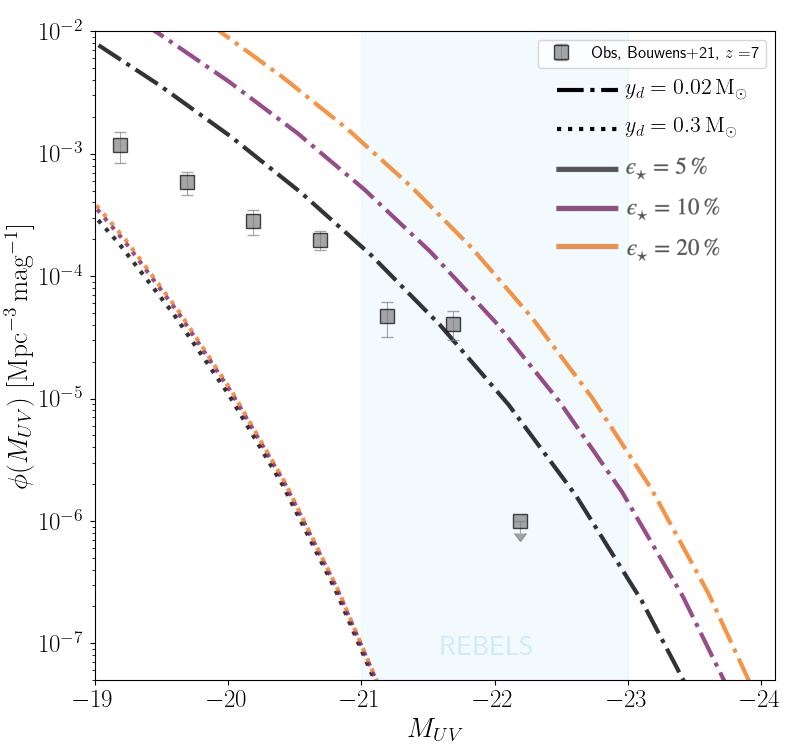}
\end{minipage}%
\hfill
\begin{minipage}[c]{0.45\textwidth}
  \centering
  \includegraphics[width=\linewidth]{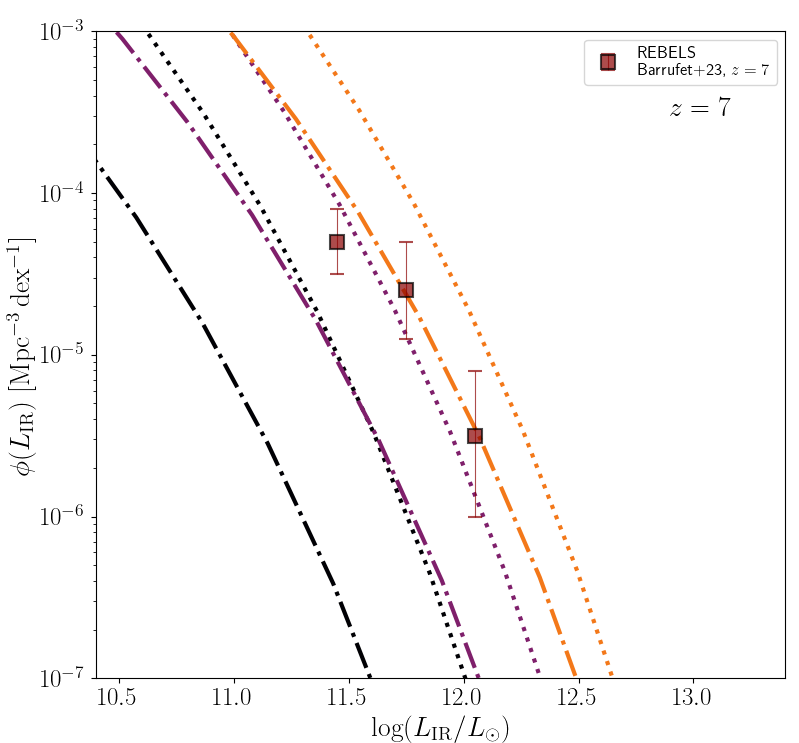}
\end{minipage}

\vspace{1.0em}

\begin{minipage}[c]{0.05\textwidth}
  \centering
  \rotatebox{90}{\textbf{Stellar Dust, Uniform ISM}}
\end{minipage}%
\hfill
\begin{minipage}[c]{0.45\textwidth}
  \centering
  \includegraphics[width=\linewidth]{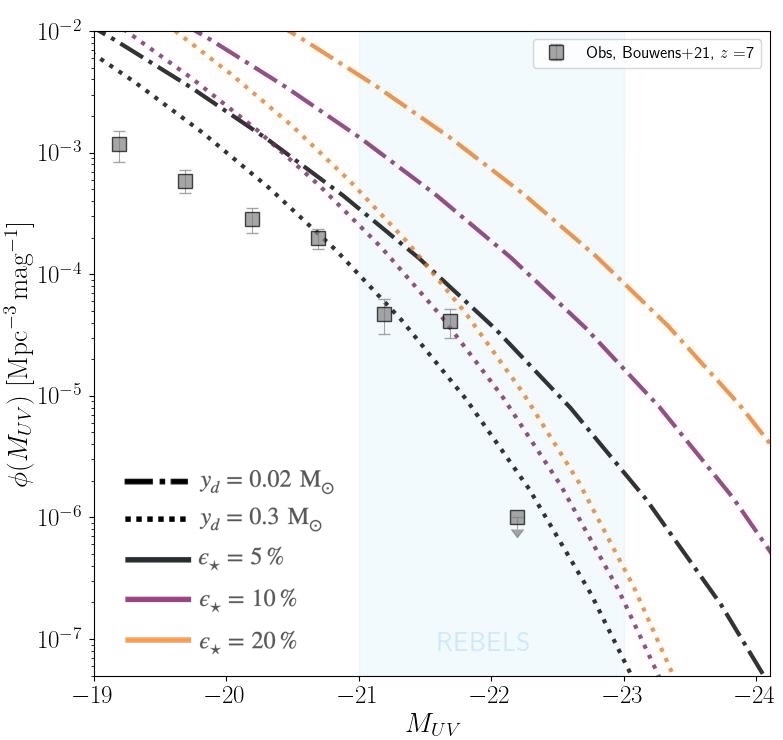}
\end{minipage}%
\hfill
\begin{minipage}[c]{0.45\textwidth}
  \centering
  \includegraphics[width=\linewidth]{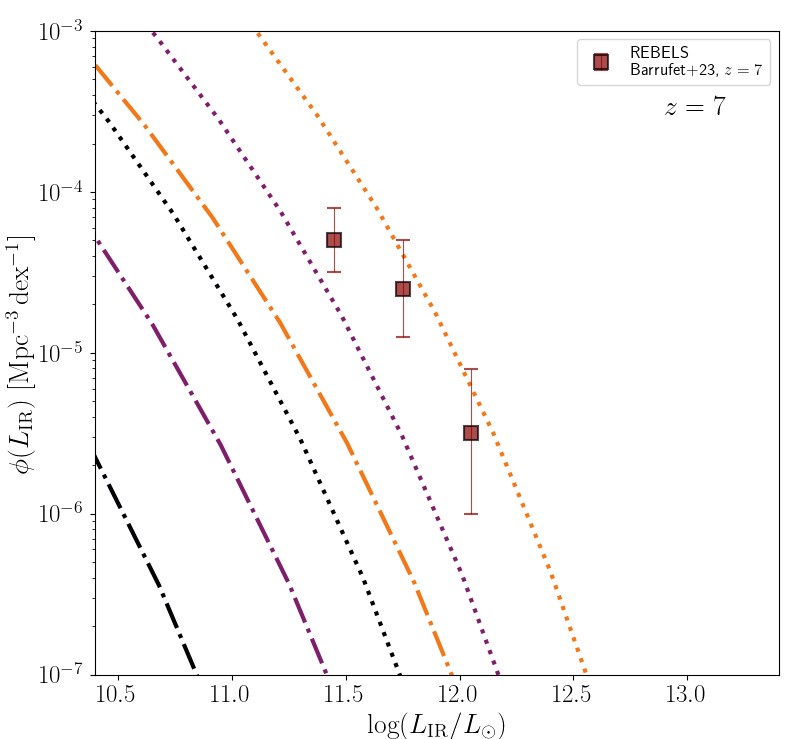}
\end{minipage}

\caption{UV (\textbf{left}) and IR (\textbf{right}) luminosity functions at $z=7$ assuming a uniform ISM. \textbf{Top row:} Results using Milky Way dust \citep{Draine03} In each panel, curves show different star–formation efficiencies (color, as in Fig.~\ref{Fig_fobs-UVLF-z1014}) and dust yields (line style), as in Fig.~\ref{Fig_fobs-UVLF-z1014}  UV and IR LFs are compared with HST \citep{Bouwens21} and ALMA data-inferred LFs \citep{Barrufet23}. \textbf{Bottom row:} Same as the top row but adopting the stellar–dust model from \citet{Hirashita19} (see also Fig.~\ref{Fig_att_curve}). \textit{Overall: for Milky Way dust in a uniform ISM, no combination of $\epsilon_\star$ and $y_d$ simultaneously reproduces the UV and IR LF; this tension is partly alleviated when considering stellar dust, where high dust yields and intermediate $\epsilon_\star$ (purple dotted line) perform noticeably better, though still not perfectly.}}
\label{fig:uv_ir_lf_z7_panels}
\end{figure*}

We now move from individual galaxies to population–level statistics by constructing the UV and IR luminosity functions across redshift. For each redshift, we consider four dust/ISM treatments:  
(i) Milky Way dust with a uniform ISM,  
(ii) stellar dust with a uniform ISM, and  
(iii)-(iv) a porous/turbulent ISM, adopting both Milky Way and stellar dust compositions.

Once the intrinsic and attenuated UV magnitudes and IR luminosities of individual galaxies have been computed (Secs.~\ref{sect:model_mass},\ref{subsec:dust}), a uniform ISM yields a one–to–one relation between halo mass and emergent luminosity. We approximate the impact of halo-spin variations by sampling a small set of stratified spin quantiles and combining the resulting transmissions into an effective $\left< T_{1500}(M_h)\right>$, which smooths size--induced fluctuations in the attenuation without explicitly propagating the full spin-induced scatter into the luminosity function.
The luminosity functions then follow directly from the Jacobian of the halo mass function. For the UV,
\begin{equation}
\phi(\MUV)
=
\frac{dn}{d\log M_h}
\left|
\frac{d\MUV}{d\log M_h}
\right|^{-1},
\end{equation}
and for the IR LF we use the identical expression but with $\MUV$ replaced by $\log L_{\rm IR}$.  
In practice, we evaluate $\MUV(M_h)$ and $L_{\rm IR}(M_h)$ on our halo mass grid, smooth the resulting curves, and compute the derivatives via finite differences.

As discussed in the previous section, in the turbulent ISM the emergent UV magnitude at fixed halo mass is no longer unique. Each halo hosts a distribution of dust surface densities along different sightlines, arising from the combined effects of geometric (size–driven) variations and turbulence–induced fluctuations. We model this by constructing, for each halo mass $M_h$, a single effective probability distribution $P(\Sigma_d\,|\,M_h)$ in dust surface density, obtained by convolving the variances associated with geometry (size) and turbulence in log–space (Sec.~\ref{sect:turbulent_ism}). 

Because attenuation is now stochastic, the usual Jacobian expression for the UV LF cannot be applied. Instead, for each halo mass $M_h$ we compute the fraction of sightlines that remain brighter than a given observed $M_{\rm UV}$. For the corresponding
required transmitted fraction, $T_{1500}$, we invert the radiative–transfer solution $T_{1500}(\tau_{1500})$ to obtain the maximum allowed optical depth, $\tau_{\max}(T_{1500})$, and therefore the surface–density threshold
\begin{equation}
    \Sigma_{d,\max}(M_{\rm UV}, M_h)
    = \frac{\tau_{\max}}{\kappa_{1500}} .
\end{equation}
The fraction of sightlines capable of producing an observed magnitude brighter than $M_{\rm UV}$ is then
\begin{equation}
    f(M_{\rm UV}\,|\,M_h)
    = \int_0^{\Sigma_{d,\max}}
      P(\Sigma_d\,|\,M_h)\, d\Sigma_d ,
\end{equation}
where ${\rm PDF}(\Sigma_d\,|\,M_h)$ is the lognormal distribution of $\Sigma_d$ accounting for both turbulence and variations in $\Lambda$ introduced in Sec.~\ref{sect:turbulent_ism}.
The number density of galaxies brighter than $M_{\rm UV}$ is
\begin{equation}
    N(<M_{\rm UV}) =
    \int_{M_h(M_{\rm UV}^{\rm int})}^{\infty}
    \frac{dN}{dM_h}\,
    f(M_{\rm UV}\,|\,M_h)\, dM_h ,
\end{equation}
where $M_h(M_{\rm UV}^{\rm int})$ is the inverse of the intrinsic magnitude–mass relation.  
Differentiating with respect to $M_{\rm UV}$ yields the UV luminosity function:
\begin{equation}
    \phi(M_{\rm UV}) =
    \int_{M_h(M_{\rm UV}^{\rm int})}^{\infty}
    \frac{dN}{dM_h}\,
    \left.
    \frac{\partial f(M_{\rm UV}'\,|\,M_h)}
         {\partial M_{\rm UV}'}
    \right|_{M_{\rm UV}' = M_{\rm UV}}
    dM_h .
\end{equation}
The derivative in the integrand can be written in closed form as
\begin{equation}
    \frac{\partial f}{\partial M_{\rm UV}} =
    P\!\left(\Sigma_{d,\max}\,|\,M_h\right)\,
    \frac{\partial \Sigma_{d,\max}}{\partial M_{\rm UV}},
\end{equation}
and is evaluated numerically in our implementation. This formalism correctly accounts for the stochastic, multi–sightline nature of dust attenuation and avoids the pathologies that arise when attenuation is treated as a deterministic function of halo mass.

For the IR LF in the turbulent ISM case we again use the same dust–surface–density distribution, but in contrast to the UV case all sightlines contribute to the emergent IR luminosity, due to the lack of attenuation. For each halo mass $M_h$ we thus compute the mean absorbed fraction by  $(1 - T_{1500}(\tau_{1500,abs}))$ over the lognormal $P(\Sigma_d\,|\,M_h)$, as already defined in eq.~\ref{eq_LIR_Mach}. This yields a single deterministic mapping $L_{\rm IR}(M_h)$, which remains monotonic. The IR LF can therefore be obtained with the standard Jacobian formalism,
\begin{equation}
    \phi(L_{\rm IR}) =
    \frac{dN}{dM_h}
    \left|\frac{dM_h}{dL_{\rm IR}}\right|,
\end{equation}
evaluated at the implicit inverse relation $M_h(L_{\rm IR})$.

We present the resulting UV and IR LFs for cases~(i)--(iv) across redshifts $7 \leq z \leq 14$ in the following.

\begin{figure*}[!t]
\centering

\begin{minipage}[c]{0.04\textwidth}
  \centering
  \rotatebox{90}{\textbf{Turbulent ISM, MW dust}}
\end{minipage}%
\hfill
\begin{minipage}[c]{0.89\textwidth}
  \centering
  \includegraphics[width=0.95\linewidth]{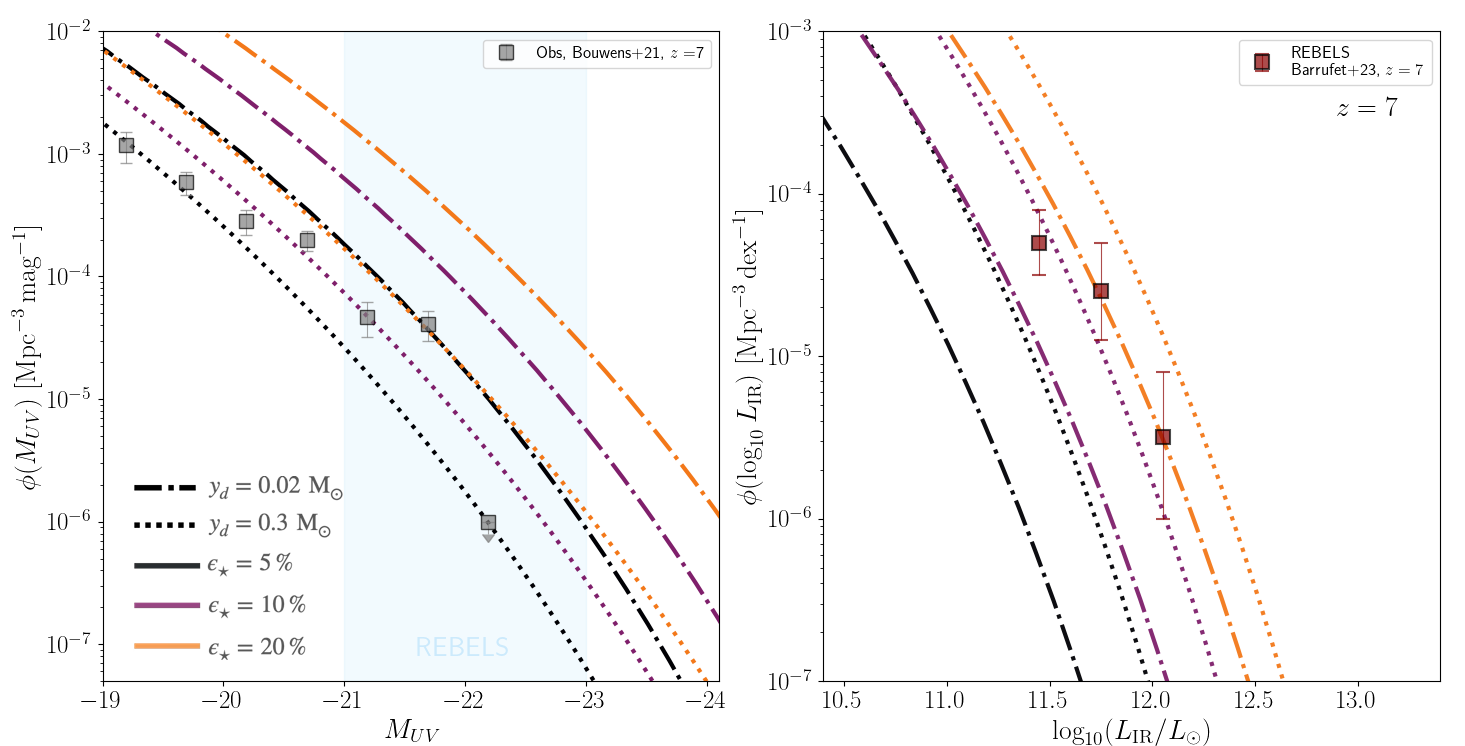}
\end{minipage}

\vspace{1.2em}

\begin{minipage}[c]{0.04\textwidth}
  \centering
  \rotatebox{90}{\textbf{Turbulent ISM, Stellar Dust}}
\end{minipage}%
\hfill
\begin{minipage}[c]{0.89\textwidth}
  \centering
  \includegraphics[width=0.95\linewidth]{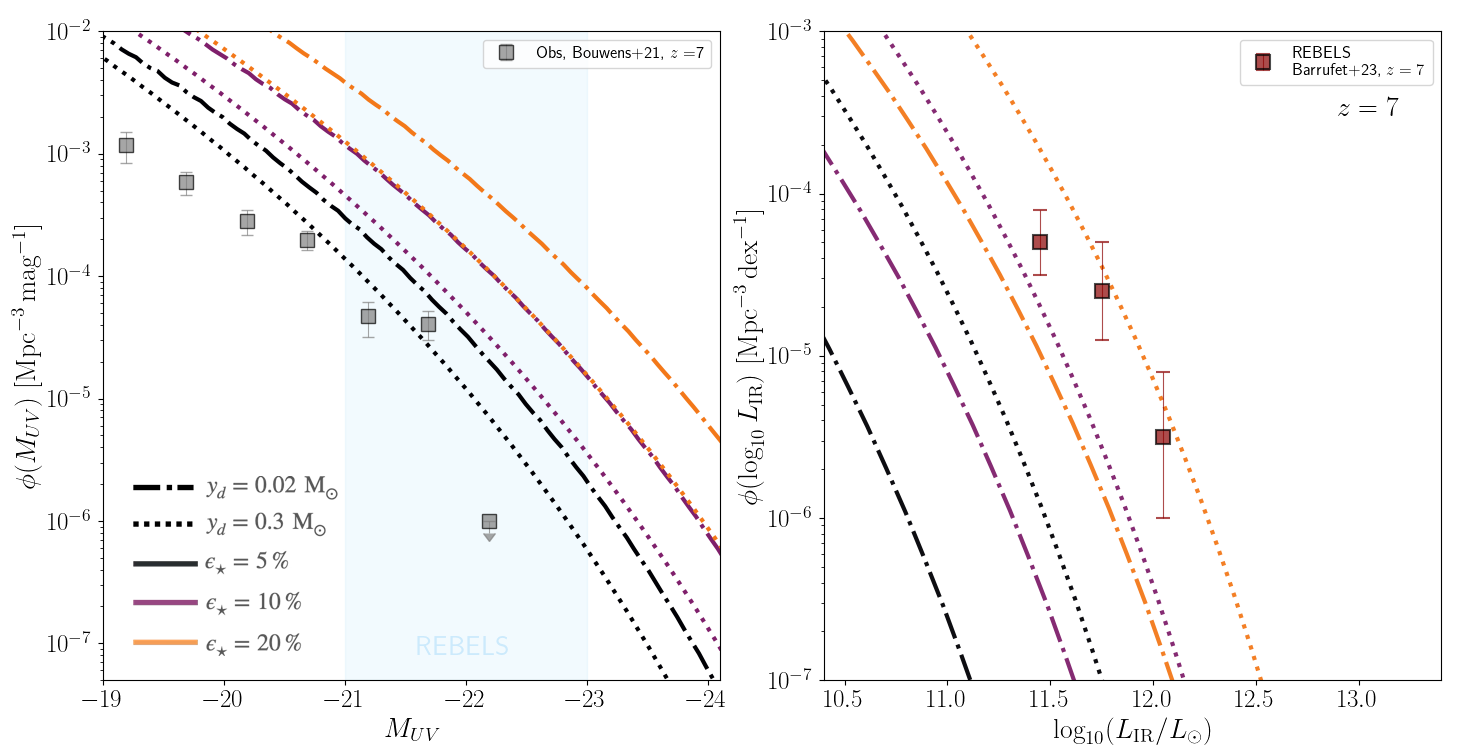}
\end{minipage}

\caption{UV and IR luminosity functions at $z=7$, plotted as in Fig.~\ref{fig:uv_ir_lf_z7_panels} but instead of a uniform dust surface density, here we consider a turbulent ISM $\mathcal{M}=30$. See text for further details on the implementation. \textit{Overall: Once both radiative transfer and ISM turbulence/porosity are included, stellar dust is no longer strictly required to reconcile the bright-end UV and IR emission: with high dust yields and moderate star-formation efficiency (purple dotted line), one can qualitatively reproduce both LFs using Milky Way dust.}}
\label{fig:LF_z7_turbulent}
\end{figure*}

\subsection{LFs for Uniform ISM case}
We begin by discussing \(z=10\), the redshift regime we have focused on so far. The comparison between our models and the observed UV LFs for this regime are shown in the top–right panel of Fig.~\ref{Fig_fobs-UVLF-z1014}. At \(z=10\), we find that observations at the bright-end of the UV LF can be reproduced without requiring an increase in star formation efficiency beyond \(\sfe = 0.05\), as long as the dust yield remains \(\leq0.02\,M_\odot\) per SN. Larger dust yields over-attenuate galaxies and suppress the number of sources brighter than \(M_{\mathrm{UV}} \sim -19\) (see dotted lines in Fig.~\ref{Fig_fobs-UVLF-z1014}).  
As shown in the lower panels of Fig.~\ref{Fig_fobs-UVLF-z1014}, at \(z=12\) the UV magnitude range $-21 \simlt \MUV\simlt -19 $ is well matched for \(\sfe = 0.05\), but reproducing galaxies brighter than \(M_{\mathrm{UV}} \sim -21\) requires higher efficiencies. At \(z=14\), this threshold shifts further, with \(\sfe > 0.05\) needed to match the observed counts already for galaxies brighter than \(M_{\mathrm{UV}} \sim -20\).

At lower redshift we can also compare our predictions with observed IR luminosity functions. The UV and IR LFs at \(z=7\) are shown in Fig.~\ref{fig:uv_ir_lf_z7_panels}. As already seen at \(z > 10\) the bright end of the UV LF (\(M_{\mathrm{UV}} \lesssim -21\)) can be reproduced assuming a fiducial star formation efficiency \(\sfe = 0.05\) and relatively low dust yields \(\yd < 0.1\,M_\odot\). However, the same set of assumptions dramatically underpredicts the IR LF at \(z=7\). Matching the IR number counts of the REBELS sample requires either adopting the lowest dust yield \(\yd = 0.02\,M_\odot\) along with a high star formation efficiency \(\sfe = 0.2\), or a moderate SF efficiency \(\sfe = 0.1\) but maximally efficient dust production \(\yd = 0.3\,M_\odot\). This tension highlights the challenge of simultaneously satisfying both UV and IR constraints within physically motivated parameter ranges.

We note that our predicted luminosity functions tend to be systematically steeper than observed at the faint end, both in the UV and IR\footnote{For the IR LF, it is worth mentioning that only a few REBELS galaxies have multiple FIR–continuum observations \citep[REBELS-25, 38, 12;][]{Algera24_REB25, Algera24}. As a result, their IR luminosities are typically inferred by rescaling a single detection and assuming a fixed dust temperature (based on models such as \citealt{Sommovigo21, Ferrara22a}), and are therefore highly uncertain (for a detailed discussion, see \citealt{Sommovigo25_Td}). Upcoming ALMA Band 8 observations (project 2024.1.00406.S, PI Algera) will help reduce these uncertainties by providing a shorter-wavelength probe—adding rest-frame 88 $\rm{\mu m}$ continuum to the already 158 $\rm{\mu m}$ continuum detected REBELS galaxies.}. The turbulence prescription partially alleviates this steepness. A bursty star-formation prescription would likely have a similar effect, and will be explored in forthcoming work.
This is a well-known feature of semi-analytical models employing halo-mass-based star formation prescriptions. A steeper LF can be alleviated by introducing an explicit mass dependence in the star formation efficiency, such as \(\epsilon_{\star} = \epsilon_{\star,10} \left(M_{\mathrm{h}}/10^{10}\,M_\odot\right)^\alpha\) \citep[see e.g.][]{Ferrara23a,Park18}. However, as already discussed in the context of the galaxy size modeling, we intentionally avoid introducing such empirical tuning in this first exploratory work.

\begin{figure*}
    \centering
    \vspace{6pt}
    \parbox[c]{0.45\textwidth}{\centering \textbf{Turbulent ISM, MW dust}}%
    \parbox[c]{0.45\textwidth}{\centering \textbf{Turbulent ISM, Stellar Dust}}%

    \vspace{2pt}
    \includegraphics[width=0.45\textwidth]{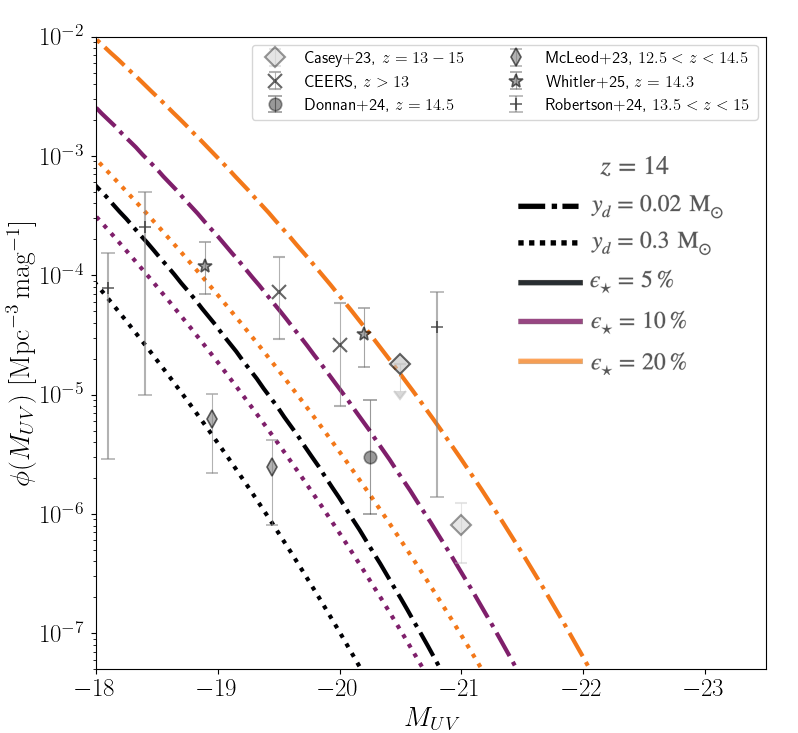}
    \includegraphics[width=0.45\textwidth]{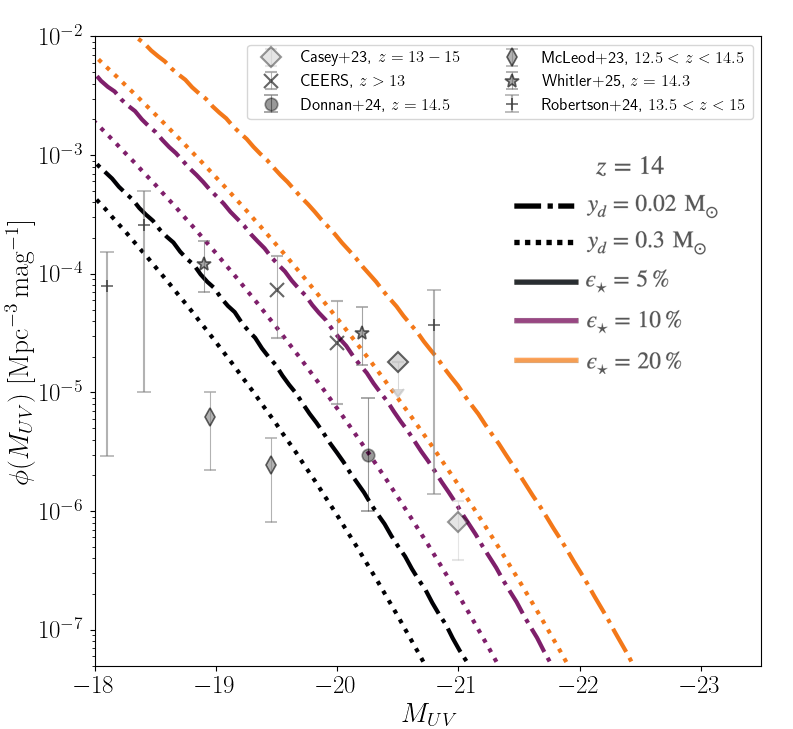}
    \caption{UV luminosity functions at $z=14$, plotted as in the bottom-right panel of Fig.~\ref{Fig_fobs-UVLF-z1014}, but including the effect of a turbulent ISM ($\mathcal{M}=30$). The left panel shows Milky Way dust, and the right panel stellar dust. See text for further details on the implementation. \textit{Overall: At $z=14$, however, ISM porosity alone is no longer sufficient: reproducing the observed bright end ($M_{\rm UV}<-20$) requires a star-formation efficiency twice as high ($\sfe=20\%$), and either stellar dust (dotted orange line, right panel) or reduced dust yields (orange dotted–dashed lines in both panels).} }
    \label{Fig_UVLF-z14-Mach-Hdust}
\end{figure*}

\subsection{The effect of: Stellar dust}
We treat the stellar--dust and Milky Way--dust models as limiting cases that should bracket the more complex (and likely evolving) nature of dust at high-redshift. In particular, adopting pure stellar dust should be regarded as an extreme scenario, since grain growth in the ISM is expected to operate on short timescales in dense, metal-enriched regions. Under these assumptions, we find that replacing MW dust with stellar dust (bottom panels of Fig.~\ref{fig:uv_ir_lf_z7_panels}) permits even high-yield models ($\yd = 0.3\,M_\odot$) with moderate star-formation efficiencies ($\sfe = 0.05$--0.1) to match the bright end of the $z=7$ UV LF at $\MUV\lesssim -20$ \citep[see also][]{Narayanan25}.

However, the lower UV opacities of pure stellar dust reduces the absorbed UV fraction and therefore the re-emitted IR luminosity, leading to a systematic underprediction of the IR LF relative to both the MW-dust case and the REBELS measurements (bottom-right panel of Fig.~\ref{fig:uv_ir_lf_z7_panels}). Thus, even this extreme \quotes{large-grain--dominated} scenario softens but does not eliminate the UV--IR tension over $z\sim7$--14.

It is also worth noting that attenuation curves based on $T_{1500,{\rm mix}}$ (but not $T_{1500,{\rm ps}}$) naturally flatten at high dust surface densities--even for MW-like dust-- as discussed in Sec.~\ref{subsec:dust}. This renders the impact of a pure stellar-dust distribution on the emergent luminosities less dramatic than naive extinction-curve arguments might suggest. Finally, tentative detections of PAH features in some $z\sim6$ galaxies indicate that ISM processing may operate more rapidly than predicted by the \citet{Hirashita19} model, possibly challenging the assumption that early galaxies are dominated by unprocessed stellar dust.

\subsection{The effect of: a Porous ISM}\label{subsect:turbulent_ism_LF}

We now incorporate the turbulence–driven porous ISM model into the luminosity–function calculation at $z\simeq7$--14. 

We show the result in Fig.~\ref{fig:LF_z7_turbulent} at $z=7$. We find that large $\mathcal{M}$ substantially boosts the probability of low–attenuation sightlines; thus, for fixed $(\sfe,\,y_{\rm d})$, the UV LF rises while the IR LF is controlled by the average absorbed fraction. With a large but plausible $\mathcal{M}=30$, a moderate efficiency $\sfe=0.1$ and high yield $y_{\rm d}=0.3\,M_\odot$ per SN we can reproduce the UV LF and come within $1\sigma $ of the IR LF. Using stellar dust instead of MW dust improves the UV agreement even for the lowest $\sfe=0.05$, but --as in the uniform case-- leads to a systematic underprediction of the IR LF.

At $z\simeq14$, turbulence continues to alleviate attenuation at the bright end, but only to an extent. For moderate star-formation efficiencies ($\sfe \lesssim 0.1$), both MW--like and large-grain dust models underpredict the brightest sources (roughly beyond $\MUV\simeq-19$ and $-20$, respectively) when high dust yields are assumed. Simply increasing $\sfe$ is not a sufficient remedy: even for strong turbulence ($\mathcal{M}=30$), we need to invoke low dust yields ($y_d<0.02\ \rm{M_{\odot}}$) or high $\sfe$ to reproduce the high number counts at $\MUV<-20$ at $z=14$. This tension for MW--like dust and high dust yields is also qualitatively consistent with the IR SED constraints presented in Appendix~\ref{App_IR_z12}. For $\yd \gtrsim 0.1\,M_\odot$, MW opacities tend to push the predicted IR fluxes close to current upper limits for $z\gtrsim10$ systems, whereas lower effective yields or stellar dust opacities remain more easily compatible with the data.

In conclusion, two broad classes of explanations remain consistent with the data under the moderate $\sfe$ assumption: (i) Supernova dust yields may be low ($\yd \ll 0.3\,M_\odot$), with grain growth becoming important by $z\simeq7$. In this case, even if the total dust mass becomes large at later times, a porous ISM can allow substantial UV escape, independently of the grain size distribution.  
(ii) Alternatively, galaxies may remain UV–bright if dust is efficiently ejected or destroyed by stellar feedback, as proposed in the AFM model \citep{Ferrara24a}. In this picture, high SN dust yields ($\yd \approx 0.3\,M_\odot$) are compatible with the observations because the most UV–luminous systems at $z>10$ would clear their ISM of dust, resulting in negligible attenuation.

\section{Discussion and Caveats}\label{sect:Cav_Disc}
While the framework presented here captures the key connections between halo growth, star formation, and dust attenuation at high redshift, several simplifying assumptions remain. We do not yet model explicit outflows, bursty star formation, IMF variations, or dust destruction and growth in the ISM. Below we outline the possible implications of neglecting such physical processes and prospects for future development. 

\paragraph{Outflows and Star Formation Regulation}
Our model currently lacks an explicit treatment of feedback-driven outflows or dust clearing. Outflows would facilitate higher UV luminosities for fixed SFE/dust yield, while having a more modest impact on the IR luminosities since feedback preferentially ejects lower column density sight lines in a turbulent medium \citep{Thompson16, Menon_25}. 
More generally, in our framework the net effect of feedback-driven dust removal can be interpreted, to zeroth order, as a reduction in the effective dust yield $y_{\rm d}$, which already encapsulates net dust production after unresolved destruction and removal processes. In addition, feedback can redistribute metals and dust to larger radii, lowering central column densities and therefore the effective attenuation at fixed dust mass. Part of this qualitative effect is indirectly captured by the broad distribution of effective dust radii arising from halo spin variations. As shown in Fig.~\ref{Fig_Muv-Mstar_att}, this produces a $\sim$2 mag spread in $\MUV$ at fixed mass, with an even larger spread in the point-source transmission case, where size variations shift systems across the optically thick threshold.

It is worth noting that, following \citet{Ferrara25}, none of our model galaxies reach the critical specific star formation rate ($\mathrm{sSFR}_{\rm crit} \approx 25\,\mathrm{Gyr^{-1}}$) required for large-scale dust expulsion\footnote{In the stellar dust model, a $z = 14$ galaxy with intrinsic $\MUV \simeq -20$ has $M_\star \sim 10^{8.4}\,M_\odot$ for $\sfe = 0.1$, corresponding to $\mathrm{sSFR} \simeq 10.4\,\mathrm{Gyr^{-1}}$, well below the threshold for efficient dust clearing in the AFM scenario}. This reflects our choice to tie the star-formation timescale to the halo accretion rate rather than to the gas free-fall time, an approach that successfully reproduces the observed evolution of dust temperatures \citep{Sommovigo22_z7,Sommovigo22_z57}. 

While UV-bright galaxies in the EoR do show evidence for dusty, metal-enriched outflows driven by young starbursts \citep[see e.g.][]{Marques-Chaves_2025}, the presence of dust within these outflows indicates that feedback does not straightforwardly imply dust-free systems. A self-consistent treatment of star formation, feedback \citep[see e.g.][]{Ferrara23a,Somerville25}, turbulence, and dust production is therefore required to assess whether outflows can alleviate the UV--IR tension, and is deferred to future work.

\paragraph{Dust Evolution and Attenuation Modeling}
We currently neglect dust growth and destruction, which
would require a self-consistent treatment of metal enrichment and the introduction of several poorly constrained parameters \citep{Popping17,Dayal22,Ferrara25}. To preserve a minimal and interpretable model, we instead treat $y_{\rm d}$ as the net dust production efficiency. Including dust growth could naturally keep dust-to-stellar mass ratios low at the highest redshifts, while allowing them to increase toward $z \sim 7$ as metallicity builds up.

Our attenuation-curve treatment is also idealized. Real curves depend on the evolving grain-size distribution, dust--star geometry, line-of-sight column density variations, and differential obscuration \citep{Narayanan18,Matsumoto25}. High-resolution simulations \citep[e.g.][]{Choban24,Dubois24,Matsumoto25} show rapid changes in UV slope and in the strength of the 2175\,\AA\ feature resulting from these effects. The noticeable differences in our model outcomes with MW-dust and pure stellar dust reinforces the need to understand the nature of dust and its attenuation properties at these epochs. However,  currently available simulations are not fully adequate for calibrating more detailed prescriptions due to limited resolution, post-processed dust, and their focus on the local Universe \citep[see e.g.][]{Sommovigo25_TNG}. Forthcoming high-redshift simulations with on-the-fly dust physics, combined with JWST attenuation-curve constraints \citep{Shivaei25,Markov24,Fisher25}, will be essential for future progress.

\paragraph{Bursty SF and IMF Variations}
We assume smooth star formation histories and a fixed Salpeter IMF (1--100~$\mathrm{M_\odot}$) at a stellar metallicity of $Z_\star = 0.1\,Z_\odot$. Current spectroscopic constraints indicate that these systems do not exhibit extremely low gas-phase metallicities \citep{curti2022,Curti24,Ferrara25}, making this assumption reasonable for the present analysis. A more detailed treatment of metallicity effects is deferred to future work.
Variations in the IMF or burstier star formation histories could increase the intrinsic UV luminosity \citep[e.g.][]{Trinca24}, but would also enhance dust production, making their net impact on the UV and IR luminosity functions non-trivial. As a result, it is unclear whether IMF variations alone would substantially alter the UV--IR tension discussed here, although a shallower IMF could help reconcile the brightest UV sources at $z \sim 14$, where IR constraints are currently lacking.

We do not introduce burstiness phenomenologically in the current model. Instead, it will be incorporated self-consistently once metal production and outflows are included, allowing feedback-driven variability to emerge naturally. Existing work shows that these processes are tightly coupled: strong SFR fluctuations reproduce bright JWST sources \citep{Pallottini24} but overproduce metallicity scatter relative to observations \citep{Nakajima23,Curti24,Chemerynska24b}. Hydrodynamical simulations likewise link burstiness to broadened metallicity distributions \citep{Sun23b,Marszewski24,Kravtsov24}. Reconciling UV-LF enhancements with chemical constraints requires following metals and feedback self-consistently.

\paragraph{Multi-Phase ISM}
We treat the ISM as a single turbulent region, effectively encoding an evolving covering fraction rather than a full multi-phase medium. This assumption is reasonable for extremely compact systems--- such as the \quotes{super clouds} associated with Blue Monsters \citep{Ziparo23}---but is oversimplified for more extended $z\sim7$ galaxies with likely distinct UV- and FIR-emitting phases.

UV--IR non-co-spatiality provides a distinct and complementary solution to the UV--IR luminosity function tension explored in this work, as it allows the UV and IR emission to arise from different ISM phases with different filling factors, potentially reconciling both luminosity functions without requiring changes to the global dust yield or star-formation efficiency. Such non-co-spatiality is well established locally \citep{Hoyer_23_Phangs,Kingfish_2011}, observed at high redshift through HST--ALMA offsets \citep{Hodge20,Bowler18,Carniani18,Laporte17,Inami22}, explains deviations from the IRX--$\beta$ relation in dusty systems \citep{Casey14,Elbaz18,Guijarro18}, and is predicted theoretically in EoR galaxies \citep{Faisst17,behrens18,Liang19,Sommovigo20,Sommovigo22_z57,Dayal22}.

\subsection{Comparison to previous literature}
Proposed astrophysical models -- i.e. that do not require altering the $\Lambda$CDM cosmological paradigm -- to explain the large number of UV bright galaxies unveiled by JWST at $z>10$ generally fall into three broad categories: (i) changes in galaxy feedback physics, (ii) modifications to star formation efficiency under extreme ISM conditions, and (iii) variations in dust physics. 

\paragraph{(i) Feedback-based scenarios}  
In the \emph{Feedback-Free Burst} (FFB) model of \citet{Dekel23} and its extensions by \citet{Li24}, star formation proceeds at extremely high efficiency in halos whose densities exceed a critical threshold, before feedback has time to act. Such scenario reproduces bright UV continua, but come at the cost of very high instantaneous efficiencies. In such cases, the resulting large dust masses inevitably raise the UV attenuation, unless the dust is expelled to large radii. This is the solution proposed in the \emph{attenuation-free model} (AFM) of \citet{Ferrara23a,Ferrara24a,Ferrara24b}, which suggests that dust is normally produced in vigorously star forming high-z systems, but rapidly expelled due to radiation pressure once the specific SFR exceeds a critical value, leaving galaxies effectively transparent in the UV. This introduces a fine balance between pushing the dust to achieve low attenuation and simultaneously preserving the IR emission observed in REBELS-like galaxies at $z\!\sim\!7$.

In this context, it is instructive to consider the relevant physical timescales governing dust production, removal, and reaccretion. A typical core-collapse supernova explodes after $\sim3-40$~Myr, while local giant molecular clouds (GMCs) disperse on timescales of $\sim2$--10~Myr \citep{Sommovigo20,Chevance2023}. Once supernova feedback launches galactic winds, a characteristic velocity of $v_{\rm w}\!\sim\!200$~km~s$^{-1}$ implies a clearing time of only $\sim3$~Myr for a galaxy of radius $\sim660$~pc. Thus, dust \quotes{production} and \quotes{clearing} processes act on timescales that are orders of magnitude shorter than the $\sim289$~Myr cosmic time interval between $z=10$ and $z=7$. 
By contrast, the baryonic accretion timescale for a halo of $M_{\rm h}\!\approx\!10^{10.2}\,{\rm M_\odot}$ hosting a $M_\star\!\approx\!10^{9}\,{\rm M_\odot}$ galaxy at $z=7$ is $\tau_{\rm acc}\!\sim\!295$~Myr—comparable to the time elapsed between these two epochs. This means that even if dust is efficiently expelled or partially destroyed at very high redshift, halos have ample time to re-form and/or re-accrete metal-enriched gas, retaining it due to their higher potential wells (i.e. lower sSFR) at $z\sim 7$. Thus, a turbulent ISM may still need to be invoked to reproduce the UV and IR LFs at $z=7$.

Finally, it is worth mentioning the recent SPICE \citep{Bhagwat23} radiation–hydrodynamic simulations \citep{Basu25}, which explicitly demonstrate how different supernova feedback prescriptions modulate the \emph{bursty} nature of star formation and thereby the $M_h$–$\MUV$ relation. While their simulation volume does not extend to the most massive systems uncovered by JWST, extrapolations from their models suggest that variability driven by different modes of supernova feedback could directly influence the bright end of the UV luminosity function at $z>10$.

\paragraph{(ii) ISM–conditions–driven star–formation regulation}  
A second family of models attributes the early UV–bright population to how star formation couples to the dense, turbulent ISM at high redshift. The \emph{Density–Modulated SFE} (DMSFE) framework of \citet{Somerville25} links $\sfe$ to the gas surface density, boosting the efficiency integrated over a galaxy's lifetime (though not per free–fall time), while FIRE–2 simulations \citep[e.g.][]{Sun23,Gelli24} show that bursty, feedback–regulated star formation can transiently enhance $\MUV$ without altering the long–term average efficiency. Our results echo the lack of need for a systematically higher star–formation efficiency, although the impact of short–term variability has not been explored here. This is consistent with recent observational analyses that find no evidence for elevated SFEs in the most massive $z>8$ galaxies \citep[e.g.][]{Chworowsky24}.

\paragraph{(iii) Dust physics}  
Finally, several recent studies have highlighted the role of dust grain properties. \citet{Narayanan25} suggest that large-grain–dominated dust distributions ($a\!\sim\!0.2\,\mu$m) can lower UV opacity and produce blue slopes even for moderately dusty systems -- a result consistent with the dynamical dust model of \citet{Hirashita19}. We find a similar effect: adopting such stellar dust reduces the optical depth at 1500\,\AA{} by a factor of $\sim$8.5, helping to reconcile UV and IR constraints but not fully matching the IR luminosities of the most dust-rich $z\!\sim\!7$ galaxies, a point which had not been raised so far. Further, we show that adopting a porous ISM, and accounting for the flattening of the attenuation curves due to radiative transfer effects leads to qualitatively similar results (see Fig.\ref{Fig_att_curve}). 

Observationally, attenuation curves do appear to flatten toward earlier epochs—consistent with large-grain populations—as indicated by studies of \citet{Markov24} and \citet{Shivaei25}, although few sources still present more MW-like attenuation curves \citep{Fisher25,Ormerod25}. Yet detections of PAH emission at $z\!\approx\!6$ suggest that small grains and complex dust re-processing in the ISM already exists \citep{Witstok23,Ormerod25,Lin25}, implying that a uniform large-grain population cannot be the whole story.  

In summary, our model complements these frameworks by demonstrating that the observed properties of UV-bright, dust-free systems can emerge from geometry and dust physics alone, without invoking drastic changes to feedback, IMF, or global efficiency. 

\section{Summary} \label{sect:summary}

In this work, we construct a minimal analytical framework to assess whether the dust-free, UV-bright galaxies revealed by \textit{JWST} at $z>10$ can be simulatnously reconciled with the dusty, IR-luminous systems detected by \textit{ALMA} at $z\approx7$. Our model intentionally isolates the key physical parameters that regulate the UV and IR continuum emission of massive early galaxies—namely, the star formation efficiency ($\sfe$) and the dust yield per Type~II SN ($y_{\rm d}$)—and tests their consistency against observations both at the individual-galaxy level and across entire luminosity functions.

We begin with a \quotes{vanilla} configuration adopting Milky Way--like dust \citep{Weingartner01} in a uniform ISM geometry. In addition, we explore the implications of considering a large-grain pure-stellar dust model, and incorporating a turbulent ISM. 

\begin{itemize}[topsep=2pt, itemsep=4pt, leftmargin=15pt, label=$\diamond$]

    \item \textbf{Impact of disk size.}  
    Variations in the spin parameter produce large changes in galaxy (and dust) radii, generating substantial scatter in attenuation. Even for fixed parameters ($\sfe$, $y_{\rm d}$) = (0.1, 0.1), a $M_\star = 10^{8.5}\,{\rm M_\odot}$ galaxy at $z=10$ spans $\MUV=-17$ to $-19$ purely due to size variation.

    \item \textbf{UV--IR tension in the uniform ISM.}  
    Reproducing the bright-end UV luminosity functions at $z>10$ requires modest efficiencies ($\sfe=0.05$--0.1) and \emph{low} dust yields ($y_{\rm d}<0.01\,{\rm M_\odot}$).  
    At $z=7$, however, the IR luminosity functions demand the opposite: \emph{high} dust yields ($y_{\rm d}\sim0.3\,{\rm M_\odot}$) for typical $\sfe$.  
    No single parameter choice fits both epochs, and higher efficiencies overproduce the UV LF by generating a large hidden population of dust-obscured galaxies (missed by current UV observations).

    \item \textbf{Large-grain stellar dust reduces UV opacity.}  
    Adopting a grain-size distribution peaked at $a\sim0.2\,\mu$m \citep{Hirashita19}, with negligibly small grains, reduces the 1500\,\AA\ optical depth by a factor $\sim8.5$. This alleviates—but does not eliminate—the UV--IR tension at $z=7$, as the suppressed attenuation also lowers the re-emitted IR luminosity. Moreover, stellar dust should be regarded as an extreme limiting case as PAH (i.e. small grains) features are already tentatively detected at $z\sim6$.

    \item \textbf{Turbulent ISM: low-$\Sigma_d$ sightlines and high IR output.}  
    Introducing ISM turbulence, modeled as a lognormal column-density distribution whose width depends on the Mach number, allows UV photons to escape along low-$\Sigma_{\rm d}$ sightlines while leaving the total absorbed energy (and thus the IR luminosity) nearly unchanged. A high turbulence ($\mathcal{M}\!\gg\!10$) brings the $z=7$ UV and IR LFs into better agreement for high dust yields even without modifying the grain size distribution. This is true only when radiative transfer effects are included (i.e. the flattening of the attenuation curve at high dust surface densities), rather than using a fixed attenuation/extinction curve template.
    
    \item \textbf{Residual tension at $z\simeq14$.}
    At the earliest epochs even strong ISM turbulence fails to reproduce the abundance of very bright galaxies
    ($\MUV\lesssim -20$) if SN dust yields are high
    ($y_{\rm d}\gtrsim 0.1\,{\rm M_\odot}$) and star–formation efficiencies remain moderate (5--10\%). Under these assumptions, matching the bright end of the UV LF at $z=14$ would require higher star–formation efficiencies ($\sfe>0.2$), stellar dust and/or lower dust yields ($y_d\sim 0.01\,{\rm M_\odot}$).     
\end{itemize}

If the high dust–to–stellar mass ratios inferred at $z\simeq7$ are primarily produced by rapid grain growth in the ISM \citep{Algera25}, then SNe can remain relatively inefficient dust producers at all times. Early galaxies ($z\gtrsim10$) would then remain intrinsically dust-poor and UV bright, while the bulk of the dust mass builds up later.  

Even in this grain–growth scenario, however, accounting for the porosity of the ISM is essential. A turbulent, clumpy ISM allows galaxies to be simultaneously UV bright and IR bright at $z\simeq7$, a requirement that cannot be met in a uniform screen model and that highlights the importance of incorporating IR constraints. Once ISM porosity is included, no change in the underlying grain size distribution is needed to reconcile UV and IR luminosities.  

A complementary possibility is that dust is efficiently removed at early times by strong outflows \citep{Ferrara23a}, keeping the dust content of $z>10$ galaxies low even if SN yields are intrinsically high. Disentangling these effects—grain growth, ISM porosity, and feedback-driven dust removal—will require future work.

\acknowledgments{}
We thank Stefano Carniani, Mike Grudic, Irene Shivaei, Amiel Sternberg, and Aaron Yung for insightful discussions and valuable feedback that helped improve this work. 
L.S. additionally thanks Agnes Valenti and Ulrich Steinwandel for conceptual discussions that helped shape the overall direction of the study. 
Finally, we thank Bruce Draine for his pioneering commitment to open data releases; the publicly available optical–constant tables on his website have been essential for ensuring full reproducibility and accuracy in dust–physics calculations for many works over the last twenty years.

\appendix

\subsection{Predictions for Massive Galaxies at $z\simeq7$}
\label{build-up_z7_appx} 
\begin{figure*}
    \centering   \includegraphics[width=0.91\textwidth]{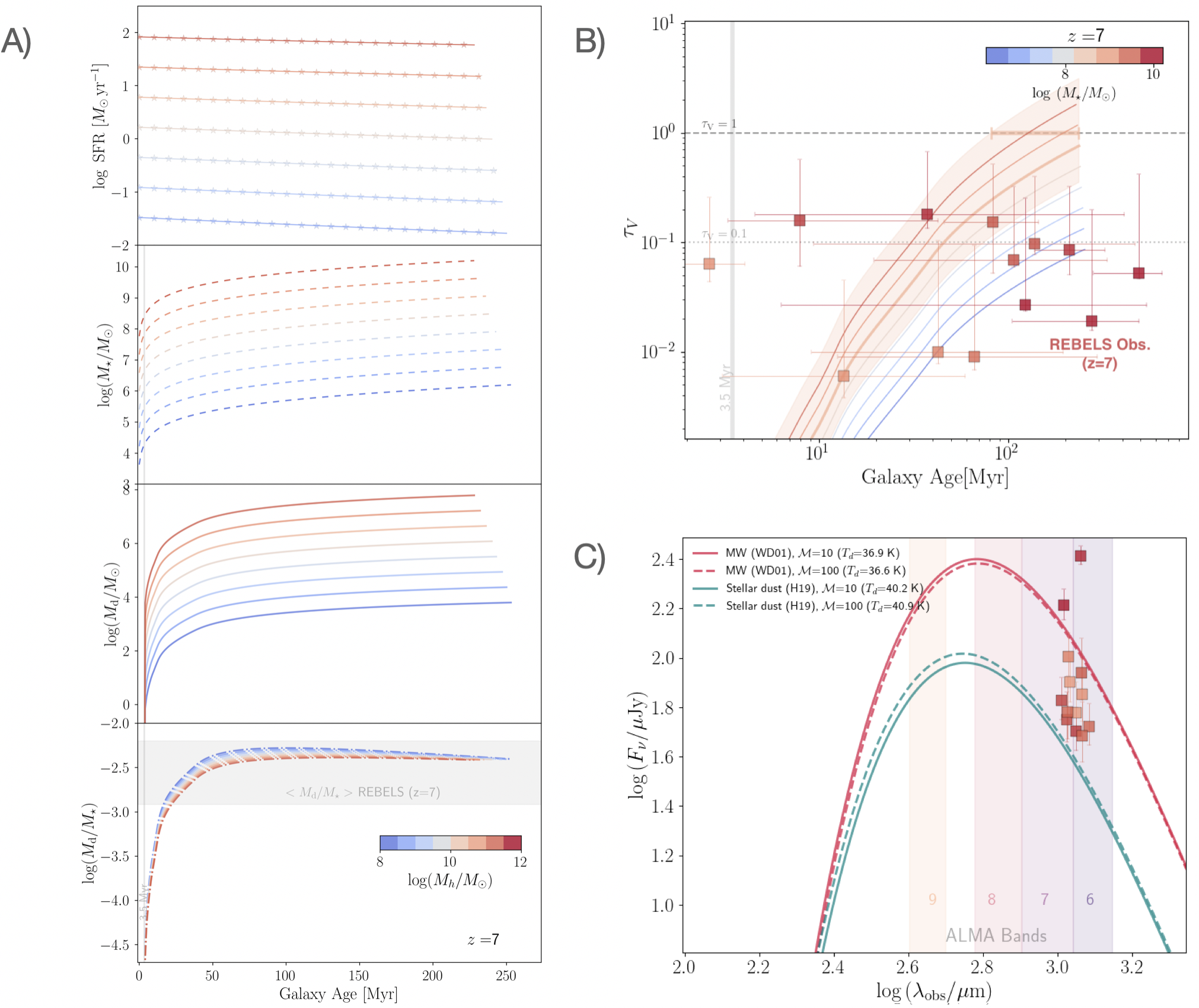}
    \caption{
    (\textbf{A}) Same as Fig.~\ref{Fig1}, but at redshift $z=7$, showing the mass, dust, and SFR evolution for the reference halo sample. 
    (\textbf{B}) Same as Fig.~\ref{Fig_opacity_tau}, but at $z=7$. We compare our predicted $\tau_V$ with optical depths inferred from SED fitting of dust–continuum detected REBELS galaxies \citep{Bouwens21}, which also underpin the IR luminosity function in Figs.~\ref{fig:uv_ir_lf_z7_panels} and~\ref{fig:LF_z7_turbulent} as derived by \citet{Barrufet23}. 
    (\textbf{C}) IR SED predicted by our model for a galaxy with stellar mass $\log (M_{\star}/M_{\odot})\simeq 9.8$ at $z=7$, corresponding to the median stellar mass of the continuum-detected $z\sim7$ REBELS galaxies assuming a non-parametric SFH \citep{topping2022}. We show MW dust (pink) and stellar dust (teal); REBELS continuum fluxes at rest-frame $158\,\mu$m \citep{Inami22} are overplotted and colour-coded by stellar mass.
    }
    \label{Fig_z7}
\end{figure*}

In Fig.~\ref{Fig_z7} we show what our framework predicts for massive galaxies at $z\!\sim\!7$ when applying exactly the same modelling described in Sec.~\ref{sect:model_mass} for the $z=10$ population. The qualitative behaviour remains unchanged: once the first supernovae explode, the dust mass rises rapidly 
and quickly approaches the dust--to--stellar mass ratio set by the adopted stellar dust yield. Here we show the fiducial case with $\sfe=0.1$ and $\yd=0.1$. The resulting optical depths are slightly lower than in the $z=10$ case because galaxies at $z\!\sim\!7$ are physically larger; for example, a halo of mass 
$\log(M_h/M_\odot)=10.9$ lies only marginally above the optically--thick threshold even in the uniform--ISM, high--$\yd$ limit.

Because many UV--bright $z\!\sim\!7$ systems have now been detected in the far--infrared (most notably those from the \textsc{rebels} survey), we also compute their expected IR spectral energy distributions. We assume a single--temperature greybody, following eq.~8 of \citet{Sommovigo21} (see also \citealt{daCunha13}), using the IR luminosities obtained from eqs.~\ref{eq_LIR_unif} and \ref{eq_LIR_Mach} together with the dust mass from eq.~\ref{md}. The dust temperature is derived from the standard greybody relation \citep{Ferrara22a}:
\begin{equation}
    \bar{T}_d 
    = \left( \frac{L_{\rm IR}}{\Theta\,M_d} \right)^{1/(4+\beta_d)},
\end{equation}
where $\beta_d$ is the dust emissivity index, and
\begin{equation}
    \Theta 
    = \frac{8\pi}{c^2}\, \kappa_{158,abs}\, \nu_{158}^{\,\beta_d}\,
      \frac{k_B^{4+\beta_d}}{h^{3+\beta_d}}\,
      \zeta(4+\beta_d)\,\Gamma(4+\beta_d)\ .
\end{equation}
Here $c$ is the speed of light, $k_B$ the Boltzmann constant, $h$ the Planck constant, $\nu_{158}$ the rest-frame frequency corresponding to $\lambda=158\,\mu$m, and $\kappa_{158,{\rm abs}}$ the mass absorption coefficient at this wavelength. The latter is computed self-consistently for each dust model from the adopted grain-size distribution, yielding $\kappa_{158,{\rm abs}}=12.79\ \mathrm{cm^2\,g^{-1}}$ for Milky Way dust and $\kappa_{158,{\rm abs}}=3.19\ \mathrm{cm^2\,g^{-1}}$ for the stellar-dust model.

As shown in Fig.~\ref{Fig_att_curve}, the stellar-dust model exhibits systematically lower opacities than the MW model across the entire wavelength range, with the largest differences occurring in the UV. At long wavelengths the contrast between the two models is weaker, though still significant.

The differences in the IR SEDs are therefore smaller than those seen in the UV, but not negligible: the total IR luminosities differ by up to $\sim0.3$--$0.4$ dex. This does not contradict the much larger discrepancy in UV attenuation. The observed UV magnitude depends exponentially on the transmission, $T_{1500} \approx e^{-\kappa_{1500}\Sigma_d}$,
so even a moderate change in $\kappa_{1500}$ produces a large shift in $\MUV$. By contrast, the absorbed fraction $f_{\rm abs}=1-T_{1500}$ saturates once the UV optical depth exceeds unity, as is the case for the massive halo considered here. Consequently, the total absorbed UV luminosity—and thus $L_{\rm IR}$—varies only at the factor-of-few level between the two dust models.

The resulting dust temperatures remain even more similar ($\Delta T_d \sim 4$ K). This is because the lower UV opacity in the stellar-dust model reduces $L_{\rm IR}$, but this effect is partially compensated by its $\sim \times 4$ lower far-IR mass absorption coefficient, which increases the equilibrium temperature required to reproduce a given far-IR emission at fixed dust mass. The combination of saturation in $f_{\rm abs}$ and the weaker contrast in far-IR opacities therefore explains why the IR SED shapes and temperatures remain comparable despite the much larger differences in the UV attenuation.

Finally, we note that for the median stellar mass of the \textsc{rebels} sample, our predicted rest--frame 158\,$\mu$m fluxes agree well with the measurements presented in \citet{Inami22}. The corresponding dust temperatures are also consistent with those obtained for the few galaxies with multi--band FIR detections \citep{Algera25}, although they lie slightly below the mean value inferred using the [C\,\textsc{ii}]-based model of \citet{Sommovigo22_z7} ($T_{\rm d}=47\pm7$\,K). 

\begin{figure}
    \centering
    \includegraphics[width=0.5\linewidth]{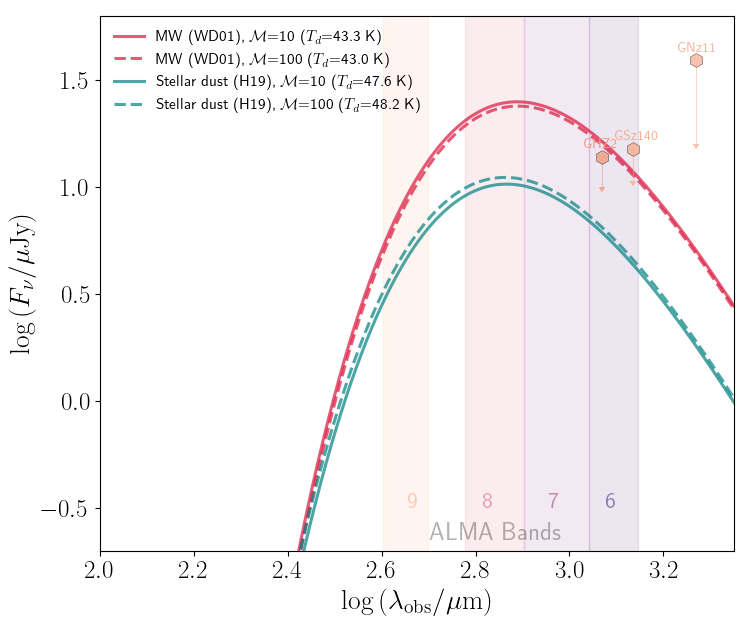}
    \caption{IR SED predicted by our model for a galaxy with stellar mass $\log (M_{\star}/M_{\odot}) = 8.87$ at $z=12.37$, corresponding to the mean stellar mass of the $z\gtrsim10$ sources with sub-mm follow-up (GNz11, GSz14, and GHZ2; \citealt{Fudamoto23, Schouws24, Bakx22}; see also \citealt{Carniani24_z14, Popping22}). As in panel C of Fig.~\ref{Fig_z7}, we show MW dust (pink) and stellar dust (teal) for $\yd=0.1$ and $\sfe=0.1$. The shaded regions indicate the ALMA bands. Observational IR flux measurements and upper limits are overplotted.}
    \label{FIG_ir_sed-z14}
\end{figure}

\subsection{IR SED and Sub-mm Constraints at $z\geq10$}\label{App_IR_z12}

Following the same formalism described above, we compute the IR SED at $z=12.37$ for a representative galaxy with $\log (M_\star/{\rm M_\odot}) = 8.87$, corresponding to the mean stellar mass of the $z\gtrsim10$ sources with sub-mm follow-up with ALMA and NOEMA (GNz11, GSz14, and GHZ2; \citealt{Fudamoto23, Schouws24, Bakx22}; see also \citealt{Carniani24_z14, Popping22}). We adopt the same fiducial parameters as in the $z=7$ case (Fig.~\ref{Fig_z7}C), namely $\yd=0.1$ and $\sfe=0.1$.

The dust temperatures are modestly higher than at $z=7$ ($T_d \sim 43$ K for MW dust and $\sim 48$ K for stellar dust), as expected from the shorter gas depletion times and correspondingly stronger radiation fields at earlier epochs in a single-zone equilibrium framework (\citealt{Sommovigo22_z7,Sommovigo22_z57}).
The difference between MW and stellar dust remains moderate but visible in the SED normalization, with the MW model brighter by $\sim0.3$ dex near the peak. The MW dust case lies close to current IR upper limits for GHZ2 and GSz14, whereas the stellar-dust model remains comfortably below them. While 
these constraints are based on non-detections, they already disfavour scenarios combining high dust yields with MW-like opacities at $z>10$, in line with the UV LF constraints discussed in Sec.~\ref{subsect:turbulent_ism_LF}.

We emphasize that modeling the IR emission with a single-temperature greybody is an idealization—though commonly adopted at very high redshift, where only one or two FIR continuum points are typically available (see \citealt{Sommovigo25_Td}). Moreover, stellar mass estimates for these 
$z>10$ systems remain uncertain (\citealt{Popping22, Cochrane25, Narayanan24_Mstar}), directly affecting the inferred dust masses and SED normalization. Given that present measurements are largely upper limits, the IR constraints should therefore be interpreted with caution.
Finally, the spatial distribution of dust is critical. If dust is transported to sufficiently large radii—e.g., via outflows, high halo spin, or radiation-pressure-driven redistribution—the resulting weaker radiation field can drive dust temperatures close to the CMB, rendering the emission effectively undetectable irrespective of the total dust mass (\citealt{Ferrara24c}).

\begin{figure}
    \centering
    \vspace{6pt}
    \includegraphics[width=0.5\textwidth]{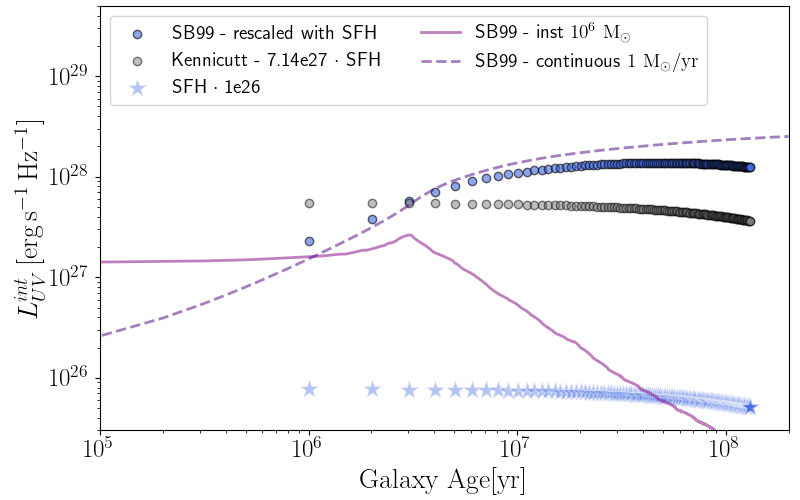}
    \caption{Time evolution of the intrinsic UV luminosity at $\rm 1500\ \dot{A}$, $L_{\rm UV}^{\rm int}$, for different models. Blue circles show $L_{\rm UV}$ computed by convolving the SB99 SSP with the time-dependent SFH from our reference model. Grey circles represent the luminosity estimated using the KS+98 calibration (see text. Light blue stars show the SFH described in eq.~\ref{sfr}, scaled by $10^{26}$ for comparison. For reference, purple lines show the SB99 outputs for an instantaneous burst of $10^6\ \rm{M_\odot}$ (solid) and for a continuous SFR of $1\ \rm{M_\odot}$/yr (dashed).}
    \label{Fig_LUV-time}
\end{figure}

\subsection{Impact of SFH Modelling on UV Luminosity Estimates} \label{SF_to_LUV_appx} 
We compare with what we should have obtained assuming a constant SFH and the simplistic conversion between SFR and UV luminosity relying on the empirical conversion $(\rm{L_{UV}}/\mathrm{erg\ s^{-1} Hz^{-1}})=7.14 \cdot 10^{27}\ (\rm{SFR}/\rm{M_{\odot} yr^{-1}})$ \citep{Kennicutt98}. 
As already discussed following Eq.~\ref{SFH_const_mst}, the SFH adopted in this work declines only mildly with time, resulting in a modulated UV luminosity ($\rm L^{int}_{UV}$) that closely resembles the one obtained under the assumption of a constant SFH, shown by the dashed purple line in Fig.~\ref{Fig_LUV-time} for an SFR of $1\ \mathrm{M_{\odot}\ yr^{-1}}$. In contrast, estimating $\rm L^{int}_{UV}$ by applying the empirical SFR–$\rm L_{UV}$ conversion separately in each time bin—without summing the UV light produced by stars formed at earlier times—leads to an underestimation of the galaxy’s total $\rm L_{UV}$, as illustrated by the difference between the blue and grey points in the figure.

We retain the binned SFH implementation, despite its smooth time evolution, because
its discrete structure makes it straightforward to introduce stochastic
perturbations and model bursty star formation episodes in future extensions of this
framework.

\bibliographystyle{aasjournal}
\bibliography{bibliography}

\end{document}

%% file: definitions.tex
\def\be{\begin{equation}}
\def\ee{\end{equation}}

\newcommand\quotes[1]{``{#1}"}
\def\gsim{\lower.5ex\hbox{\gtsima}} 
\def\lsim{\lower.5ex\hbox{\ltsima}} 
\def\gtsima{$\; \buildrel > \over \sim \;$} 
\def\ltsima{$\; \buildrel < \over \sim \;$} \def\gsim{\lower.5ex\hbox{\gtsima}} 
\def\lsim{\lower.5ex\hbox{\ltsima}} 
\def\simgt{\lower.5ex\hbox{\gtsima}} 
\def\simlt{\lower.5ex\hbox{\ltsima}}

\def\msun{{\rm M}_{\odot}}

\def\S*{$\Sigma_{\rm SFR}$}

\definecolor{apcolor}{HTML}{b3003b}
\definecolor{afcolor}{HTML}{800080}
\definecolor{lvcolor}{HTML}{DF7401}
\definecolor{mdcolor}{HTML}{01abdf} 
\definecolor{cbcolor}{HTML}{ff0000}
\definecolor{sccolor}{HTML}{cc5500} 
\definecolor{sgcolor}{HTML}{00cc7a}

%% file: main.bbl
\begin{thebibliography}{}
\expandafter\ifx\csname natexlab\endcsname\relax\def\natexlab#1{#1}\fi
\providecommand{\url}[1]{\href{#1}{#1}}
\providecommand{\dodoi}[1]{doi:~\href{http://doi.org/#1}{\nolinkurl{#1}}}
\providecommand{\doeprint}[1]{\href{http://ascl.net/#1}{\nolinkurl{http://ascl.net/#1}}}
\providecommand{\doarXiv}[1]{\href{https://arxiv.org/abs/#1}{\nolinkurl{https://arxiv.org/abs/#1}}}

\bibitem[{{Algera} {et~al.}(2024{\natexlab{a}}){Algera}, {Inami}, {De Looze}, {Ferrara}, {Hirashita}, {Aravena}, {Bakx}, {Bouwens}, {Bowler}, {Da Cunha}, {Dayal}, {Fudamoto}, {Hodge}, {Hygate}, {van Leeuwen}, {Nanayakkara}, {Palla}, {Pallottini}, {Rowland}, {Smit}, {Sommovigo}, {Stefanon}, {Vijayan}, \& {van der Werf}}]{Algera24_REB25}
{Algera}, H. S.~B., {Inami}, H., {De Looze}, I., {et~al.} 2024{\natexlab{a}}, \mnras, 533, 3098, \dodoi{10.1093/mnras/stae1994}

\bibitem[{{Algera} {et~al.}(2024{\natexlab{b}}){Algera}, {Inami}, {Sommovigo}, {Fudamoto}, {Schneider}, {Graziani}, {Dayal}, {Bouwens}, {Aravena}, {da Cunha}, {Ferrara}, {Hygate}, {van Leeuwen}, {De Looze}, {Palla}, {Pallottini}, {Smit}, {Stefanon}, {Topping}, \& {van der Werf}}]{Algera24}
{Algera}, H. S.~B., {Inami}, H., {Sommovigo}, L., {et~al.} 2024{\natexlab{b}}, \mnras, 527, 6867, \dodoi{10.1093/mnras/stad3111}

\bibitem[{{Algera} {et~al.}(2025{\natexlab{a}}){Algera}, {Rowland}, {Stefanon}, {Palla}, {Sommovigo}, {Inami}, {Bouwens}, {Aravena}, {Bowler}, {Dayal}, {De Looze}, {Ferrara}, {Fisher}, {Graziani}, {Gulis}, {Heintz}, {Hodge}, {Laza-Ramos}, {van Leeuwen}, {Pallottini}, {Phillips}, {Schouws}, {Smit}, {Stark}, \& {van der Werf}}]{Algera25}
{Algera}, H. S.~B., {Rowland}, L., {Stefanon}, M., {et~al.} 2025{\natexlab{a}}, \mnras, \dodoi{10.1093/mnras/staf1897}

\bibitem[{{Algera} {et~al.}(2025{\natexlab{b}}){Algera}, {Weaver}, {Bakx}, {Aravena}, {Bouwens}, {Cescon}, {Chen}, {da Cunha}, {Dayal}, {Faisst}, {Ferrara}, {Fujimoto}, {Hashimoto}, {Heintz}, {Herrera-Camus}, {Hodge}, {Inami}, {Inoue}, {Matthee}, {Meyer}, {Mizukoshi}, {Mondal}, {Nanayakkara}, {Oesch}, {Pallottini}, {R{\"o}ttgering}, {Rowland}, {Schouws}, {Smit}, {Sommovigo}, {Stark}, {Sugahara}, {Vallini}, {Vijarnwannaluk}, {van der Werf}, {Werner}, {Witstok}, \& {Xiao}}]{Algera25b}
{Algera}, H. S.~B., {Weaver}, J.~R., {Bakx}, T. J.~L.~C., {et~al.} 2025{\natexlab{b}}, arXiv e-prints, arXiv:2512.14486, \dodoi{10.48550/arXiv.2512.14486}

\bibitem[{{Arrabal Haro} {et~al.}(2023){Arrabal Haro}, {Dickinson}, {Finkelstein}, {Fujimoto}, {Fern{\'a}ndez}, {Kartaltepe}, {Jung}, {Cole}, {Burgarella}, {Chworowsky}, {Hutchison}, {Morales}, {Papovich}, {Simons}, {Amor{\'\i}n}, {Backhaus}, {Bagley}, {Bisigello}, {Calabr{\`o}}, {Castellano}, {Cleri}, {Dav{\'e}}, {Dekel}, {Ferguson}, {Fontana}, {Gawiser}, {Giavalisco}, {Harish}, {Hathi}, {Hirschmann}, {Holwerda}, {Huertas-Company}, {Koekemoer}, {Larson}, {Lucas}, {Mobasher}, {P{\'e}rez-Gonz{\'a}lez}, {Pirzkal}, {Rose}, {Santini}, {Trump}, {de la Vega}, {Wang}, {Weiner}, {Wilkins}, {Yang}, {Yung}, \& {Zavala}}]{Arrabal23}
{Arrabal Haro}, P., {Dickinson}, M., {Finkelstein}, S.~L., {et~al.} 2023, arXiv e-prints, arXiv:2304.05378, \dodoi{10.48550/arXiv.2304.05378}

\bibitem[{Atek {et~al.}(2022)Atek, Shuntov, Furtak, Richard, Kneib, Zitrin, \& Charlot}]{Atek22}
Atek, H., Shuntov, M., Furtak, L.~J., {et~al.} 2022, Revealing Galaxy Candidates out to $z \sim 16$ with JWST Observations of the Lensing Cluster SMACS0723,  arXiv, \dodoi{10.48550/ARXIV.2207.12338}

\bibitem[{{Baggen} {et~al.}(2023){Baggen}, {van Dokkum}, {Labb{\'e}}, {Brammer}, {Miller}, {Bezanson}, {Leja}, {Wang}, {Whitaker}, {Suess}, \& {Nelson}}]{Baggen_2023}
{Baggen}, J. F.~W., {van Dokkum}, P., {Labb{\'e}}, I., {et~al.} 2023, \apjl, 955, L12, \dodoi{10.3847/2041-8213/acf5ef}

\bibitem[{{Bakx} {et~al.}(2020){Bakx}, {Tamura}, {Hashimoto}, {Inoue}, {Lee}, {Mawatari}, {Ota}, {Umehata}, {Zackrisson}, {Hatsukade}, {Kohno}, {Matsuda}, {Matsuo}, {Okamoto}, {Shibuya}, {Shimizu}, {Taniguchi}, \& {Yoshida}}]{Bakx20}
{Bakx}, T. J.~L.~C., {Tamura}, Y., {Hashimoto}, T., {et~al.} 2020, \mnras, 493, 4294, \dodoi{10.1093/mnras/staa509}

\bibitem[{{Bakx} {et~al.}(2022){Bakx}, {Zavala}, {Mitsuhashi}, {Treu}, {Fontana}, {Tadaki}, {Casey}, {Castellano}, {Glazebrook}, {Hagimoto}, {Ikeda}, {Jones}, {Leethochawalit}, {Mason}, {Morishita}, {Nanayakkara}, {Pentericci}, {Roberts-Borsani}, {Santini}, {Serjeant}, {Tamura}, {Trenti}, \& {Vanzella}}]{Bakx22}
{Bakx}, T. J.~L.~C., {Zavala}, J.~A., {Mitsuhashi}, I., {et~al.} 2022, arXiv e-prints, arXiv:2208.13642.
\newblock \doarXiv{2208.13642}

\bibitem[{Bakx {et~al.}(2025{\natexlab{a}})Bakx, Sommovigo, Tamura, Smit, Ferrara, Algera, Aalto, Bossion, Carniani, Esmerian, Hagimoto, Hashimoto, Hatsukade, Ibar, Inami, Inoue, Knudsen, Laporte, Mawatari, Molina, Nyman, Okamoto, Pallottini, Sameera, Umehata, Vlemmings, \& Yoshida}]{Bakx25}
Bakx, T. J. L.~C., Sommovigo, L., Tamura, Y., {et~al.} 2025{\natexlab{a}}, A warm ultra-luminous infrared galaxy just 600 million years after the Big Bang.
\newblock \doarXiv{2511.08327}

\bibitem[{Bakx {et~al.}(2025{\natexlab{b}})Bakx, Algera, Jolly, Esmerian, Knudsen, Sommovigo, Witstok, Carniani, Chen, Eales, Ferrara, Fudamoto, Hagimoto, Hashimoto, Inami, Inoue, Khouri, Mitsuhashi, Nyman, Olander, Serjeant, Smit, Yoon, Zavala, Aalto, Casey, Tamura, \& Vlemmings}]{Bakx25_PIXIE}
Bakx, T. J. L.~C., Algera, H. S.~B., Jolly, J.-B., {et~al.} 2025{\natexlab{b}}, Probing Infrared eXcess to Investigate Early-Universe Dust (PIXIEDust).
\newblock \doarXiv{2512.07964}

\bibitem[{{Bari{\v{s}}i{\'c}} {et~al.}(2020){Bari{\v{s}}i{\'c}}, {Pacifici}, {van der Wel}, {Straatman}, {Bell}, {Bezanson}, {Brammer}, {D'Eugenio}, {Franx}, {van Houdt}, {Maseda}, {Muzzin}, {Sobral}, \& {Wu}}]{Barisic2020}
{Bari{\v{s}}i{\'c}}, I., {Pacifici}, C., {van der Wel}, A., {et~al.} 2020, \apj, 903, 146, \dodoi{10.3847/1538-4357/abba37}

\bibitem[{{Barrufet} {et~al.}(2023){Barrufet}, {Oesch}, {Bouwens}, {Inami}, {Sommovigo}, {Algera}, {da Cunha}, {Aravena}, {Dayal}, {Ferrara}, {Fudamoto}, {Gonzalez}, {Graziani}, {Hygate}, {de Looze}, {Nanayakkara}, {Pallottini}, {Schneider}, {Stefanon}, {Topping}, \& {van der Werf}}]{Barrufet23}
{Barrufet}, L., {Oesch}, P.~A., {Bouwens}, R., {et~al.} 2023, \mnras, 522, 3926, \dodoi{10.1093/mnras/stad1259}

\bibitem[{{Basu} {et~al.}(2025){Basu}, {Bhagwat}, {Ciardi}, \& {Costa}}]{Basu25}
{Basu}, A., {Bhagwat}, A., {Ciardi}, B., \& {Costa}, T. 2025, arXiv e-prints, arXiv:2501.18559, \dodoi{10.48550/arXiv.2501.18559}

\bibitem[{{Battisti} {et~al.}(2016){Battisti}, {Calzetti}, \& {Chary}}]{Battisti2016}
{Battisti}, A.~J., {Calzetti}, D., \& {Chary}, R.~R. 2016, \apj, 818, 13, \dodoi{10.3847/0004-637X/818/1/13}

\bibitem[{{Behrens} {et~al.}(2018){Behrens}, {Pallottini}, {Ferrara}, {Gallerani}, \& {Vallini}}]{behrens18}
{Behrens}, C., {Pallottini}, A., {Ferrara}, A., {Gallerani}, S., \& {Vallini}, L. 2018, \mnras, 477, 552, \dodoi{10.1093/mnras/sty552}

\bibitem[{{Bhagwat} {et~al.}(2024){Bhagwat}, {Costa}, {Ciardi}, {Pakmor}, \& {Garaldi}}]{Bhagwat23}
{Bhagwat}, A., {Costa}, T., {Ciardi}, B., {Pakmor}, R., \& {Garaldi}, E. 2024, \mnras, 531, 3406, \dodoi{10.1093/mnras/stae1125}

\bibitem[{{Bocchio} {et~al.}(2016){Bocchio}, {Marassi}, {Schneider}, {Bianchi}, {Limongi}, \& {Chieffi}}]{Bocchio16}
{Bocchio}, M., {Marassi}, S., {Schneider}, R., {et~al.} 2016, \aap, 587, A157, \dodoi{10.1051/0004-6361/201527432}

\bibitem[{{Bouwens} {et~al.}(2021){Bouwens}, {Oesch}, {Stefanon}, {Illingworth}, {Labb{\'e}}, {Reddy}, {Atek}, {Montes}, {Naidu}, {Nanayakkara}, {Nelson}, \& {Wilkins}}]{Bouwens21}
{Bouwens}, R.~J., {Oesch}, P.~A., {Stefanon}, M., {et~al.} 2021, \aj, 162, 47, \dodoi{10.3847/1538-3881/abf83e}

\bibitem[{{Bowler} {et~al.}(2018){Bowler}, {Bourne}, {Dunlop}, {McLure}, \& {McLeod}}]{Bowler18}
{Bowler}, R.~A.~A., {Bourne}, N., {Dunlop}, J.~S., {McLure}, R.~J., \& {McLeod}, D.~J. 2018, \mnras, 481, 1631, \dodoi{10.1093/mnras/sty2368}

\bibitem[{{Boylan-Kolchin}(2023)}]{Boylan23}
{Boylan-Kolchin}, M. 2023, Nature Astronomy, 7, 731, \dodoi{10.1038/s41550-023-01937-7}

\bibitem[{{Bunker} {et~al.}(2023){Bunker}, {Saxena}, {Cameron}, {Willott}, {Curtis-Lake}, {Jakobsen}, {Carniani}, {Smit}, {Maiolino}, {Witstok}, {Curti}, {D'Eugenio}, {Jones}, {Ferruit}, {Arribas}, {Charlot}, {Chevallard}, {Giardino}, {de Graaff}, {Looser}, {Luetzgendorf}, {Maseda}, {Rawle}, {Rix}, {Rodriguez Del Pino}, {Alberts}, {Egami}, {Eisenstein}, {Endsley}, {Hainline}, {Hausen}, {Johnson}, {Rieke}, {Rieke}, {Robertson}, {Shivaei}, {Stark}, {Sun}, {Tacchella}, {Tang}, {Williams}, {Willmer}, {Baker}, {Baum}, {Bhatawdekar}, {Bowler}, {Boyett}, {Chen}, {Circosta}, {Helton}, {Ji}, {Lyu}, {Nelson}, {Parlanti}, {Perna}, {Sandles}, {Scholtz}, {Suess}, {Topping}, {Uebler}, {Wallace}, \& {Whitler}}]{Bunker23}
{Bunker}, A.~J., {Saxena}, A., {Cameron}, A.~J., {et~al.} 2023, arXiv e-prints, arXiv:2302.07256, \dodoi{10.48550/arXiv.2302.07256}

\bibitem[{{Carniani} {et~al.}(2018){Carniani}, {Maiolino}, {Amorin}, {Pentericci}, {Pallottini}, {Ferrara}, {Willott}, {Smit}, {Matthee}, {Sobral}, {Santini}, {Castellano}, {De Barros}, {Fontana}, {Grazian}, \& {Guaita}}]{Carniani18}
{Carniani}, S., {Maiolino}, R., {Amorin}, R., {et~al.} 2018, \mnras, 478, 1170, \dodoi{10.1093/mnras/sty1088}

\bibitem[{{Carniani} {et~al.}(2024){Carniani}, {Hainline}, {D'Eugenio}, {Eisenstein}, {Jakobsen}, {Witstok}, {Johnson}, {Chevallard}, {Maiolino}, {Helton}, {Willott}, {Robertson}, {Alberts}, {Arribas}, {Baker}, {Bhatawdekar}, {Boyett}, {Bunker}, {Cameron}, {Cargile}, {Charlot}, {Curti}, {Curtis-Lake}, {Egami}, {Giardino}, {Isaak}, {Ji}, {Jones}, {Kumari}, {Maseda}, {Parlanti}, {P{\'e}rez-Gonz{\'a}lez}, {Rawle}, {Rieke}, {Rieke}, {Del Pino}, {Saxena}, {Scholtz}, {Smit}, {Sun}, {Tacchella}, {{\"U}bler}, {Venturi}, {Williams}, \& {Willmer}}]{Carniani24a}
{Carniani}, S., {Hainline}, K., {D'Eugenio}, F., {et~al.} 2024, \nat, 633, 318, \dodoi{10.1038/s41586-024-07860-9}

\bibitem[{Carniani {et~al.}(2024)Carniani, D'Eugenio, Ji, Parlanti, Scholtz, Sun, Venturi, Bakx, Curti, Maiolino, Tacchella, Zavala, Hainline, Witstok, Johnson, Alberts, Bunker, Charlot, Eisenstein, Helton, Jakobsen, Kumari, Robertson, Saxena, Übler, Williams, Willmer, \& Willott}]{Carniani24_z14}
Carniani, S., D'Eugenio, F., Ji, X., {et~al.} 2024, The eventful life of a luminous galaxy at z = 14: metal enrichment, feedback, and low gas fraction?
\newblock \doarXiv{2409.20533}

\bibitem[{{Carniani} {et~al.}(2024){Carniani}, {Venturi}, {Parlanti}, {de Graaff}, {Maiolino}, {Arribas}, {Bonaventura}, {Boyett}, {Bunker}, {Cameron}, {Charlot}, {Chevallard}, {Curti}, {Curtis-Lake}, {Eisenstein}, {Giardino}, {Hausen}, {Kumari}, {Maseda}, {Nelson}, {Perna}, {Rix}, {Robertson}, {Del Pino}, {Sandles}, {Scholtz}, {Simmonds}, {Smit}, {Tacchella}, {{\"U}bler}, {Williams}, {Willott}, \& {Witstok}}]{Carniani24b}
{Carniani}, S., {Venturi}, G., {Parlanti}, E., {et~al.} 2024, \aap, 685, A99, \dodoi{10.1051/0004-6361/202347230}

\bibitem[{Casey {et~al.}(2014)Casey, Narayanan, \& Cooray}]{Casey14}
Casey, C.~M., Narayanan, D., \& Cooray, A. 2014, Physics Reports, 541, 45, \dodoi{https://doi.org/10.1016/j.physrep.2014.02.009}

\bibitem[{{Castellano} {et~al.}(2024){Castellano}, {Napolitano}, {Fontana}, {Roberts-Borsani}, {Treu}, {Vanzella}, {Zavala}, {Arrabal Haro}, {Calabr{\`o}}, {Llerena}, {Mascia}, {Merlin}, {Paris}, {Pentericci}, {Santini}, {Bakx}, {Bergamini}, {Cupani}, {Dickinson}, {Filippenko}, {Glazebrook}, {Grillo}, {Kelly}, {Malkan}, {Mason}, {Morishita}, {Nanayakkara}, {Rosati}, {Sani}, {Wang}, \& {Yoon}}]{Castellano24}
{Castellano}, M., {Napolitano}, L., {Fontana}, A., {et~al.} 2024, arXiv e-prints, arXiv:2403.10238, \dodoi{10.48550/arXiv.2403.10238}

\bibitem[{{Chemerynska} {et~al.}(2024){Chemerynska}, {Atek}, {Dayal}, {Furtak}, {Feldmann}, {Greene}, {Maseda}, {Nanayakkara}, {Oesch}, {Fujimoto}, {Labb{\'e}}, {Bezanson}, {Brammer}, {Cutler}, {Leja}, {Pan}, {Price}, {Wang}, {Weaver}, \& {Whitaker}}]{Chemerynska24b}
{Chemerynska}, I., {Atek}, H., {Dayal}, P., {et~al.} 2024, \apjl, 976, L15, \dodoi{10.3847/2041-8213/ad8dc9}

\bibitem[{{Chevallard} {et~al.}(2013){Chevallard}, {Charlot}, {Wandelt}, \& {Wild}}]{Chevallard13}
{Chevallard}, J., {Charlot}, S., {Wandelt}, B., \& {Wild}, V. 2013, \mnras, 432, 2061, \dodoi{10.1093/mnras/stt523}

\bibitem[{Chevance {et~al.}(2023)Chevance, Krumholz, McLeod, Ostriker, Rosolowsky, \& Sternberg}]{Chevance2023}
Chevance, M., Krumholz, M.~R., McLeod, A.~F., {et~al.} 2023, Protostars Planets VII, 534, 1, \dodoi{10.48550/arXiv.2203.09570}

\bibitem[{{Choban} {et~al.}(2024){Choban}, {Salim}, {Kere{\v{s}}}, {Hayward}, \& {Sandstrom}}]{Choban24}
{Choban}, C.~R., {Salim}, S., {Kere{\v{s}}}, D., {Hayward}, C.~C., \& {Sandstrom}, K.~M. 2024, arXiv e-prints, arXiv:2408.08962, \dodoi{10.48550/arXiv.2408.08962}

\bibitem[{{Chworowsky} {et~al.}(2024){Chworowsky}, {Finkelstein}, {Boylan-Kolchin}, {McGrath}, {Iyer}, {Papovich}, {Dickinson}, {Taylor}, {Yung}, {Arrabal Haro}, {Bagley}, {Backhaus}, {Bhatawdekar}, {Cheng}, {Cleri}, {Cole}, {Cooper}, {Costantin}, {Dekel}, {Franco}, {Fujimoto}, {Hayward}, {Holwerda}, {Huertas-Company}, {Hirschmann}, {Hutchison}, {Koekemoer}, {Larson}, {Li}, {Long}, {Lucas}, {Pirzkal}, {Rodighiero}, {Somerville}, {Vanderhoof}, {de la Vega}, {Wilkins}, {Yang}, \& {Zavala}}]{Chworowsky24}
{Chworowsky}, K., {Finkelstein}, S.~L., {Boylan-Kolchin}, M., {et~al.} 2024, \aj, 168, 113, \dodoi{10.3847/1538-3881/ad57c1}

\bibitem[{Cochrane {et~al.}(2024)Cochrane, Anglés-Alcázar, Cullen, \& Hayward}]{Cochrane_2024}
Cochrane, R.~K., Anglés-Alcázar, D., Cullen, F., \& Hayward, C.~C. 2024, The Astrophysical Journal, 961, 37, \dodoi{10.3847/1538-4357/ad02f8}

\bibitem[{Cochrane {et~al.}(2025)Cochrane, Katz, Begley, Hayward, \& Best}]{Cochrane25}
Cochrane, R.~K., Katz, H., Begley, R., Hayward, C.~C., \& Best, P.~N. 2025, The Astrophysical Journal Letters, 978, L42, \dodoi{10.3847/2041-8213/ad9a4d}

\bibitem[{{Code}(1973)}]{Code73}
{Code}, A.~D. 1973, in Interstellar Dust and Related Topics, ed. J.~M. {Greenberg} \& H.~C. {van de Hulst}, Vol.~52, 505

\bibitem[{{Correa} {et~al.}(2015){Correa}, {Wyithe}, {Schaye}, \& {Duffy}}]{Correa15}
{Correa}, C.~A., {Wyithe}, J. S.~B., {Schaye}, J., \& {Duffy}, A.~R. 2015, \mnras, 450, 1521, \dodoi{10.1093/mnras/stv697}

\bibitem[{{Cullen} {et~al.}(2024){Cullen}, {McLeod}, {McLure}, {Dunlop}, {Donnan}, {Carnall}, {Keating}, {Magee}, {Arellano-Cordova}, {Bowler}, {Begley}, {Flury}, {Hamadouche}, \& {Stanton}}]{Cullen_2024}
{Cullen}, F., {McLeod}, D.~J., {McLure}, R.~J., {et~al.} 2024, \mnras, 531, 997, \dodoi{10.1093/mnras/stae1211}

\bibitem[{{Curti} {et~al.}(2022){Curti}, {D'Eugenio}, {Carniani}, {Maiolino}, {Sandles}, {Witstok}, {Baker}, {Bennett}, {Piotrowska}, {Tacchella}, {Charlot}, {Nakajima}, {Maheson}, {Mannucci}, {Amiri}, {Arribas}, {Belfiore}, {Bonaventura}, {Bunker}, {Chevallard}, {Cresci}, {Curtis-Lake}, {Hayden-Pawson}, {Jones}, {Kumari}, {Laseter}, {Looser}, {Marconi}, {Maseda}, {Scholtz}, {Smit}, {{\"U}bler}, \& {Wallace}}]{curti2022}
{Curti}, M., {D'Eugenio}, F., {Carniani}, S., {et~al.} 2022, \mnras, \dodoi{10.1093/mnras/stac2737}

\bibitem[{{Curti} {et~al.}(2024){Curti}, {Maiolino}, {Curtis-Lake}, {Chevallard}, {Carniani}, {D'Eugenio}, {Looser}, {Scholtz}, {Charlot}, {Cameron}, {{\"U}bler}, {Witstok}, {Boyett}, {Laseter}, {Sandles}, {Arribas}, {Bunker}, {Giardino}, {Maseda}, {Rawle}, {Rodr{\'\i}guez Del Pino}, {Smit}, {Willott}, {Eisenstein}, {Hausen}, {Johnson}, {Rieke}, {Robertson}, {Tacchella}, {Williams}, {Willmer}, {Baker}, {Bhatawdekar}, {Egami}, {Helton}, {Ji}, {Kumari}, {Perna}, {Shivaei}, \& {Sun}}]{Curti24}
{Curti}, M., {Maiolino}, R., {Curtis-Lake}, E., {et~al.} 2024, \aap, 684, A75, \dodoi{10.1051/0004-6361/202346698}

\bibitem[{{Curtis-Lake} {et~al.}(2023){Curtis-Lake}, {Carniani}, {Cameron}, {Charlot}, {Jakobsen}, {Maiolino}, {Bunker}, {Witstok}, {Smit}, {Chevallard}, {Willott}, {Ferruit}, {Arribas}, {Bonaventura}, {Curti}, {D'Eugenio}, {Franx}, {Giardino}, {Looser}, {L{\"u}tzgendorf}, {Maseda}, {Rawle}, {Rix}, {Rodr{\'\i}guez del Pino}, {{\"U}bler}, {Sirianni}, {Dressler}, {Egami}, {Eisenstein}, {Endsley}, {Hainline}, {Hausen}, {Johnson}, {Rieke}, {Robertson}, {Shivaei}, {Stark}, {Tacchella}, {Williams}, {Willmer}, {Bhatawdekar}, {Bowler}, {Boyett}, {Chen}, {de Graaff}, {Helton}, {Hviding}, {Jones}, {Kumari}, {Lyu}, {Nelson}, {Perna}, {Sandles}, {Saxena}, {Suess}, {Sun}, {Topping}, {Wallace}, \& {Whitler}}]{Curtis23}
{Curtis-Lake}, E., {Carniani}, S., {Cameron}, A., {et~al.} 2023, Nature Astronomy, \dodoi{10.1038/s41550-023-01918-w}

\bibitem[{Da~Cunha {et~al.}(2013)Da~Cunha, Groves, Walter, Decarli, Weiss, Bertoldi, Carilli, Daddi, Elbaz, Ivison, {et~al.}}]{daCunha13}
Da~Cunha, E., Groves, B., Walter, F., {et~al.} 2013, The Astrophysical Journal, 766, 13

\bibitem[{{Dayal} {et~al.}(2022){Dayal}, {Ferrara}, {Sommovigo}, {Bouwens}, {Oesch}, {Smit}, {Gonzalez}, {Schouws}, {Stefanon}, {Kobayashi}, {Bremer}, {Algera}, {Aravena}, {Bowler}, {da Cunha}, {Fudamoto}, {Graziani}, {Hodge}, {Inami}, {De Looze}, {Pallottini}, {Riechers}, {Schneider}, {Stark}, \& {Endsley}}]{Dayal22}
{Dayal}, P., {Ferrara}, A., {Sommovigo}, L., {et~al.} 2022, \mnras, 512, 989, \dodoi{10.1093/mnras/stac537}

\bibitem[{{Dekel} {et~al.}(2023){Dekel}, {Sarkar}, {Birnboim}, {Mandelker}, \& {Li}}]{Dekel23}
{Dekel}, A., {Sarkar}, K.~C., {Birnboim}, Y., {Mandelker}, N., \& {Li}, Z. 2023, \mnras, 523, 3201, \dodoi{10.1093/mnras/stad1557}

\bibitem[{{Di Mascia} {et~al.}(2024){Di Mascia}, {Pallottini}, {Sommovigo}, \& {Decataldo}}]{DiMascia24}
{Di Mascia}, F., {Pallottini}, A., {Sommovigo}, L., \& {Decataldo}, D. 2024, arXiv e-prints, arXiv:2407.01662, \dodoi{10.48550/arXiv.2407.01662}

\bibitem[{{Draine}(2003)}]{Draine03}
{Draine}, B.~T. 2003, \araa, 41, 241, \dodoi{10.1146/annurev.astro.41.011802.094840}

\bibitem[{{Dubois} {et~al.}(2024){Dubois}, {Rodr{\'\i}guez Montero}, {Guerra}, {Trebitsch}, {Han}, {Beckmann}, {Yi}, {Lewis}, \& {Jang}}]{Dubois24}
{Dubois}, Y., {Rodr{\'\i}guez Montero}, F., {Guerra}, C., {et~al.} 2024, \aap, 687, A240, \dodoi{10.1051/0004-6361/202449784}

\bibitem[{{Elbaz} {et~al.}(2018){Elbaz}, {Leiton}, {Nagar}, {Okumura}, {Franco}, {Schreiber}, {Pannella}, {Wang}, {Dickinson}, {D{\'\i}az-Santos}, {Ciesla}, {Daddi}, {Bournaud}, {Magdis}, {Zhou}, \& {Rujopakarn}}]{Elbaz18}
{Elbaz}, D., {Leiton}, R., {Nagar}, N., {et~al.} 2018, \aap, 616, A110, \dodoi{10.1051/0004-6361/201732370}

\bibitem[{{Faisst} {et~al.}(2017){Faisst}, {Capak}, {Yan}, {Pavesi}, {Riechers}, {Bari{\v{s}}i{\'c}}, {Cooke}, {Kartaltepe}, \& {Masters}}]{Faisst17}
{Faisst}, A.~L., {Capak}, P.~L., {Yan}, L., {et~al.} 2017, \apj, 847, 21, \dodoi{10.3847/1538-4357/aa886c}

\bibitem[{{Faisst} {et~al.}(2025){Faisst}, {Fujimoto}, {Tsujita}, {Wang}, {Khosravaninezhad}, {Loiacono}, {{\"U}bler}, {B{\'e}thermin}, {Dessauges-Zavadsky}, {Herrera-Camus}, {Schaerer}, {Silverman}, {Yan}, {Aravena}, {De Looze}, {F{\"o}rster Schreiber}, {Gonz{\'a}lez-L{\'o}pez}, {Spilker}, {Tadaki}, {Casey}, {Franco}, {Harish}, {McCracken}, {Kartaltepe}, {Koekemoer}, {Khostovan}, {Liu}, {Rhodes}, {Robertson}, {Amorin}, {Assef}, {Battisti}, {Birkin}, {Boquien}, {Da Cunha}, {Dam}, {Davies}, {G{\'o}mez-Espinoza}, {Ferrara}, {Fudamoto}, {Gillman}, {Ginolfi}, {Gozaliasl}, {Gruppioni}, {Hadi}, {Hathi}, {Ibar}, {Ikeda}, {Inami}, {Jones}, {Kohandel}, {Li}, {Lin}, {Liu}, {Liu}, {Long}, {Magdis}, {Maraston}, {Martin}, {Mitsuhashi}, {Mobasher}, {Molina}, {Nanni}, {Palla}, {Pallottini}, {Pozzi}, {Relano}, {Ren}, {Riechers}, {Romano}, {Sanders}, {Sawant}, {Shuntov}, {Smit}, {Sommovigo}, {Talia}, {Vallini}, {Veraldi}, {Vergani}, {Vijayan}, {Villanueva}, \& {Zamorani}}]{Faisst25}
{Faisst}, A.~L., {Fujimoto}, S., {Tsujita}, A., {et~al.} 2025, arXiv e-prints, arXiv:2510.16111, \dodoi{10.48550/arXiv.2510.16111}

\bibitem[{Fakhouri {et~al.}(2010)Fakhouri, Ma, \& Boylan-Kolchin}]{Fakhouri10}
Fakhouri, O., Ma, C.-P., \& Boylan-Kolchin, M. 2010, Monthly Notices of the Royal Astronomical Society, 406, 2267, \dodoi{10.1111/j.1365-2966.2010.16859.x}

\bibitem[{{Federrath} {et~al.}(2010){Federrath}, {Roman-Duval}, {Klessen}, {Schmidt}, \& {Mac Low}}]{Federrath2010}
{Federrath}, C., {Roman-Duval}, J., {Klessen}, R.~S., {Schmidt}, W., \& {Mac Low}, M.-M. 2010, \aap, 512, A81, \dodoi{10.1051/0004-6361/200912437}

\bibitem[{{Ferrara}(2024{\natexlab{a}})}]{Ferrara24a}
{Ferrara}, A. 2024{\natexlab{a}}, \aap, 684, A207, \dodoi{10.1051/0004-6361/202348321}

\bibitem[{{Ferrara}(2024{\natexlab{b}})}]{Ferrara24b}
---. 2024{\natexlab{b}}, arXiv e-prints, arXiv:2405.20370, \dodoi{10.48550/arXiv.2405.20370}

\bibitem[{{Ferrara} {et~al.}(2024){Ferrara}, {Carniani}, {di Mascia}, {Bouwens}, {Oesch}, \& {Schouws}}]{Ferrara24c}
{Ferrara}, A., {Carniani}, S., {di Mascia}, F., {et~al.} 2024, arXiv e-prints, arXiv:2409.17223.
\newblock \doarXiv{2409.17223}

\bibitem[{{Ferrara} {et~al.}(2023){Ferrara}, {Pallottini}, \& {Dayal}}]{Ferrara23a}
{Ferrara}, A., {Pallottini}, A., \& {Dayal}, P. 2023, \mnras, 522, 3986, \dodoi{10.1093/mnras/stad1095}

\bibitem[{{Ferrara} {et~al.}(2025){Ferrara}, {Pallottini}, \& {Sommovigo}}]{Ferrara25}
{Ferrara}, A., {Pallottini}, A., \& {Sommovigo}, L. 2025, \aap, 694, A286, \dodoi{10.1051/0004-6361/202452707}

\bibitem[{{Ferrara} \& {Peroux}(2021)}]{Ferrara21}
{Ferrara}, A., \& {Peroux}, C. 2021, \mnras, 503, 4537, \dodoi{10.1093/mnras/stab761}

\bibitem[{{Ferrara} {et~al.}(2000){Ferrara}, {Pettini}, \& {Shchekinov}}]{Ferrara00b}
{Ferrara}, A., {Pettini}, M., \& {Shchekinov}, Y. 2000, \mnras, 319, 539, \dodoi{10.1046/j.1365-8711.2000.03857.x}

\bibitem[{{Ferrara} \& {Tolstoy}(2000)}]{Ferrara00}
{Ferrara}, A., \& {Tolstoy}, E. 2000, \mnras, 313, 291, \dodoi{10.1046/j.1365-8711.2000.03209.x}

\bibitem[{{Ferrara} {et~al.}(2016){Ferrara}, {Viti}, \& {Ceccarelli}}]{Ferrara16}
{Ferrara}, A., {Viti}, S., \& {Ceccarelli}, C. 2016, \mnras, 463, L112, \dodoi{10.1093/mnrasl/slw165}

\bibitem[{{Ferrara} {et~al.}(2022){Ferrara}, {Sommovigo}, {Dayal}, {Pallottini}, {Bouwens}, {Gonzalez}, {Inami}, {Smit}, {Bowler}, {Endsley}, {Oesch}, {Schouws}, {Stark}, {Stefanon}, {Aravena}, {da Cunha}, {De Looze}, {Fudamoto}, {Graziani}, {Hodge}, {Riechers}, {Schneider}, {Algera}, {Barrufet}, {Hygate}, {Labb{\'e}}, {Li}, {Nanayakkara}, {Topping}, \& {van der Werf}}]{Ferrara22a}
{Ferrara}, A., {Sommovigo}, L., {Dayal}, P., {et~al.} 2022, \mnras, 512, 58, \dodoi{10.1093/mnras/stac460}

\bibitem[{{Finkelstein} {et~al.}(2023){Finkelstein}, {Leung}, {Bagley}, {Dickinson}, {Ferguson}, {Papovich}, {Akins}, {Arrabal Haro}, {Dave}, {Dekel}, {Kartaltepe}, {Kocevski}, {Koekemoer}, {Pirzkal}, {Somerville}, {Yung}, {Amorin}, {Backhaus}, {Behroozi}, {Bisigello}, {Bromm}, {Casey}, {Chavez Ortiz}, {Cheng}, {Chworowsky}, {Cleri}, {Cooper}, {Davis}, {de la Vega}, {Elbaz}, {Franco}, {Fontana}, {Fujimoto}, {Giavalisco}, {Grogin}, {Holwerda}, {Huertas-Company}, {Hirschmann}, {Iyer}, {Jogee}, {Jung}, {Larson}, {Lucas}, {Mobasher}, {Morales}, {Morley}, {Mukherjee}, {Perez-Gonzalez}, {Ravindranath}, {Rodighiero}, {Rowland}, {Tacchella}, {Taylor}, {Trump}, \& {Wilkins}}]{Finkelstein23}
{Finkelstein}, S.~L., {Leung}, G. C.~K., {Bagley}, M.~B., {et~al.} 2023, arXiv e-prints, arXiv:2311.04279, \dodoi{10.48550/arXiv.2311.04279}

\bibitem[{{Fisher} {et~al.}(2025){Fisher}, {Bowler}, {Stefanon}, {Rowland}, {Algera}, {Aravena}, {Bouwens}, {Dayal}, {Ferrara}, {Fudamoto}, {Gulis}, {Hodge}, {Inami}, {Ormerod}, {Pallottini}, {Phillips}, {Sartorio}, {Schouws}, {Smit}, {Sommovigo}, {Stark}, \& {van der Werf}}]{Fisher25}
{Fisher}, R., {Bowler}, R.~A.~A., {Stefanon}, M., {et~al.} 2025, \mnras, 539, 109, \dodoi{10.1093/mnras/staf485}

\bibitem[{{Fudamoto} {et~al.}(2023){Fudamoto}, {Oesch}, {Walter}, {Decarli}, {Carilli}, {Ferrara}, {Barrufet}, {Bouwens}, {Dessauges-Zavadsky}, {Nelson}, {Dannerbauer}, {Illingworth}, {Inoue}, {Marques-Chaves}, {P{\'e}rez-Fournon}, {Riechers}, {Schaerer}, {Smit}, {Sugahara}, \& {van der Werf}}]{Fudamoto23}
{Fudamoto}, Y., {Oesch}, P.~A., {Walter}, F., {et~al.} 2023, arXiv e-prints, arXiv:2309.02493, \dodoi{10.48550/arXiv.2309.02493}

\bibitem[{{Fujimoto} {et~al.}(2022){Fujimoto}, {Finkelstein}, {Burgarella}, {Carilli}, {Buat}, {Casey}, {Ciesla}, {Tacchella}, {Zavala}, {Brammer}, {Fudamoto}, {Ouchi}, {Valentino}, {Cooper}, {Dickinson}, {Franco}, {Giavalisco}, {Hutchison}, {Kartaltepe}, {Koekemoer}, {Kojima}, {Larson}, {Murphy}, {Papovich}, {P{\'e}rez-Gonz{\'a}lez}, {Somerville}, {Yoon}, {Wilkins}, {Yung}, {Akins}, {Amor{\'\i}n}, {Arrabal Haro}, {Bagley}, {Chworowsky}, {Cleri}, {Cooper}, {Costantin}, {Daddi}, {Ferguson}, {Grogin}, {Jim{\'e}nez-Andrade}, {Juneau}, {Kirkpatrick}, {Kocevski}, {Le Bail}, {Long}, {Lucas}, {Magnelli}, {McKinney}, {Rose}, {Seill{\'e}}, {Simons}, \& {Weiner}}]{Fujimoto22}
{Fujimoto}, S., {Finkelstein}, S.~L., {Burgarella}, D., {et~al.} 2022, arXiv e-prints, arXiv:2211.03896, \dodoi{10.48550/arXiv.2211.03896}

\bibitem[{{Fujimoto} {et~al.}(2023{\natexlab{a}}){Fujimoto}, {Wang}, {Weaver}, {Kokorev}, {Atek}, {Bezanson}, {Labbe}, {Brammer}, {Greene}, {Chemerynska}, {Dayal}, {de Graaff}, {Furtak}, {Oesch}, {Setton}, {Price}, {Miller}, {Williams}, {Whitaker}, {Zitrin}, {Cutler}, {Leja}, {Pan}, {Coe}, {van Dokkum}, {Feldmann}, {Fudamoto}, {Goulding}, {Khullar}, {Marchesini}, {Maseda}, {Nanayakkara}, {Nelson}, {Smit}, {Stefanon}, \& {Weibel}}]{Fujimoto23b}
{Fujimoto}, S., {Wang}, B., {Weaver}, J., {et~al.} 2023{\natexlab{a}}, arXiv e-prints, arXiv:2308.11609, \dodoi{10.48550/arXiv.2308.11609}

\bibitem[{{Fujimoto} {et~al.}(2023{\natexlab{b}}){Fujimoto}, {Bezanson}, {Labbe}, {Brammer}, {Price}, {Wang}, {Weaver}, {Fudamoto}, {Oesch}, {Williams}, {Dayal}, {Feldmann}, {Greene}, {Leja}, {Whitaker}, {Zitrin}, {Cutler}, {Furtak}, {Pan}, {Chemerynska}, {Kokorev}, {Miller}, {Atek}, {van Dokkum}, {Juneau}, {Kassin}, {Khullar}, {Marchesini}, {Maseda}, {Nelson}, {Setton}, \& {Smit}}]{Fujimoto23}
{Fujimoto}, S., {Bezanson}, R., {Labbe}, I., {et~al.} 2023{\natexlab{b}}, arXiv e-prints, arXiv:2309.07834, \dodoi{10.48550/arXiv.2309.07834}

\bibitem[{{Gallagher} {et~al.}(2012){Gallagher}, {Everett}, {Keating}, {Hill}, \& {Deo}}]{Gallagher12}
{Gallagher}, S.~C., {Everett}, J.~E., {Keating}, S.~K., {Hill}, A.~R., \& {Deo}, R.~P. 2012, in Astronomical Society of the Pacific Conference Series, Vol. 460, AGN Winds in Charleston, ed. G.~{Chartas}, F.~{Hamann}, \& K.~M. {Leighly}, 199, \dodoi{10.48550/arXiv.1201.5018}

\bibitem[{{Gelli} {et~al.}(2024){Gelli}, {Salvadori}, {Ferrara}, \& {Pallottini}}]{Gelli24}
{Gelli}, V., {Salvadori}, S., {Ferrara}, A., \& {Pallottini}, A. 2024, \apj, 964, 76, \dodoi{10.3847/1538-4357/ad23ec}

\bibitem[{{Ginzburg} {et~al.}(2022){Ginzburg}, {Dekel}, {Mandelker}, \& {Krumholz}}]{Ginzburg22}
{Ginzburg}, O., {Dekel}, A., {Mandelker}, N., \& {Krumholz}, M.~R. 2022, \mnras, 513, 6177, \dodoi{10.1093/mnras/stac1324}

\bibitem[{{Gomez} {et~al.}(2012){Gomez}, {Krause}, {Barlow}, {Swinyard}, {Owen}, {Clark}, {Matsuura}, {Gomez}, {Rho}, {Besel}, {Bouwman}, {Gear}, {Henning}, {Ivison}, {Polehampton}, \& {Sibthorpe}}]{Gomez12}
{Gomez}, H.~L., {Krause}, O., {Barlow}, M.~J., {et~al.} 2012, \apj, 760, 96, \dodoi{10.1088/0004-637X/760/1/96}

\bibitem[{{G{\'o}mez-Guijarro} {et~al.}(2018){G{\'o}mez-Guijarro}, {Toft}, {Karim}, {Magnelli}, {Magdis}, {Jim{\'e}nez-Andrade}, {Capak}, {Fraternali}, {Fujimoto}, {Riechers}, {Schinnerer}, {Smol{\v{c}}i{\'c}}, {Aravena}, {Bertoldi}, {Cortzen}, {Hasinger}, {Hu}, {Jones}, {Koekemoer}, {Lee}, {McCracken}, {Micha{\l}owski}, {Navarrete}, {Povi{\'c}}, {Puglisi}, {Romano-D{\'\i}az}, {Sheth}, {Silverman}, {Staguhn}, {Steinhardt}, {Stockmann}, {Tanaka}, {Valentino}, {van Kampen}, \& {Zirm}}]{Guijarro18}
{G{\'o}mez-Guijarro}, C., {Toft}, S., {Karim}, A., {et~al.} 2018, \apj, 856, 121, \dodoi{10.3847/1538-4357/aab206}

\bibitem[{{Helton} {et~al.}(2024){Helton}, {Rieke}, {Alberts}, {Wu}, {Eisenstein}, {Hainline}, {Carniani}, {Ji}, {Baker}, {Bhatawdekar}, {Bunker}, {Cargile}, {Charlot}, {Chevallard}, {D'Eugenio}, {Egami}, {Johnson}, {Jones}, {Lyu}, {Maiolino}, {P{\'e}rez-Gonz{\'a}lez}, {Rieke}, {Robertson}, {Saxena}, {Scholtz}, {Shivaei}, {Sun}, {Tacchella}, {Whitler}, {Williams}, {Willmer}, {Willott}, {Witstok}, \& {Zhu}}]{Helton24}
{Helton}, J.~M., {Rieke}, G.~H., {Alberts}, S., {et~al.} 2024, arXiv e-prints, arXiv:2405.18462, \dodoi{10.48550/arXiv.2405.18462}

\bibitem[{{Hirashita} \& {Aoyama}(2019)}]{Hirashita19}
{Hirashita}, H., \& {Aoyama}, S. 2019, \mnras, 482, 2555, \dodoi{10.1093/mnras/sty2838}

\bibitem[{{Hodge} \& {da Cunha}(2020)}]{Hodge20}
{Hodge}, J.~A., \& {da Cunha}, E. 2020, Royal Society Open Science, 7, 200556, \dodoi{10.1098/rsos.200556}

\bibitem[{{Hoyer} {et~al.}(2023){Hoyer}, {Pinna}, {Kamlah}, {Nogueras-Lara}, {Feldmeier-Krause}, {Neumayer}, {Sormani}, {Boquien}, {Emsellem}, {Seth}, {Klessen}, {Williams}, {Schinnerer}, {Barnes}, {Leroy}, {Bonoli}, {Kruijssen}, {Neumann}, {S{\'a}nchez-Bl{\'a}zquez}, {Dale}, {Watkins}, {Thilker}, {Rosolowsky}, {Bigiel}, {Grasha}, {Egorov}, {Liu}, {Sandstrom}, {Larson}, {Blanc}, \& {Hassani}}]{Hoyer_23_Phangs}
{Hoyer}, N., {Pinna}, F., {Kamlah}, A. W.~H., {et~al.} 2023, \apjl, 944, L25, \dodoi{10.3847/2041-8213/aca53e}

\bibitem[{{Hsiao} {et~al.}(2023){Hsiao}, {Abdurro'uf}, {Coe}, {Larson}, {Jung}, {Mingozzi}, {Dayal}, {Kumari}, {Kokorev}, {Vikaeus}, {Brammer}, {Furtak}, {Adamo}, {Andrade-Santos}, {Antwi-Danso}, {Bradac}, {Bradley}, {Broadhurst}, {Carnall}, {Conselice}, {Diego}, {Donahue}, {Eldridge}, {Fujimoto}, {Henry}, {Hernandez}, {Hutchison}, {James}, {Norman}, {Park}, {Pirzkal}, {Postman}, {Ricotti}, {Rigby}, {Vanzella}, {Welch}, {Wilkins}, {Windhorst}, {Xu}, {Zackrisson}, \& {Zitrin}}]{Hsiao23}
{Hsiao}, T. Y.-Y., {Abdurro'uf}, {Coe}, D., {et~al.} 2023, arXiv e-prints, arXiv:2305.03042, \dodoi{10.48550/arXiv.2305.03042}

\bibitem[{{Inami} {et~al.}(2022){Inami}, {Algera}, {Schouws}, {Sommovigo}, {Bouwens}, {Smit}, {Stefanon}, {Bowler}, {Endsley}, {Ferrara}, {Oesch}, {Stark}, {Aravena}, {Barrufet}, {da Cunha}, {Dayal}, {De Looze}, {Fudamoto}, {Gonzalez}, {Graziani}, {Hodge}, {Hygate}, {Nanayakkara}, {Pallottini}, {Riechers}, {Schneider}, {Topping}, \& {van der Werf}}]{Inami22}
{Inami}, H., {Algera}, H., {Schouws}, S., {et~al.} 2022, \mnras, \dodoi{10.1093/mnras/stac1779}

\bibitem[{{Inayoshi} {et~al.}(2022){Inayoshi}, {Harikane}, {Inoue}, {Li}, \& {Ho}}]{Inayoshi22}
{Inayoshi}, K., {Harikane}, Y., {Inoue}, A.~K., {Li}, W., \& {Ho}, L.~C. 2022, \apjl, 938, L10, \dodoi{10.3847/2041-8213/ac9310}

\bibitem[{{Inserra} {et~al.}(2011){Inserra}, {Turatto}, {Pastorello}, {Benetti}, {Cappellaro}, {Pumo}, {Zampieri}, {Agnoletto}, {Bufano}, {Botticella}, {Della Valle}, {Elias Rosa}, {Iijima}, {Spiro}, \& {Valenti}}]{Inserra11}
{Inserra}, C., {Turatto}, M., {Pastorello}, A., {et~al.} 2011, \mnras, 417, 261, \dodoi{10.1111/j.1365-2966.2011.19128.x}

\bibitem[{{Kaasinen} {et~al.}(2022){Kaasinen}, {van Marrewijk}, {Popping}, {Ginolfi}, {Di Mascolo}, {Mroczkowski}, {Concas}, {Di Cesare}, {Killi}, \& {Langan}}]{Kaasinen22}
{Kaasinen}, M., {van Marrewijk}, J., {Popping}, G., {et~al.} 2022, arXiv e-prints, arXiv:2210.03754.
\newblock \doarXiv{2210.03754}

\bibitem[{{Kazandjian} {et~al.}(2016){Kazandjian}, {Pelupessy}, {Meijerink}, {Israel}, {Coppola}, {Rosenberg}, \& {Spaans}}]{Kazandjian16}
{Kazandjian}, M.~V., {Pelupessy}, I., {Meijerink}, R., {et~al.} 2016, \aap, 595, A124, \dodoi{10.1051/0004-6361/201424594}

\bibitem[{{Kennicutt}(1998)}]{Kennicutt98}
{Kennicutt}, Robert~C., J. 1998, \araa, 36, 189, \dodoi{10.1146/annurev.astro.36.1.189}

\bibitem[{{Kotak} {et~al.}(2009){Kotak}, {Meikle}, {Farrah}, {Gerardy}, {Foley}, {Van Dyk}, {Fransson}, {Lundqvist}, {Sollerman}, {Fesen}, {Filippenko}, {Mattila}, {Silverman}, {Andersen}, {H{\"o}flich}, {Pozzo}, \& {Wheeler}}]{Kotak09}
{Kotak}, R., {Meikle}, W.~P.~S., {Farrah}, D., {et~al.} 2009, \apj, 704, 306, \dodoi{10.1088/0004-637X/704/1/306}

\bibitem[{{Kravtsov} \& {Belokurov}(2024)}]{Kravtsov24}
{Kravtsov}, A., \& {Belokurov}, V. 2024, arXiv e-prints, arXiv:2405.04578, \dodoi{10.48550/arXiv.2405.04578}

\bibitem[{Krumholz {et~al.}(2018)Krumholz, Burkhart, Forbes, \& Crocker}]{Krumholz18}
Krumholz, M.~R., Burkhart, B., Forbes, J.~C., \& Crocker, R.~M. 2018, Monthly Notices of the Royal Astronomical Society, 477, 2716, \dodoi{10.1093/mnras/sty852}

\bibitem[{{Laporte} {et~al.}(2017){Laporte}, {Ellis}, {Boone}, {Bauer}, {Qu{\'e}nard}, {Roberts-Borsani}, {Pell{\'o}}, {P{\'e}rez-Fournon}, \& {Streblyanska}}]{Laporte17}
{Laporte}, N., {Ellis}, R.~S., {Boone}, F., {et~al.} 2017, \apjl, 837, L21, \dodoi{10.3847/2041-8213/aa62aa}

\bibitem[{{Leitherer} {et~al.}(1999){Leitherer}, {Schaerer}, {Goldader}, {Delgado}, {Robert}, {Kune}, {de Mello}, {Devost}, \& {Heckman}}]{Leitherer99}
{Leitherer}, C., {Schaerer}, D., {Goldader}, J.~D., {et~al.} 1999, \apjs, 123, 3, \dodoi{10.1086/313233}

\bibitem[{{Li} {et~al.}(2024){Li}, {Dekel}, {Sarkar}, {Aung}, {Giavalisco}, {Mandelker}, \& {Tacchella}}]{Li24}
{Li}, Z., {Dekel}, A., {Sarkar}, K.~C., {et~al.} 2024, \aap, 690, A108, \dodoi{10.1051/0004-6361/202348727}

\bibitem[{{Liang} {et~al.}(2019){Liang}, {Feldmann}, {Kere{\v{s}}}, {Scoville}, {Hayward}, {Faucher-Gigu{\`e}re}, {Schreiber}, {Ma}, {Hopkins}, \& {Quataert}}]{Liang19}
{Liang}, L., {Feldmann}, R., {Kere{\v{s}}}, D., {et~al.} 2019, \mnras, 489, 1397, \dodoi{10.1093/mnras/stz2134}

\bibitem[{{Lin} {et~al.}(2025){Lin}, {Yang}, {Li}, \& {Witstok}}]{Lin25}
{Lin}, Q., {Yang}, X., {Li}, A., \& {Witstok}, J. 2025, \aap, 694, A84, \dodoi{10.1051/0004-6361/202452372}

\bibitem[{{Lin} {et~al.}(2021){Lin}, {Hirashita}, {Camps}, \& {Baes}}]{Lin21}
{Lin}, Y.-H., {Hirashita}, H., {Camps}, P., \& {Baes}, M. 2021, \mnras, 507, 2755, \dodoi{10.1093/mnras/stab2242}

\bibitem[{{Liu} \& {Bromm}(2023)}]{Liu23}
{Liu}, B., \& {Bromm}, V. 2023, arXiv e-prints, arXiv:2312.04085, \dodoi{10.48550/arXiv.2312.04085}

\bibitem[{{Markov} {et~al.}(2024){Markov}, {Gallerani}, {Ferrara}, {Pallottini}, {Parlanti}, {Di Mascia}, {Sommovigo}, \& {Kohandel}}]{Markov24}
{Markov}, V., {Gallerani}, S., {Ferrara}, A., {et~al.} 2024, arXiv e-prints, arXiv:2402.05996, \dodoi{10.48550/arXiv.2402.05996}

\bibitem[{{Marques-Chaves} {et~al.}(2025){Marques-Chaves}, {Schaerer}, {Dessauges-Zavadsky}, {{\'A}lvarez-M{\'a}rquez}, {Hashimoto}, {Colina}, {Inoue}, {Blanco-Prieto}, {Nakazato}, {Costantin}, {Arribas}, {Bakx}, {Ceverino}, {Crespo G{\'o}mez}, {Fudamoto}, {Hagimoto}, {Hamada}, {Matsuoka}, {Mawatari}, {Onoue}, {Osone}, {Ren}, {Sugahara}, {Terui}, \& {Yoshida}}]{Marques-Chaves_2025}
{Marques-Chaves}, R., {Schaerer}, D., {Dessauges-Zavadsky}, M., {et~al.} 2025, arXiv e-prints, arXiv:2510.12411, \dodoi{10.48550/arXiv.2510.12411}

\bibitem[{{Marszewski} {et~al.}(2024){Marszewski}, {Sun}, {Faucher-Gigu{\`e}re}, {Hayward}, \& {Feldmann}}]{Marszewski24}
{Marszewski}, A., {Sun}, G., {Faucher-Gigu{\`e}re}, C.-A., {Hayward}, C.~C., \& {Feldmann}, R. 2024, \apjl, 967, L41, \dodoi{10.3847/2041-8213/ad4cee}

\bibitem[{{Mason} {et~al.}(2023){Mason}, {Trenti}, \& {Treu}}]{Mason23}
{Mason}, C.~A., {Trenti}, M., \& {Treu}, T. 2023, \mnras, 521, 497, \dodoi{10.1093/mnras/stad035}

\bibitem[{{Matsumoto} {et~al.}(2025){Matsumoto}, {Sommovigo}, {Gebek}, {Nagamine}, {Nersesian}, {Baes}, {De Looze}, {van der Wel}, {Somerville}, {Romano}, \& {Cochrane}}]{Matsumoto25}
{Matsumoto}, K., {Sommovigo}, L., {Gebek}, A., {et~al.} 2025, arXiv e-prints, arXiv:2508.21157, \dodoi{10.48550/arXiv.2508.21157}

\bibitem[{{Matsuura} {et~al.}(2015){Matsuura}, {Dwek}, {Barlow}, {Babler}, {Baes}, {Meixner}, {Cernicharo}, {Clayton}, {Dunne}, {Fransson}, {Fritz}, {Gear}, {Gomez}, {Groenewegen}, {Indebetouw}, {Ivison}, {Jerkstrand}, {Lebouteiller}, {Lim}, {Lundqvist}, {Pearson}, {Roman-Duval}, {Royer}, {Staveley-Smith}, {Swinyard}, {van Hoof}, {van Loon}, {Verstappen}, {Wesson}, {Zanardo}, {Blommaert}, {Decin}, {Reach}, {Sonneborn}, {Van de Steene}, \& {Yates}}]{Matsuura15}
{Matsuura}, M., {Dwek}, E., {Barlow}, M.~J., {et~al.} 2015, \apj, 800, 50, \dodoi{10.1088/0004-637X/800/1/50}

\bibitem[{{Matteri} {et~al.}(2025){Matteri}, {Pallottini}, \& {Ferrara}}]{Matteri25}
{Matteri}, A., {Pallottini}, A., \& {Ferrara}, A. 2025, \aap, 697, A65, \dodoi{10.1051/0004-6361/202553701}

\bibitem[{{Menon} {et~al.}(2025){Menon}, {Burkhart}, {Somerville}, {Thompson}, \& {Sternberg}}]{Menon_25}
{Menon}, S.~H., {Burkhart}, B., {Somerville}, R.~S., {Thompson}, T.~A., \& {Sternberg}, A. 2025, \apj, 987, 12, \dodoi{10.3847/1538-4357/add2f9}

\bibitem[{Morales {et~al.}(2024)Morales, Finkelstein, Leung, Bagley, Cleri, Dave, Dickinson, Ferguson, Hathi, Jones, Koekemoer, Papovich, Pérez-González, Pirzkal, Smith, Wilkins, \& Yung}]{Morales_2024}
Morales, A.~M., Finkelstein, S.~L., Leung, G. C.~K., {et~al.} 2024, The Astrophysical Journal Letters, 964, L24, \dodoi{10.3847/2041-8213/ad2de4}

\bibitem[{{Morishita} {et~al.}(2024){Morishita}, {Stiavelli}, {Chary}, {Trenti}, {Bergamini}, {Chiaberge}, {Leethochawalit}, {Roberts-Borsani}, {Shen}, \& {Treu}}]{Morishita_2024}
{Morishita}, T., {Stiavelli}, M., {Chary}, R.-R., {et~al.} 2024, \apj, 963, 9, \dodoi{10.3847/1538-4357/ad1404}

\bibitem[{Naidu {et~al.}(2022)Naidu, Oesch, van Dokkum, Nelson, Suess, Whitaker, Allen, Bezanson, Bouwens, Brammer, Conroy, Illingworth, Labbe, Leja, Leonova, Matthee, Price, Setton, Strait, Stefanon, Tacchella, Toft, Weaver, \& Weibel}]{Naidu22}
Naidu, R.~P., Oesch, P.~A., van Dokkum, P., {et~al.} 2022, Two Remarkably Luminous Galaxy Candidates at $z\approx11-13$ Revealed by JWST,  arXiv, \dodoi{10.48550/ARXIV.2207.09434}

\bibitem[{{Naidu} {et~al.}(2025){Naidu}, {Oesch}, {Brammer}, {Weibel}, {Li}, {Matthee}, {Chisholm}, {Pollock}, {Heintz}, {Johnson}, {Shen}, {Hviding}, {Leja}, {Tacchella}, {Ganguly}, {Witten}, {Atek}, {Belli}, {Bose}, {Bouwens}, {Dayal}, {Decarli}, {de Graaff}, {Fudamoto}, {Giovinazzo}, {Greene}, {Illingworth}, {Inoue}, {Kane}, {Labbe}, {Leonova}, {Marques-Chaves}, {Meyer}, {Nelson}, {Roberts-Borsani}, {Schaerer}, {Simcoe}, {Stefanon}, {Sugahara}, {Toft}, {van der Wel}, {van Dokkum}, {Walter}, {Watson}, {Weaver}, \& {Whitaker}}]{Naidu25}
{Naidu}, R.~P., {Oesch}, P.~A., {Brammer}, G., {et~al.} 2025, arXiv e-prints, arXiv:2505.11263, \dodoi{10.48550/arXiv.2505.11263}

\bibitem[{{Nakajima} {et~al.}(2023){Nakajima}, {Ouchi}, {Isobe}, {Harikane}, {Zhang}, {Ono}, {Umeda}, \& {Oguri}}]{Nakajima23}
{Nakajima}, K., {Ouchi}, M., {Isobe}, Y., {et~al.} 2023, \apjs, 269, 33, \dodoi{10.3847/1538-4365/acd556}

\bibitem[{{Narayanan} {et~al.}(2018){Narayanan}, {Conroy}, {Dav{\'e}}, {Johnson}, \& {Popping}}]{Narayanan18}
{Narayanan}, D., {Conroy}, C., {Dav{\'e}}, R., {Johnson}, B.~D., \& {Popping}, G. 2018, \apj, 869, 70, \dodoi{10.3847/1538-4357/aaed25}

\bibitem[{{Narayanan} {et~al.}(2024){Narayanan}, {Lower}, {Torrey}, {Brammer}, {Cui}, {Dav{\'e}}, {Iyer}, {Li}, {Lovell}, {Sales}, {Stark}, {Marinacci}, \& {Vogelsberger}}]{Narayanan24_Mstar}
{Narayanan}, D., {Lower}, S., {Torrey}, P., {et~al.} 2024, \apj, 961, 73, \dodoi{10.3847/1538-4357/ad0966}

\bibitem[{{Narayanan} {et~al.}(2025){Narayanan}, {Torrey}, {Stark}, {Chisholm}, {Finkelstein}, {Garcia}, {Kelley-Derzon}, {Marinacci}, {Sales}, {Savitch}, {Vogelsberger}, \& {Zimmerman}}]{Narayanan25}
{Narayanan}, D., {Torrey}, P., {Stark}, D., {et~al.} 2025, arXiv e-prints, arXiv:2509.18266, \dodoi{10.48550/arXiv.2509.18266}

\bibitem[{{Nozawa} {et~al.}(2010){Nozawa}, {Kozasa}, {Tominaga}, {Maeda}, {Umeda}, {Nomoto}, \& {Krause}}]{Nozawa10}
{Nozawa}, T., {Kozasa}, T., {Tominaga}, N., {et~al.} 2010, \apj, 713, 356, \dodoi{10.1088/0004-637X/713/1/356}

\bibitem[{{Ormerod} {et~al.}(2025){Ormerod}, {Witstok}, {Smit}, {de Graaff}, {Helton}, {Maseda}, {Shivaei}, {Bunker}, {Carniani}, {D'Eugenio}, {Bhatawdekar}, {Chevallard}, {Franx}, {Kumari}, {Maiolino}, {Rinaldi}, {Robertson}, \& {Tacchella}}]{Ormerod25}
{Ormerod}, K., {Witstok}, J., {Smit}, R., {et~al.} 2025, \mnras, 542, 1136, \dodoi{10.1093/mnras/staf1228}

\bibitem[{Ostriker {et~al.}(2001)Ostriker, Stone, \& Gammie}]{Ostriker2001}
Ostriker, E.~C., Stone, J.~M., \& Gammie, C.~F. 2001, The Astrophysical Journal, 546, 980, \dodoi{10.1086/318290}

\bibitem[{{Pallottini} \& {Ferrara}(2023)}]{Pallottini23}
{Pallottini}, A., \& {Ferrara}, A. 2023, \aap, 677, L4, \dodoi{10.1051/0004-6361/202347384}

\bibitem[{{Pallottini} {et~al.}(2024){Pallottini}, {Ferrara}, {Gallerani}, {Sommovigo}, {Carniani}, {Vallini}, {Kohandel}, \& {Venturi}}]{Pallottini24}
{Pallottini}, A., {Ferrara}, A., {Gallerani}, S., {et~al.} 2024, arXiv e-prints, arXiv:2408.00061, \dodoi{10.48550/arXiv.2408.00061}

\bibitem[{{Pallottini} {et~al.}(2022){Pallottini}, {Ferrara}, {Gallerani}, {Behrens}, {Kohandel}, {Carniani}, {Vallini}, {Salvadori}, {Gelli}, {Sommovigo}, {D'Odorico}, {Di Mascia}, \& {Pizzati}}]{Pallottini22}
---. 2022, \mnras, 513, 5621, \dodoi{10.1093/mnras/stac1281}

\bibitem[{{Park} {et~al.}(2019){Park}, {Mesinger}, {Greig}, \& {Gillet}}]{Park18}
{Park}, J., {Mesinger}, A., {Greig}, B., \& {Gillet}, N. 2019, \mnras, 484, 933, \dodoi{10.1093/mnras/stz032}

\bibitem[{{Pathak} {et~al.}(2024){Pathak}, {Leroy}, {Thompson}, {Lopez}, {Belfiore}, {Boquien}, {Dale}, {Glover}, {Klessen}, {Koch}, {Rosolowsky}, {Sandstrom}, {Schinnerer}, {Smith}, {Sun}, {Sutter}, {Williams}, {Bigiel}, {Cao}, {Chastenet}, {Chevance}, {Chown}, {Emsellem}, {Faesi}, {Larson}, {Lee}, {Meidt}, {Ostriker}, {Ramambason}, {Sarbadhicary}, \& {Thilker}}]{Pathak_2024}
{Pathak}, D., {Leroy}, A.~K., {Thompson}, T.~A., {et~al.} 2024, \aj, 167, 39, \dodoi{10.3847/1538-3881/ad110d}

\bibitem[{{Popping}(2022)}]{Popping22}
{Popping}, G. 2022, arXiv e-prints, arXiv:2208.13072.
\newblock \doarXiv{2208.13072}

\bibitem[{{Popping} {et~al.}(2017){Popping}, {Somerville}, \& {Galametz}}]{Popping17}
{Popping}, G., {Somerville}, R.~S., \& {Galametz}, M. 2017, \mnras, 471, 3152, \dodoi{10.1093/mnras/stx1545}

\bibitem[{{Raskutti} {et~al.}(2017){Raskutti}, {Ostriker}, \& {Skinner}}]{Raskutti_2017}
{Raskutti}, S., {Ostriker}, E.~C., \& {Skinner}, M.~A. 2017, \apj, 850, 112, \dodoi{10.3847/1538-4357/aa965e}

\bibitem[{{Rizzo} {et~al.}(2024){Rizzo}, {Bacchini}, {Kohandel}, {Di Mascolo}, {Fraternali}, {Roman-Oliveira}, {Zanella}, {Popping}, {Valentino}, {Magdis}, \& {Whitaker}}]{Rizzo24}
{Rizzo}, F., {Bacchini}, C., {Kohandel}, M., {et~al.} 2024, \aap, 689, A273, \dodoi{10.1051/0004-6361/202450455}

\bibitem[{{Robertson} {et~al.}(2024){Robertson}, {Johnson}, {Tacchella}, {Eisenstein}, {Hainline}, {Arribas}, {Baker}, {Bunker}, {Carniani}, {Cargile}, {Carreira}, {Charlot}, {Chevallard}, {Curti}, {Curtis-Lake}, {D'Eugenio}, {Egami}, {Hausen}, {Helton}, {Jakobsen}, {Ji}, {Jones}, {Maiolino}, {Maseda}, {Nelson}, {P{\'e}rez-Gonz{\'a}lez}, {Pusk{\'a}s}, {Rieke}, {Smit}, {Sun}, {{\"U}bler}, {Whitler}, {Williams}, {Willmer}, {Willott}, \& {Witstok}}]{Robertson24}
{Robertson}, B., {Johnson}, B.~D., {Tacchella}, S., {et~al.} 2024, \apj, 970, 31, \dodoi{10.3847/1538-4357/ad463d}

\bibitem[{{Robertson} {et~al.}(2023){Robertson}, {Tacchella}, {Johnson}, {Hainline}, {Whitler}, {Eisenstein}, {Endsley}, {Rieke}, {Stark}, {Alberts}, {Dressler}, {Egami}, {Hausen}, {Rieke}, {Shivaei}, {Williams}, {Willmer}, {Arribas}, {Bonaventura}, {Bunker}, {Cameron}, {Carniani}, {Charlot}, {Chevallard}, {Curti}, {Curtis-Lake}, {D'Eugenio}, {Jakobsen}, {Looser}, {L{\"u}tzgendorf}, {Maiolino}, {Maseda}, {Rawle}, {Rix}, {Smit}, {{\"U}bler}, {Willott}, {Witstok}, {Baum}, {Bhatawdekar}, {Boyett}, {Chen}, {de Graaff}, {Florian}, {Helton}, {Hviding}, {Ji}, {Kumari}, {Lyu}, {Nelson}, {Sandles}, {Saxena}, {Suess}, {Sun}, {Topping}, \& {Wallace}}]{Robertson23}
{Robertson}, B.~E., {Tacchella}, S., {Johnson}, B.~D., {et~al.} 2023, Nature Astronomy, 7, 611, \dodoi{10.1038/s41550-023-01921-1}

\bibitem[{{Rowland} {et~al.}(2025){Rowland}, {Stefanon}, {Bouwens}, {Hodge}, {Algera}, {Fisher}, {Dayal}, {Pallottini}, {Stark}, {Heintz}, {Aravena}, {Bowler}, {Cescon}, {Endsley}, {Ferrara}, {Gonzalez}, {Graziani}, {Gulis}, {Herard-Demanche}, {Inami}, {Laza-Ramos}, {van Leeuwen}, {de Looze}, {Nanayakkara}, {Oesch}, {Ormerod}, {Sartorio}, {Schouws}, {Smit}, {Sommovigo}, {Toft}, {Weaver}, \& {van der Werf}}]{Rowland25}
{Rowland}, L.~E., {Stefanon}, M., {Bouwens}, R., {et~al.} 2025, arXiv e-prints, arXiv:2501.10559, \dodoi{10.48550/arXiv.2501.10559}

\bibitem[{{Salim} {et~al.}(2018){Salim}, {Boquien}, \& {Lee}}]{Salim18}
{Salim}, S., {Boquien}, M., \& {Lee}, J.~C. 2018, \apj, 859, 11, \dodoi{10.3847/1538-4357/aabf3c}

\bibitem[{{Salim} \& {Narayanan}(2020)}]{Salim20}
{Salim}, S., \& {Narayanan}, D. 2020, \araa, 58, 529, \dodoi{10.1146/annurev-astro-032620-021933}

\bibitem[{{Sarangi}(2022)}]{Sarangi22}
{Sarangi}, A. 2022, \aap, 668, A57, \dodoi{10.1051/0004-6361/202244391}

\bibitem[{{Sarangi} {et~al.}(2018){Sarangi}, {Matsuura}, \& {Micelotta}}]{Sarangi_18}
{Sarangi}, A., {Matsuura}, M., \& {Micelotta}, E.~R. 2018, \ssr, 214, 63, \dodoi{10.1007/s11214-018-0492-7}

\bibitem[{Sarangi {et~al.}(2025)Sarangi, Zsíros, Szalai, Martinez, Shahbandeh, Fox, Van~Dyk, Filippenko, Bersten, De~Looze, Ashall, Temim, Jencson, Rest, Milisavljevic, Dessart, Dwek, Smith, Tinyanont, Brink, Zheng, Clayton, \& Andrews}]{Sarangi_2025}
Sarangi, A., Zsíros, S., Szalai, T., {et~al.} 2025, The Astrophysical Journal, 993, 94, \dodoi{10.3847/1538-4357/ae0645}

\bibitem[{{Schneider} {et~al.}(2004){Schneider}, {Ferrara}, \& {Salvaterra}}]{Schneider04}
{Schneider}, R., {Ferrara}, A., \& {Salvaterra}, R. 2004, \mnras, 351, 1379, \dodoi{10.1111/j.1365-2966.2004.07876.x}

\bibitem[{{Schneider} \& {Maiolino}(2023)}]{Schneider23}
{Schneider}, R., \& {Maiolino}, R. 2023, arXiv e-prints, arXiv:2310.00053, \dodoi{10.48550/arXiv.2310.00053}

\bibitem[{{Schneider} \& {Maiolino}(2024)}]{Schneider24}
---. 2024, \aapr, 32, 2, \dodoi{10.1007/s00159-024-00151-2}

\bibitem[{Schouws {et~al.}(2024)Schouws, Bouwens, Ormerod, Smit, Algera, Sommovigo, Hodge, Ferrara, Oesch, Rowland, van Leeuwen, Stefanon, Herard-Demanche, Fudamoto, Röttgering, \& van~der Werf}]{Schouws24}
Schouws, S., Bouwens, R.~J., Ormerod, K., {et~al.} 2024, Detection of [OIII]88$\mu$m in JADES-GS-z14-0 at z=14.1793.
\newblock \doarXiv{2409.20549}

\bibitem[{{Seon} \& {Draine}(2016)}]{SeonDraine16}
{Seon}, K.-I., \& {Draine}, B.~T. 2016, \apj, 833, 201, \dodoi{10.3847/1538-4357/833/2/201}

\bibitem[{{Shahbandeh} {et~al.}(2023){Shahbandeh}, {Sarangi}, {Temim}, {Szalai}, {Fox}, {Tinyanont}, {Dwek}, {Dessart}, {Filippenko}, {Brink}, {Foley}, {Jencson}, {Pierel}, {Zs{\'\i}ros}, {Rest}, {Zheng}, {Andrews}, {Clayton}, {De}, {Engesser}, {Gezari}, {Gomez}, {Gonzaga}, {Johansson}, {Kasliwal}, {Lau}, {De Looze}, {Marston}, {Milisavljevic}, {O'Steen}, {Siebert}, {Skrutskie}, {Smith}, {Strolger}, {Van Dyk}, {Wang}, {Williams}, {Williams}, {Xiao}, \& {Yang}}]{Shahbandeh23}
{Shahbandeh}, M., {Sarangi}, A., {Temim}, T., {et~al.} 2023, \mnras, 523, 6048, \dodoi{10.1093/mnras/stad1681}

\bibitem[{{Shivaei} {et~al.}(2025){Shivaei}, {Naidu}, {Rodr{\'\i}guez Montero}, {Matsumoto}, {Leja}, {Matthee}, {Johnson}, {Oesch}, {Chevallard}, {Adamo}, {Bodansky}, {Bunker}, {Covelo Paz}, {Di Cesare}, {Egami}, {Furtak}, {Heintz}, {Kramarenko}, {Meyer}, {Reddy}, {Rinaldi}, {Tacchella}, {Torralba}, {Witstok}, {Wozniak}, \& {Xiao}}]{Shivaei25}
{Shivaei}, I., {Naidu}, R.~P., {Rodr{\'\i}guez Montero}, F., {et~al.} 2025, arXiv e-prints, arXiv:2509.01795, \dodoi{10.48550/arXiv.2509.01795}

\bibitem[{{Skibba} {et~al.}(2011){Skibba}, {Engelbracht}, {Dale}, {Hinz}, {Zibetti}, {Crocker}, {Groves}, {Hunt}, {Johnson}, {Meidt}, {Murphy}, {Appleton}, {Armus}, {Bolatto}, {Brandl}, {Calzetti}, {Croxall}, {Galametz}, {Gordon}, {Kennicutt}, {Koda}, {Krause}, {Montiel}, {Rix}, {Roussel}, {Sandstrom}, {Sauvage}, {Schinnerer}, {Smith}, {Walter}, {Wilson}, \& {Wolfire}}]{Kingfish_2011}
{Skibba}, R.~A., {Engelbracht}, C.~W., {Dale}, D., {et~al.} 2011, \apj, 738, 89, \dodoi{10.1088/0004-637X/738/1/89}

\bibitem[{{Slavin} {et~al.}(2020){Slavin}, {Dwek}, {Mac Low}, \& {Hill}}]{Slavin20}
{Slavin}, J.~D., {Dwek}, E., {Mac Low}, M.-M., \& {Hill}, A.~S. 2020, \apj, 902, 135, \dodoi{10.3847/1538-4357/abb5a4}

\bibitem[{{Somerville} {et~al.}(2025){Somerville}, {Yung}, {Lancaster}, {Menon}, {Sommovigo}, \& {Finkelstein}}]{Somerville25}
{Somerville}, R.~S., {Yung}, L.~Y.~A., {Lancaster}, L., {et~al.} 2025, \mnras, \dodoi{10.1093/mnras/staf1824}

\bibitem[{{Sommovigo} \& {Algera}(2025)}]{Sommovigo25_Td}
{Sommovigo}, L., \& {Algera}, H. 2025, \mnras, 540, 3693, \dodoi{10.1093/mnras/staf897}

\bibitem[{{Sommovigo} {et~al.}(2021){Sommovigo}, {Ferrara}, {Carniani}, {Zanella}, {Pallottini}, {Gallerani}, \& {Vallini}}]{Sommovigo21}
{Sommovigo}, L., {Ferrara}, A., {Carniani}, S., {et~al.} 2021, \mnras, 503, 4878, \dodoi{10.1093/mnras/stab720}

\bibitem[{{Sommovigo} {et~al.}(2020){Sommovigo}, {Ferrara}, {Pallottini}, {Carniani}, {Gallerani}, \& {Decataldo}}]{Sommovigo20}
{Sommovigo}, L., {Ferrara}, A., {Pallottini}, A., {et~al.} 2020, \mnras, 497, 956, \dodoi{10.1093/mnras/staa1959}

\bibitem[{{Sommovigo} {et~al.}(2022{\natexlab{a}}){Sommovigo}, {Ferrara}, {Pallottini}, {Dayal}, {Bouwens}, {Smit}, {da Cunha}, {De Looze}, {Bowler}, {Hodge}, {Inami}, {Oesch}, {Endsley}, {Gonzalez}, {Schouws}, {Stark}, {Stefanon}, {Aravena}, {Graziani}, {Riechers}, {Schneider}, {van der Werf}, {Algera}, {Barrufet}, {Fudamoto}, {Hygate}, {Labb{\'e}}, {Li}, {Nanayakkara}, \& {Topping}}]{Sommovigo22_z7}
---. 2022{\natexlab{a}}, \mnras, 513, 3122, \dodoi{10.1093/mnras/stac302}

\bibitem[{{Sommovigo} {et~al.}(2022{\natexlab{b}}){Sommovigo}, {Ferrara}, {Carniani}, {Pallottini}, {Dayal}, {Pizzati}, {Ginolfi}, {Markov}, \& {Faisst}}]{Sommovigo22_z57}
{Sommovigo}, L., {Ferrara}, A., {Carniani}, S., {et~al.} 2022{\natexlab{b}}, \mnras, 517, 5930, \dodoi{10.1093/mnras/stac2997}

\bibitem[{{Sommovigo} {et~al.}(2025){Sommovigo}, {Cochrane}, {Somerville}, {Hayward}, {Lovell}, {Starkenburg}, {Popping}, {Iyer}, {Gabrielpillai}, {Ho}, {Steinwandel}, \& {Perez}}]{Sommovigo25_TNG}
{Sommovigo}, L., {Cochrane}, R.~K., {Somerville}, R.~S., {et~al.} 2025, \apj, 990, 114, \dodoi{10.3847/1538-4357/addec1}

\bibitem[{{Sun} {et~al.}(2023{\natexlab{a}}){Sun}, {Faucher-Gigu{\`e}re}, {Hayward}, {Shen}, {Wetzel}, \& {Cochrane}}]{Sun23}
{Sun}, G., {Faucher-Gigu{\`e}re}, C.-A., {Hayward}, C.~C., {et~al.} 2023{\natexlab{a}}, \apjl, 955, L35, \dodoi{10.3847/2041-8213/acf85a}

\bibitem[{{Sun} {et~al.}(2023{\natexlab{b}}){Sun}, {Faucher-Gigu{\`e}re}, {Hayward}, {Shen}, {Wetzel}, \& {Cochrane}}]{Sun23b}
---. 2023{\natexlab{b}}, \apjl, 955, L35, \dodoi{10.3847/2041-8213/acf85a}

\bibitem[{{Szalai} {et~al.}(2025){Szalai}, {Zs{\'\i}ros}, {Jencson}, {Fox}, {Shahbandeh}, {Sarangi}, {Temim}, {De Looze}, {Smith}, {Filippenko}, {Van Dyk}, {Andrews}, {Ashall}, {Clayton}, {Dessart}, {Dulude}, {Dwek}, {Gomez}, {Johansson}, {Milisavljevic}, {Pierel}, {Rest}, {Tinyanont}, {Brink}, {De}, {Engesser}, {Foley}, {Gezari}, {Kasliwal}, {Lau}, {Marston}, {O'Steen}, {Siebert}, {Skrutskie}, {Strolger}, {Wang}, {Williams}, {Williams}, {Xiao}, \& {Zheng}}]{Szalai25}
{Szalai}, T., {Zs{\'\i}ros}, S., {Jencson}, J., {et~al.} 2025, \aap, 697, A132, \dodoi{10.1051/0004-6361/202451470}

\bibitem[{{Tacchella} {et~al.}(2023){Tacchella}, {Eisenstein}, {Hainline}, {Johnson}, {Baker}, {Helton}, {Robertson}, {Suess}, {Chen}, {Nelson}, {Pusk{\'a}s}, {Sun}, {Alberts}, {Egami}, {Hausen}, {Rieke}, {Rieke}, {Shivaei}, {Williams}, {Willmer}, {Bunker}, {Cameron}, {Carniani}, {Charlot}, {Curti}, {Curtis-Lake}, {Looser}, {Maiolino}, {Maseda}, {Rawle}, {Rix}, {Smit}, {{\"U}bler}, {Willott}, {Witstok}, {Baum}, {Bhatawdekar}, {Boyett}, {Danhaive}, {de Graaff}, {Endsley}, {Ji}, {Lyu}, {Sandles}, {Saxena}, {Scholtz}, {Topping}, \& {Whitler}}]{Tacchella23}
{Tacchella}, S., {Eisenstein}, D.~J., {Hainline}, K., {et~al.} 2023, \apj, 952, 74, \dodoi{10.3847/1538-4357/acdbc6}

\bibitem[{{Tamura} {et~al.}(2019){Tamura}, {Mawatari}, {Hashimoto}, {Inoue}, {Zackrisson}, {Christensen}, {Binggeli}, {Matsuda}, {Matsuo}, {Takeuchi}, {Asano}, {Sunaga}, {Shimizu}, {Okamoto}, {Yoshida}, {Lee}, {Shibuya}, {Taniguchi}, {Umehata}, {Hatsukade}, {Kohno}, \& {Ota}}]{Tamura19}
{Tamura}, Y., {Mawatari}, K., {Hashimoto}, T., {et~al.} 2019, \apj, 874, 27, \dodoi{10.3847/1538-4357/ab0374}

\bibitem[{{Thompson} \& {Krumholz}(2016)}]{Thompson16}
{Thompson}, T.~A., \& {Krumholz}, M.~R. 2016, \mnras, 455, 334, \dodoi{10.1093/mnras/stv2331}

\bibitem[{{Todini} \& {Ferrara}(2001)}]{Todini01}
{Todini}, P., \& {Ferrara}, A. 2001, \mnras, 325, 726, \dodoi{10.1046/j.1365-8711.2001.04486.x}

\bibitem[{{Topping} {et~al.}(2022{\natexlab{a}}){Topping}, {Stark}, {Endsley}, {Plat}, {Whitler}, {Chen}, \& {Charlot}}]{Topping_2022}
{Topping}, M.~W., {Stark}, D.~P., {Endsley}, R., {et~al.} 2022{\natexlab{a}}, \apj, 941, 153, \dodoi{10.3847/1538-4357/aca522}

\bibitem[{{Topping} {et~al.}(2022{\natexlab{b}}){Topping}, {Stark}, {Endsley}, {Bouwens}, {Schouws}, {Smit}, {Stefanon}, {Inami}, {Bowler}, {Oesch}, {Gonzalez}, {Dayal}, {da Cunha}, {Algera}, {van der Werf}, {Pallottini}, {Barrufet}, {Schneider}, {De Looze}, {Sommovigo}, {Whitler}, {Graziani}, {Fudamoto}, \& {Ferrara}}]{topping2022}
---. 2022{\natexlab{b}}, \mnras, 516, 975, \dodoi{10.1093/mnras/stac2291}

\bibitem[{{Trayford} {et~al.}(2020){Trayford}, {Lagos}, {Robotham}, \& {Obreschkow}}]{Trayford20}
{Trayford}, J.~W., {Lagos}, C. d.~P., {Robotham}, A. S.~G., \& {Obreschkow}, D. 2020, \mnras, 491, 3937, \dodoi{10.1093/mnras/stz3234}

\bibitem[{Trinca {et~al.}(2024)Trinca, Schneider, Valiante, Graziani, Ferrotti, Omukai, \& Chon}]{Trinca24}
Trinca, A., Schneider, R., Valiante, R., {et~al.} 2024, Monthly Notices of the Royal Astronomical Society, 529, 3563, \dodoi{10.1093/mnras/stae651}

\bibitem[{{Valiante} {et~al.}(2009){Valiante}, {Schneider}, {Bianchi}, \& {Andersen}}]{Valiante09}
{Valiante}, R., {Schneider}, R., {Bianchi}, S., \& {Andersen}, A.~C. 2009, \mnras, 397, 1661, \dodoi{10.1111/j.1365-2966.2009.15076.x}

\bibitem[{{V{\'a}zquez-Semadeni} \& {Garc{\'\i}a}(2001)}]{Semadeni2001}
{V{\'a}zquez-Semadeni}, E., \& {Garc{\'\i}a}, N. 2001, \apj, 557, 727, \dodoi{10.1086/321688}

\bibitem[{Városi \& Dwek(1999)}]{Városi_1999}
Városi, F., \& Dwek, E. 1999, The Astrophysical Journal, 523, 265, \dodoi{10.1086/307729}

\bibitem[{Wang {et~al.}(2023)Wang, Fujimoto, Labbé, Furtak, Miller, Setton, Zitrin, Atek, Bezanson, Brammer, Leja, Oesch, Price, Chemerynska, Cutler, Dayal, van Dokkum, Goulding, Greene, Fudamoto, Khullar, Kokorev, Marchesini, Pan, Weaver, Whitaker, \& Williams}]{Wang23}
Wang, B., Fujimoto, S., Labbé, I., {et~al.} 2023, The Astrophysical Journal Letters, 957, L34, \dodoi{10.3847/2041-8213/acfe07}

\bibitem[{{Wang} {et~al.}(2023){Wang}, {Lei}, {Yuan}, \& {Fan}}]{Wang_2023}
{Wang}, Y.-Y., {Lei}, L., {Yuan}, G.-W., \& {Fan}, Y.-Z. 2023, \apjl, 954, L48, \dodoi{10.3847/2041-8213/acf46c}

\bibitem[{{Weingartner} \& {Draine}(2001)}]{Weingartner01}
{Weingartner}, J.~C., \& {Draine}, B.~T. 2001, \apj, 548, 296, \dodoi{10.1086/318651}

\bibitem[{{Witstok} {et~al.}(2023){Witstok}, {Jones}, {Maiolino}, {Smit}, \& {Schneider}}]{Witstok23}
{Witstok}, J., {Jones}, G.~C., {Maiolino}, R., {Smit}, R., \& {Schneider}, R. 2023, \mnras, 523, 3119, \dodoi{10.1093/mnras/stad1470}

\bibitem[{Xu {et~al.}(2025)Xu, Ouchi, Nakajima, Harikane, Isobe, Ono, Umeda, \& Zhang}]{Xu25}
Xu, Y., Ouchi, M., Nakajima, K., {et~al.} 2025, The Astrophysical Journal, 984, 182, \dodoi{10.3847/1538-4357/adc733}

\bibitem[{{Yoon} {et~al.}(2022){Yoon}, {Carilli}, {Fujimoto}, {Castellano}, {Merlin}, {Santini}, {Yun}, {Murphy}, {Jung}, {Casey}, {Finkelstein}, {Papovich}, {Fontana}, {Treu}, \& {Letai}}]{Yoon22}
{Yoon}, I., {Carilli}, C.~L., {Fujimoto}, S., {et~al.} 2022, arXiv e-prints, arXiv:2210.08413.
\newblock \doarXiv{2210.08413}

\bibitem[{{Yung} {et~al.}(2024){Yung}, {Somerville}, {Finkelstein}, {Wilkins}, \& {Gardner}}]{Yung24}
{Yung}, L.~Y.~A., {Somerville}, R.~S., {Finkelstein}, S.~L., {Wilkins}, S.~M., \& {Gardner}, J.~P. 2024, \mnras, 527, 5929, \dodoi{10.1093/mnras/stad3484}

\bibitem[{{Yung} {et~al.}(2025){Yung}, {Somerville}, \& {Iyer}}]{Yung25}
{Yung}, L.~Y.~A., {Somerville}, R.~S., \& {Iyer}, K.~G. 2025, \mnras, 543, 3802, \dodoi{10.1093/mnras/staf1699}

\bibitem[{{Yung} {et~al.}(2023){Yung}, {Somerville}, {Nguyen}, {Behroozi}, {Modi}, \& {Gardner}}]{Yung23}
{Yung}, L.~Y.~A., {Somerville}, R.~S., {Nguyen}, T., {et~al.} 2023, arXiv e-prints, arXiv:2309.14408, \dodoi{10.48550/arXiv.2309.14408}

\bibitem[{{Zavala} {et~al.}(2024){Zavala}, {Castellano}, {Akins}, {Bakx}, {Burgarella}, {Casey}, {Ch{\'a}vez Ortiz}, {Dickinson}, {Finkelstein}, {Mitsuhashi}, {Nakajima}, {P{\'e}rez-Gonz{\'a}lez}, {Arrabal Haro}, {Buat}, {Backhaus}, {Calabr{\`o}}, {Cleri}, {Fern{\'a}ndez-Arenas}, {Fontana}, {Franco}, {Giavalisco}, {Grogin}, {Hathi}, {Hirschmann}, {Ikeda}, {Jung}, {Kartaltepe}, {Koekemoer}, {Larson}, {McKinney}, {Papovich}, {Saito}, {Santini}, {Terlevich}, {Terlevich}, {Treu}, \& {Yung}}]{Zavala24}
{Zavala}, J.~A., {Castellano}, M., {Akins}, H.~B., {et~al.} 2024, arXiv e-prints, arXiv:2403.10491, \dodoi{10.48550/arXiv.2403.10491}

\bibitem[{{Ziparo} {et~al.}(2023){Ziparo}, {Ferrara}, {Sommovigo}, \& {Kohandel}}]{Ziparo23}
{Ziparo}, F., {Ferrara}, A., {Sommovigo}, L., \& {Kohandel}, M. 2023, \mnras, 520, 2445, \dodoi{10.1093/mnras/stad125}

\bibitem[{Zsíros {et~al.}(2024)Zsíros, Szalai, De Looze, Sarangi, Shahbandeh, Fox, Temim, Milisavljevic, Van Dyk, Smith, Filippenko, Brink, Zheng, Dessart, Jencson, Johansson, Pierel, Rest, Tinyanont, Niculescu-Duvaz, Barlow, Wesson, Andrews, Clayton, De, Dwek, Engesser, Foley, Gezari, Gomez, Gonzaga, Kasliwal, Lau, Marston, O’Steen, Siebert, Skrutskie, Strolger, Wang, Williams, Williams, \& Xiao}]{Zsíros24}
Zsíros, S., Szalai, T., De Looze, I., {et~al.} 2024, Monthly Notices of the Royal Astronomical Society, 529, 155, \dodoi{10.1093/mnras/stae507}

\end{thebibliography}
